\documentclass{cernrep}

\usepackage{fancyhdr}
\usepackage[colorlinks=true, linkcolor=black, citecolor=black, filecolor=black, urlcolor=blue]{hyperref}
\fancyhfoffset{4 mm}
\fancypagestyle{ARTTITLE}{%
\fancyhf{} % clear all header and footer fields
\lhead{\small{Proceedings of the CERN--Accelerator--School course: \it{Introduction to Accelerator Physics}}}
\lfoot{Available online at \url{https://cas.web.cern.ch/previous-schools}}
\rfoot{\thepage\hspace*{3mm}}

}

\begin{document}
\title{Transverse Linear Beam Dynamics}
\author{W. Hillert}
\institute{University of Hamburg, Institute of Experimental Physics, Hamburg, Germany}

\begin{abstract}
The subject of this introductory course is transverse dynamics of charged particle beams in linear approximation.
Starting with a discussion of the most important types of magnets and defining their multipole strengths, the linearized equations of motion of charged particles in static magnetic fields are derived using an orthogonal reference frame following the design orbit.
Analytical solutions are determined for linear elements of a typical beam transfer line (drift, dipole and quadrupole magnets), and stepwise combined by introducing the matrix formalism in which each element's contribution is represented by a~single transfer matrix.
Application of this formalism allows to calculate single particle's trajectories in linear approximation. 
After introducing the beam emittance as the area occupied by a particle beam in phase space, a linear treatment of transverse beam dynamics based on appropriately defined optical functions is introduced. 
The formalism is applied to the concepts of both weak and strong focusing, in particular discussing the properties of the widely-used FODO cell.
Specific characteristics of transverse beam dynamics in periodic systems like circular accelerators are studied in detail, emphazising the effects of linear field errors on orbit stability and introducing the phenomena of optical resonances.
Finally, the dynamics of off-momentum particles is presented, introducing dispersion functions and explaining effects like chromaticity.
\end{abstract}


\maketitle

\thispagestyle{ARTTITLE}

%===========================================================================
%===========================================================================
\section{Introduction}
\label{sec:Introduction}
\begin{figure}[ht]
\begin{center}
\includegraphics[width=16cm]{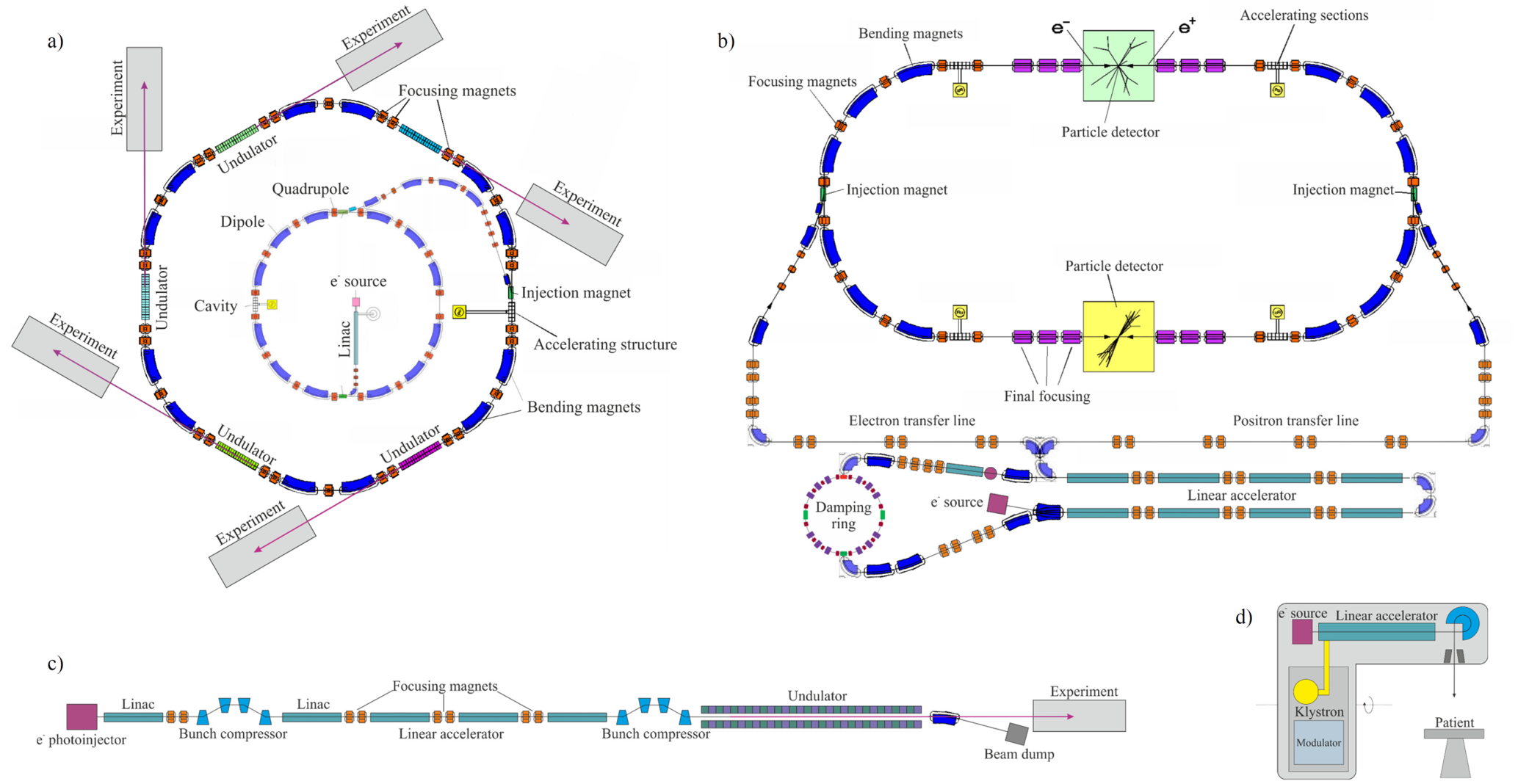}
\caption{Sketches of different types of accelerators (not to scale). a) Synchrotron radiation source. b) High energy collider. c) Free electron laser. d) Linear accelerator for medical radiotherapy.}
\label{fig:types-of-accelerators}
\end{center}
\end{figure}

%===========================================================================
%===========================================================================
We will start looking at typical large-scale accelerators such like high energy colliders, synchrotron radiation sources, free electron lasers, or compact medical accelerators used for treatments for patients with cancer
(cf.\ Fig~\ref{fig:types-of-accelerators}).
%---------------------------------------------------------------------------------
%---------------------------------------------------------------------------------
In all such accelerators a beam of charged particles (electrons, protons, ions) is generated in a dedicated source, preaccelerated in (often a series of) linear and/or circular accelerators which are connected by beam transfer lines and finally either injected into a storage ring, guided through a series of alternating dipole magnets called undulators, or directed to an external target.

In order to guide the beam along its often curved path through the accelerators and beam transfer lines, it has to be deflected by dedicated devices acting on the individual beam particles, a process which is called {\em beam bending}.
Ideally, all particles should then move on the design orbit, starting at the source and ending at the target (or being closed in a storage ring).
In fact, most particles of the beam will deviate slightly from this ideal path and therefore have to be repeatedly bend back towards the desing orbit, a~process which is called {\em beam focusing}.
The resulting movement of the particles under the influence of external forces is described by the beam dynamics.

In this introductory course, we will concentrate on the transverse beam dynamics in linear approximation, thus only treating the impact of beam bending and focusing.
We will not only present single particle's dynamics, but as well discuss the properties of macroscopic beams being formed by an~ensemble of thousands up to many billions of charged particles.
For this purpose, appropriate physical quantities like beam emittance and optical functions will be defined and used for describing the~evolution of the beam's properties.

%_______________________________________________________________________________
%---------------------------------------------------------------------------------------------------------------
\subsection{Magnetic rigidity}
\label{sec:magnetic-rigidity}
%_______________________________________________________________________________
%---------------------------------------------------------------------------------------------------------------
Both bending and focusing forces can be accomplished with electromagnetic fields.
When a particle with charge $q$ and velocity $\vec v$ is moving through electric ($\vec E$) and magnetic ($\vec B$) fields, the Lorentz force 
%------------------
\begin{equation}
\vec F = q (\vec E + \vec v \times \vec B)
\label{eq:Lorentz-force}  
\end{equation}
%------------------
is acting on the particle. 
Whereas only longitudinal electric fields $(\vec E_{\parallel} \parallel \vec v)$ will change the particle's energy, transverse electric $(\vec E_{\perp} \perp \vec v)$ and magnetic $(\vec B_{\perp} \perp \vec v)$ fields will deflect the particle without changing the absolute value of the particle's velocity $\vert \vec v \vert$.
Pure transverse fields will therefore lead to a~circular orbit with bending radius $\rho$ for which the Lorentz force has to equal the centripetal force
%------------------
\begin{equation}
m \frac{v^2}{\rho} = q (E_{\perp} + v B_{\perp})~,
\label{eq:centripetal-force}  
\end{equation}
%------------------
where $m = \gamma_r m_0$ is the relativistic mass of a particle with rest mass $m_0$.

Since ultra-relativistic particles move with a velocity very close to the speed of light ($v \approx c$) the~impact of magnetic fields is enhanced by a remarkable factor of $c = 3 \cdot 10^8\, \frac{\text m}{\text s}$.
Thus the bending force generated by a comparatively easy to produce magnetic field of $B = 1\, \text T$ would require an~equivalent electric field of $E = 300\, \frac{\text MV}{\text m}$ which is far beyond the technical limits.
Hence only transverse magnetic fields are applied for beam deflection and focusing in high energy particle accelerators.
We will hereinafter only regard transverse magnetic fields and skip the index ``{$\perp$}''. 

Introducing the particle's momentum $p = m v$, Eq.~(\ref{eq:centripetal-force}) can be written as
%------------------
\begin{equation}
p = q B \rho~,
\label{eq:magnetic-rigidity}  
\end{equation}
%------------------
giving rise to the definition of a new and important quantity, the {\em magnetic rigidity} or {\em magnetic stiffness} $B \rho$.
Given a bending magnet that is designed to bend a beam on a circular orbit with bending radius $\rho$, its $B \rho$ value indicates the corresponding beam momentum.
This can be used to characterize circular accelerators as it was done for the heavy ion synchrotrons (SIS18, SIS100) of the FAIR accelerator complex, where the numbers indicate the maximum magnetic rigidity in units Tm.

%===========================================================================
%===========================================================================
\section{Multipole expansion of magnetic fields}
\label{sec:multipole-expansion}
%===========================================================================
%===========================================================================
In the following, we will spend some efforts to classify transverse magnetic fields in a quite general way by expanding them into a series of ``multipoles''.
The aim is to gain a detailed understanding of the~impact of a given magnetic field $\vec B$ by regarding the action of its different ``multipole components'' on the beam individually.
We will restrict our treatment to transverse displacements from the design orbit which is defined by the magnetic fields.
In addition, we will assume that there is no dependence of $\vec B$ on the longitudinal coordinate inside a given accelerator magnet.

The transverse plane will be denoted as the $(x,y)$ plane.
Since we are only interested in the~displacements from the ``ideal'' orbit, the reference orbit is put to $(0,0)$.
The transverse components $B_x(x,y), B_y(x,y)$ of the magnetic field will be classified according to their dependence on $x$ and $y$, that is on the deviation from the reference orbit.
The resulting contributions (constant, linear, quadratic, etc.) are called {\em multipole components}.

Typically, the fields of accelerator magnets are generated by excitation coils which are placed outside the beam volume.
Thus, there are no excitation currents in the vicinity of the beam defining our area of interest (meaning at considerably small displacements $x$ and $y$).
Maxwell's equations then require that both the divergence and curl of $B$ will vanish
%------------------
\begin{equation}
\begin{split}
\vec{\nabla} \cdot \vec{B} = 0~, \\
\vec{\nabla} \times \vec{B} = 0~.
\end{split}
\label{eq:Maxwell2,4}  
\end{equation}
%------------------
In general, the magnetic field can be derived from a vector potential $\vec A$.
The vanishing curl of $\vec B$ in addition allows to express $\vec B$ in terms of a scalar potential $\Phi$.
Both $\vec A$ and $\Phi$ are only depending on the~transverse coordinates $x$ and $y$
%------------------
\begin{equation}
\vec B(x,y) = - \vec{\nabla} \Phi(x,y) = \vec{\nabla} \times \vec{A}(x,y)~.
\label{eq:mag-potentials}  
\end{equation}
%------------------
Since $\vec B$ is in the $(x,y)$ plane and $\vec{\nabla} \times \vec A \perp \vec A$ there is only one non-vanishing component $A_z$.
Thus we can represent the two components of the magnetic field by two other scalar quantities, $A_z$ and $\Phi$.
Evaluating Eq.~(\ref{eq:mag-potentials}) yields
%------------------
\begin{equation}
\begin{split}
B_x(x,y) = - \frac{\partial \Phi}{\partial x} = + \frac{\partial A_z}{\partial y}~, \\
B_y(x,y) = - \frac{\partial \Phi}{\partial y} = - \frac{\partial A_z}{\partial x}~.
\end{split}
\label{eq:CR-DGL}  
\end{equation}
%------------------
The two Eqs.~(\ref{eq:CR-DGL}) are the Cauchy-Riemann partial differential equations linking the real and imaginary part of a complex analytical function
%------------------
\begin{equation}
V = A_z + i \Phi~.
\label{eq:complex-potential}  
\end{equation}
%------------------
Using a complex argument $z$ which can de decomposed into cartesian $(x,y)$ or polar $(r,\varphi)$ coordinates 
%------------------
\begin{equation}
z = x + i y = r \cdot e^{i \varphi}~,
\label{eq:complex-real}  
\end{equation}
%------------------
the complex potential can be expressed as a complex power series 
%------------------
\begin{equation}
V(z) = \sum_{n=0}^{\infty}V_n(z) = \sum_{n=0}^{\infty}\left(A_{z,n}+i \Phi_n \right)
= \sum_{n=0}^{\infty}c_n \cdot z^n
= \sum_{n=0}^{\infty}\left(\lambda_n + i \mu_n \right) \cdot z^n~,
\label{eq:complex-power-series}  
\end{equation}
%------------------
where we have decomposed the complex expansion coefficients $c_n$ into their real and imaginary parts $\lambda_n$ and $\mu_n$.
The transverse magnetic field components can be calculated using Eq.~(\ref{eq:CR-DGL}) and can be represented either by their cartesian or their polar components.

%_______________________________________________________________________________
%---------------------------------------------------------------------------------------------------------------
\subsection{Expansion using polar coordinates}
\label{sec:polar-coordinates}
%_______________________________________________________________________________
%---------------------------------------------------------------------------------------------------------------
In polar coordinates we will use the coordinate vector $z = r \cdot e^{i \varphi}$ and
%------------------
\begin{equation}
z^n = r^n \cdot e^{i n \varphi} =  r^n \left(\cos(n \varphi) + i \sin(n \varphi) \right)~.
\label{eq:complex-Euler}  
\end{equation}
%------------------
We therewith can express the summands in Eq.~(\ref{eq:complex-power-series}) as
%------------------
\begin{equation}
V_n(r,\varphi) = \left( \lambda_n + i \mu_n \right) r^n \left(\cos(n \varphi) + i \sin(n \varphi) \right)~.
\label{eq:complex-potential-multipole}  
\end{equation}
%------------------
The magnetic field components are given by the gradient of the scalar potential $\Phi$:
%------------------
\begin{eqnarray}
B_r = - \frac{\partial \Phi}{\partial r} 
      = - \sum_{n=1}^{\infty} n r^{n-1} 
          \left( \mu_n \cos (n \varphi) + \lambda_n \sin (n \varphi) \right)~, \\
B_{\varphi} = - \frac{1}{r} \frac{\partial \Phi}{\partial \varphi}
                  = - \sum_{n=1}^{\infty} n r^{n-1} 
          \left( \lambda_n \cos (n \varphi) - \mu_n \sin (n \varphi) \right)~.
\end{eqnarray}
%------------------
The summands are the {\em multipole components of order $n$} of the magnetic field in polar coordinates.
It is convenient to normalize the not dimensionless coefficients $\lambda_n$ and $\mu_n$ to a ``main field'' $B_0$ and a~``reference radius'' $r_0$.
Both are in principal free parameters but typically the field configuration offers a~``natural'' or ``useful'' choice.
We therewith can define dimensionless expansion coefficients $a_n$ and $b_n$ as
%------------------
\begin{eqnarray}
a_n = \frac{n \mu_n}{B_0} r_0^{n-1}~, \\
b_n = -\frac{n \lambda_n}{B_0} r_0^{n-1}~,
\end{eqnarray}
%------------------
and finally get the conventional multipole decomposition in polar coordinates
%------------------
\begin{equation}
\boxed{
\begin{split}
B_r(r,\varphi) = \sum_{n=1}^{\infty} B_{r,n}(r,\varphi) = B_0 \sum_{n=1}^{\infty} \left( \frac {r}{r_0} \right)^{n-1} 
          \left( -a_n \cos (n \varphi) + b_n \sin (n \varphi) \right) \\
B_{\varphi}(r,\varphi) = \sum_{n=1}^{\infty} B_{\varphi,n}(r,\varphi) = B_0 \sum_{n=1}^{\infty} \left( \frac {r}{r_0} \right)^{n-1} 
          \left( b_n \cos (n \varphi) + a_n \sin (n \varphi) \right)
\end{split}
}~.
\label{eq:B-polar-norm}  
\end{equation}
%------------------
This representation offers an immediate insight into the general field pattern of a given magnetic multipole:
the field components $B_{r,n}$ and $B_{\varphi,n}$ of the $n$-th multipole have a $\frac{2\pi}{n}$ rotational symmetry whereas the magnitude of each multipole contribution $B_n$ has no azimuthal dependence, scales with $r^{n-1}$ and is proportional to the main field $B_0$
%------------------
\begin{equation}
\vert {B_n} \vert = \sqrt{B_{r,n}^2 + B_{\varphi,n}^2} 
                          = B_0 \sqrt{a_n^2 + b_n^2} \left(\frac{r}{r_0} \right)^{n-1}~.
\label{eq:multipole-magnitude}  
\end{equation}
%------------------
For reasons shown later, the $b_n$ are called the {\em normal} (or {\em upright}) components whereas the $a_n$ are called the {\em skew} components.
They represent the fraction of the multipole contribution to the $B$ field at the~reference radius $r_0$.

Since, as we will see later, the number of poles required to generate a magnetic multipole of order $n$ is $2n$, the multipoles are named as $(2n)$ which is
%------------------
\begin{table}[ht]
\begin{center}
\label{tab:polar-multipoles}
\caption{Names of different multipoles.}
\begin{tabular}{|p{2cm}|c|c|}
\hline\hline
\textbf{Order} & \textbf{Name} & \textbf{Number of poles}\\
\hline
$n = 1$          &  Dipole           & 2\\
$n = 2$          &  Quadrupole   & 4\\
$n = 3$          &  Setupole       & 6\\
$n = 4$          &  Octupole       & 8\\
$n = 5$          &  Decapole       & 10\\
$n = 6$          &  Dodecapole   & 12\\
\hline\hline
\end{tabular}

\end{center}
\end{table}
%------------------

%_______________________________________________________________________________
%---------------------------------------------------------------------------------------------------------------
\subsection{Expansion using cartesian coordinates}
\label{sec:cartesian-coordinates}
%_______________________________________________________________________________
%---------------------------------------------------------------------------------------------------------------
Again we will start from Eq.~(\ref{eq:complex-power-series}) but now using $z = x + i y$.
Taking advantage of the binomial theorem, we get for the complex contribution $V_n$
%------------------
\begin{equation}
V_n(x,y) = \left( \lambda_n + i \mu_n \right) \left( x + i y \right)^n
              = \left( \lambda_n + i \mu_n \right) \sum_{k=0}^n \binom{n}{k}x^{n-k}\left(i y\right)^k~,
\label{eq:complex-potential-cart}  
\end{equation}
%------------------
with the binomial coefficient $\binom{n}{k}$ defined by
\begin{equation}
\binom{n}{k} = \frac{n!}{(n-k)!\cdot k!}~.
\end{equation}
Since the magnetic field is derived from the gradient of the scalar potential $\Phi$, we pick the imaginary part and get the following expression for the contribution $\Phi_n(x,y)$
%------------------
\begin{equation}
\Phi_n(x,y) = \mu_n \sum_{k=0}^{n/2} \binom{n}{2k}(-1)^{k}x^{n-2k} y^{2k}
                 + \lambda_n \sum_{k=1}^{(n+1)/2} \binom{n}{2k-1}(-1)^{k-1}x^{n-2k+1} y^{2k-1}~.
\label{eq:cartesian-multipole-scalar}  
\end{equation}
%------------------ 
Building the gradient and again using the dimensionless expansion coefficients $a_n$ and $b_n$, we get for the~upright $x$ and $y$ components $B_{x,y}^{u}$ of the magnetic field
%------------------
\begin{eqnarray}
B_x^u(x,y) = B_0 \sum_{n=1}^{\infty} \frac{b_n B_0}{n r_0^{n-1}}
                     \sum_{k=1}^{n/2} \binom{n}{2k-1}(-1)^{k-1} (n-2k+1) x^{n-2k} y^{2k-1}~, \\
B_y^u(x,y) = B_0 \sum_{n=1}^{\infty} \frac{b_n B_0}{n r_0^{n-1}}
                     \sum_{k=1}^{(n+1)/2} \binom{n}{2k-1}(-1)^{k-1} (2k-1) x^{n-2k+1} y^{2k-2}~,
\end{eqnarray}
%------------------
and for the skew $x$ and $y$ components $B_{x,y}^{s}$ of the magnetic field
%------------------
\begin{eqnarray}
B_x^s(x,y) = -B_0 \sum_{n=1}^{\infty} \frac{a_n B_0}{n r_0^{n-1}}
                     \sum_{k=0}^{(n-1)/2} \binom{n}{2k}(-1)^k (n-2k) x^{n-2k-1} y^{2k}~, \\
B_y^s(x,y) = -B_0 \sum_{n=1}^{\infty} \frac{a_n B_0}{n r_0^{n-1}}
                     \sum_{k=1}^{n/2} \binom{n}{2k}(-1)^k 2k x^{n-2k} y^{2k-1}~.~~~~~~~
\end{eqnarray}
%------------------
Since the drived equations are quite bulky, it might be useful to explicitly present the first three cartesian  multipole contributions.
We obtain for the scalar potential $\Phi$
%------------------
\begin{equation}
\Phi(x,y) = B_0 \left(
                 a_1 x - b_1 y +
                 \frac{a_2 \left(x^2 - y^2 \right) -2 b_2 x y}{2 r_0} +
                 \frac{a_3 \left( x^3 - 3xy^2 \right) + b_3 \left( y^3 - 3x^2 y \right)}{3 r_0^2} + ... \right)~,
\label{eq:cartesian-potential}  
\end{equation}
%------------------
and for the cartesian components of the magnetic field $B_{x,y}$
%------------------
\begin{equation}
\boxed{
\begin{split}
B_x(x,y) = B_0 \left( - a_1 + \frac{b_2 y - a_2 x}{r_0} +
                 \frac{a_3 \left( y^2 - x^2 \right) + 2 b_3 x y}{r_0^2} + ... \right) \\
B_y(x,y) = B_0 \left( b_1 + \frac{a_2 y + b_2 x}{r_0} +
                 \frac{2 a_3 x y + b_3 \left( x^2 - y^2 \right)}{r_0^2} + ... \right)
\end{split}
}~.
\label{eq:cartesian-magfield}  
\end{equation}
%------------------
Instead of using the dimensionless expansion coefficients $a_n$ and $b_n$, in cartesian representation often so called multipole {\em strengths} are defined.
The $n$-th upright cartesian multipole strength $s_n$ is defined by the ($n$-1)-th horizontal partial derivative of the vertical magnetic field at the design orbit, normalized to particle's momentum $p$ and inverse charge $1/q$ (cf.\ sections \ref{sec:dipole-magnets} - \ref{sec:sextupole-magnets}), and is related to the coefficient $b_n$ by
%------------------
\begin{equation}
s_n = \frac{q}{p} \frac{\partial^{n-1} B_y(0,0)}{\partial x^{n-1}} = \frac{q B_0}{p r_0^{n-1}} \cdot b_n~.
\label{eq:cartesian-parameterlinkl}  
\end{equation}
%------------------
In contrast to the polar expansion coefficients which are numbered by the order of the multipole, the~cartesian multipole strengths are stated by individual symbols ($\kappa, k, m, r, ...$).
The following table summarizes the different definitions for the first four upright and skew multipole components.
The skew multipole strengths are denoted as underlined symbols.

%------------------
\begin{table}[h]
\begin{center}
\label{tab:cartesian-multipoles}
\caption{Cartesian multipole strengths and their link to the dimensionless polar expansion coefficients.}
\begin{tabular}{|p{2cm}|c|c|c|}
\hline\hline
\textbf{Order} & \textbf{Name} & \textbf{Magnetic field} & \textbf{Multipole strength} \\
\hline
\rule{0pt}{14pt}
$n = 1$  &  upright dipole           & $\vec B_1 = \frac{p}{q} \kappa \hat{e}_y$ & $\kappa = \frac{q B_0}{p} b_1$ \\
\rule{0pt}{14pt}
$n = 1$  &  skew dipole           & $\vec B_1 = -\frac{p}{q} \underline{\kappa} \hat{e}_x$ & $\underline{\kappa} = \frac{q B_0}{p} a_1$ \\
\hline
\rule{0pt}{14pt}
$n = 2$  &  upright quadrupole   & $\vec B_2 = - \frac{p}{q} k (y \hat{e}_x + x \hat{e}_y)$  & $k = -\frac{q B_0}{p r_0} b_2$\\
\rule{0pt}{14pt}
$n = 2$  &  skew quadrupole   & $\vec B_2 =  \frac{p}{q} \underline{k} (x \hat{e}_x - y \hat{e}_y)$  & $\underline{k} = -\frac{q B_0}{p r_0} a_2$\\
\hline
\rule{0pt}{14pt}
$n = 3$  &  upright setupole       & $\vec B_3 = \frac{p}{q} m \left( xy \hat{e}_x + \frac{x^2-y^2}{2} \hat{e}_y \right)$  & $m = \frac{q B_0}{p r_0^2} b_3$\\
\rule{0pt}{14pt}
$n = 3$  &  skew setupole       & $\vec B_3 = \frac{p}{q} \underline{m} \left( \frac{y^2-x^2}{2} \hat{e}_x + xy \hat{e}_y \right)$  & $\underline{m} = \frac{q B_0}{p r_0^2} a_3$\\
\hline
\rule{0pt}{14pt}
$n = 4$  &  upright octupole       & $\vec B_4 =  \frac{p}{q} r \left( \frac{3x^2y-y^3}{6} \hat{e}_x + \frac{x^3-3xy^2}{6} \hat{e}_y \right)$  & $r = \frac{q B_0}{p r_0^3} b_4$\\
\rule{0pt}{14pt}
$n = 4$  &  skew octupole       & $\vec B_4 =  \frac{p}{q} \underline{r} \left( \frac{3xy^2-x^3}{6} \hat{e}_x + \frac{3x^2y-y^3}{6} \hat{e}_y \right)$  & $\underline{r} = \frac{q B_0}{p r_0^3} a_4$\\
\hline\hline
\end{tabular}

\end{center}
\end{table}
%------------------
\noindent \underline{Remark:} {\em
It is striking that only the quadrupole strengths $k$ and $\underline{k}$ are defined using an extra minus sign.
A~priori, there exist no compelling reason for this definition.
It comes from the early days when the concept of ``weak focusing'' was applied in the first synchrotrons in which focusing is treated using the ``field index'' $n$ that is representing the normalized radial gradient of the magnetic field at the reference orbit (cf.\ Section \ref{sec:weak-focusing}).
It is noticable in this context that, when consulting different textbooks on accelerator physics, one will observe that there is no uniform definition of the quadrupole strength in the literature regarding the extra minus sign.
}

\begin{figure}[ht]
\begin{center}
\includegraphics[width=16cm]{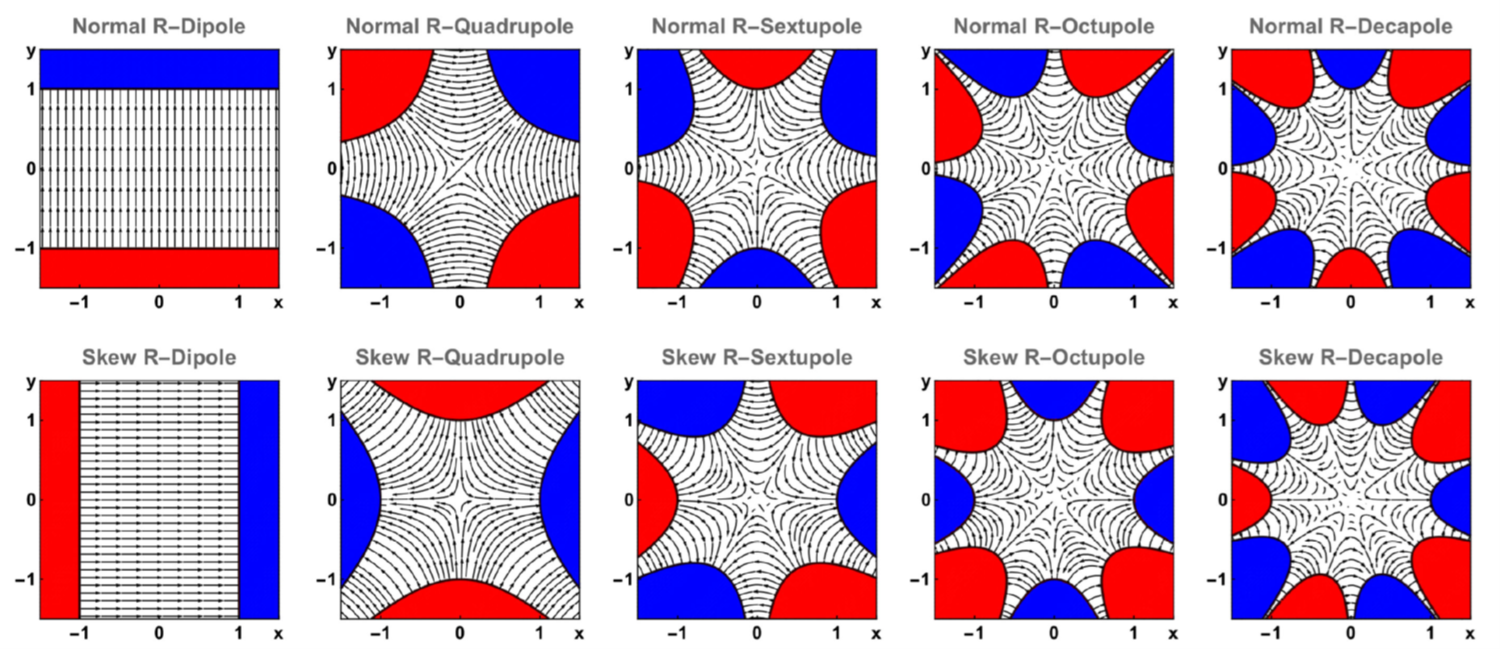}
\caption{Field patterns of different magnetic multipoles \cite{Zolkin-PRAB}.}
\label{fig:multipole-fields}
\end{center}
\end{figure}

Figure~\ref{fig:multipole-fields} shows the magnetic field lines of the first five multipole fields.
It nicely illustrates the underlying rotational $\frac{2\pi}{n}$ symmetry of a multipole $n$.
In addition, it reveals that for every upright multipole one obtains its corresponding skew component from an angular rotation of the field pattern by a rotation angle $\Delta \varphi = \frac{\pi}{2n}$ and vice versa.

%---------------------------------------------------------------------------------
%---------------------------------------------------------------------------------

Typically, accelerators are constructed as flat horizontal machines in which the design orbit stays in the horizontal plane, defined by the horizontal mid-plane $(x,0)$ of the accelerator magnets.
Since horizontal magnetic fields in this mid-plane would bend the beam out of the horizontal plane, it is therefore desireable that all horizontal magnetic fields vanish at $y=0$ ($B_x(x,0) = 0$).
This condition is fullfilled for all upright multipoles whereas for the skew multipoles all vertical fields vanish in the vertical plane $(0,y)$.
For this reason, ``normally'' upright multipoles are used in flat accelerators whereas the skew multipoles are reserved for special purposes.
This gives the reason for naming the upright multipoles as well ``normal'' multipoles.

%===========================================================================
%===========================================================================
\section{Accelerator magnets}
\label{sec:accelerator-magnets}
%===========================================================================
%===========================================================================
Accelerator magnets can be devided in two different main classes:
\begin{itemize}
\item {\em iron dominated} magnets that uses ferromagnetic material like soft iron to shape the magnetic field for generating the desired multipole component,
\item {\em current defined} magnets in which the desired magnetic multipole is determined by the current distribution defined by the conductor arrangement of the exitation coils.
\end{itemize}
Since iron dominated magnets are limited by magnetic saturation of the soft iron, magnetic fields greater than $\approx 2.5\, \text T$ can only be produced by ironless current defined magnets.
The latter typically use coils made from superconductors like NbTi to suppress the Ohmic losses which would be present in normal conducting coils.
Since a detailed discussion is outside the scope of this introductory course and will be presented in Refs.~\cite{Rijk-CAS-warm, Rijk-CAS-cold}, we will concentrate on a brief presentation of iron dominated magnets and explain how the most important first magnetic multipoles (dipole, quadrupole, sextupole) are generated.

%_______________________________________________________________________________
%---------------------------------------------------------------------------------------------------------------
\subsection{Ferromagnets}
\label{sec:ferromagnets}
%_______________________________________________________________________________
%---------------------------------------------------------------------------------------------------------------
In vacuum, the magnetic field strength $\vec H$ and the magnetic field $\vec B$ (which is often called the {\em magnetic flux density} or {\em magnetic induction}) are related by the permeability of free space $\mu_0$:
%------------------
\begin{equation}
\vec B = \mu_0 \vec H
\label{eq:B-H-vacuum} ~.
\end{equation}
%------------------
If a magnetic field penetrates a ferromagnetic material, an additional magnetic field is generated in the~material giving rise to a local enhancement of the magnetic field $\vec B$ (the {\em magnetization} $\vec M$) which for isotropic materials is described using a factor called the relative permability $\mu_r$:
%------------------
\begin{equation}
\vec B = \mu_0 \left( \vec H + \vec M \right) = \mu_0 \mu_r \vec H~.
\label{eq:B-H-magnetization}  
\end{equation}
%------------------
The magnetic field strength $\vec H$ can be computed using Ampère's law which states that the line integral of $\vec H \cdot d \vec s$ carried out along a closed path equals the enclosed electrical current $I$:
%------------------
\begin{equation}
\oint \vec H \cdot \mathrm{d} \vec s = I~.
\label{eq:Amperes-law}  
\end{equation}
%------------------
Ampère's law guarantees the continuity of $\vec H_{\parallel}$, the component of $\vec H$ parallel to the surface of the ferromagnetic material (the {\em tangential} component).
This can be derived from a line integral carried out along a rectangular loop enclosing the ferromagnet's surface in the limits $d \ll l$ and $l \to 0$ (cf.\ Fig.~\ref{fig:H-continuity}):
%---------------------------------------------------------------------------------
\begin{figure}[ht]
\begin{center}
\includegraphics[width=6cm]{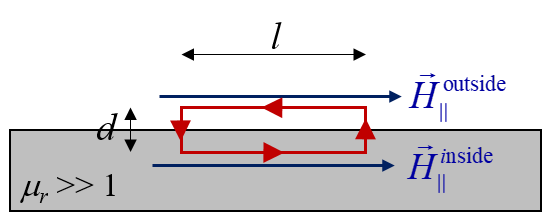}
\caption{Line integral $\oint \vec H \mathrm{d}\vec s~$ to demonstrate the continuity of the tangential component of $\vec H$.}
\label{fig:H-continuity}
\end{center}
\end{figure}
%---------------------------------------------------------------------------------
%------------------
\begin{equation}
0 = \oint \vec H \cdot \mathrm{d} \vec s \approx H_{\parallel}^{\text{inside}} \cdot l - H_{\parallel}^{\text{outside}} \cdot l ~~~~\rightarrow~~~~ H_{\parallel}^{\text{outside}} = H_{\parallel}^{\text{inside}}~.
\label{eq:H-continuity}  
\end{equation}
%------------------

\begin{figure}[ht]
\begin{center}
\includegraphics[width=6cm]{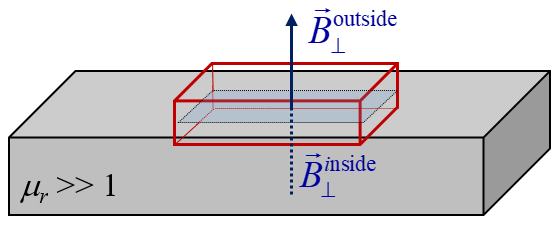}
\caption{Surface integral $\oint \vec B\cdot\mathrm{d}\vec A$ to demonstrate the continuity of the normal component of $\vec B$.}
\label{fig:B-continuity}
\end{center}
\end{figure}

Using Maxwell's second equation, the continuity of the normal component of the magnetic field $\vec B$ can be shown in the same manner.
For doing so, the rectangular loop has to be extruded to a cuboid.
The enclosed area of the rectangular loop will build one of the four faces of the cuboid perpendicular to the ferromagnet's surface (cf.\ Fig.~\ref{fig:B-continuity}).

%---------------------------------------------------------------------------------
%---------------------------------------------------------------------------------

\noindent We then get from the integral over the closed surface of the cuboid
%------------------
\begin{equation}
0 = \oint \vec B \cdot \mathrm{d} \vec A \approx B_{\perp}^{\text{outside}} \cdot A - B_{\perp}^{\text{inside}} \cdot A ~~~~\rightarrow~~~~ B_{\perp}^{\text{outside}} = B_{\perp}^{\text{inside}}~.
\label{eq:B-cloop-integral}  
\end{equation}
%------------------
Since for ferromagnets the relative permeability $\mu_r \gg 1$ is much greater than one, the tangential component of the magnetic field $\vec B$ must vanish at the ferromagnet's surface
%------------------
\begin{equation}
B_{\parallel}^{\text{outside}} = \mu_0 H_{\parallel}^{\text{outside}} = \mu_0 H_{\parallel}^{\text{inside}} \ll \mu_0 \mu_r H_{\parallel}^{\text{inside}} = B_{\parallel}^{\text{inside}}~~~~\rightarrow~~~~ B_{\parallel}^{\text{surface}} = 0~.
\label{eq:B-continuity}  
\end{equation}
%------------------
Thus, $\vec B$ is perpendicular to the surface of ferromagnets.
Since $\vec B = - \vec \nabla \Phi$, the ferromagnet's surface build an equipotential surface $\Phi = \text{const.}$ which simplifies the shaping of the magnetic field significantly.

%---------------------------------------------------------------------------------
\begin{figure}[ht]
\begin{center}
\includegraphics[width=4cm]{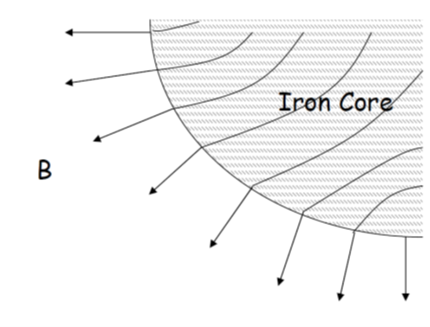}
\caption{Magnetic field $\vec B$ at the surface of a ferromagnet.}
\label{fig:equipotential-surface}
\end{center}
\end{figure}
%---------------------------------------------------------------------------------

In order to generate a specific magnetic multipole, the task reduces to build enough and appropriately shaped magnetic poles defining the equipotential surfaces for a wanted magnetic multipole.
This simplifies the precise generation of single magnetic multipole enormously and is the reason why accelerator magnets are mostly build as iron dominated magnets as long as the herewith achievable field strengths meet the requirements.
In the following, we will explicitly discuss the different types of iron dominated magnets and their action on the beam seperately and restrict ourself to the upright multipoles.
The skew multipoles can be easily obtained by a rotation of the magnet around its central symmetry axis.
When discussing the action of a given multipole on the beam, we will always use a right-handed coordinate system $(x, y, s)$ where the coordinate vector $\hat{e}_s$ points in the direction of the reference particle's velocity, $\hat{e}_y$ points vertically and $\hat{e}_x = \hat{e}_y \times \hat{e}_s$ is in the horizontal plane (see Section \ref{sec:coordinate-system}).

%_______________________________________________________________________________
%---------------------------------------------------------------------------------------------------------------
\subsection{Dipole magnets and beam deflection}
\label{sec:dipole-magnets}
\begin{figure}[ht]
\begin{center}
\includegraphics[width=14cm]{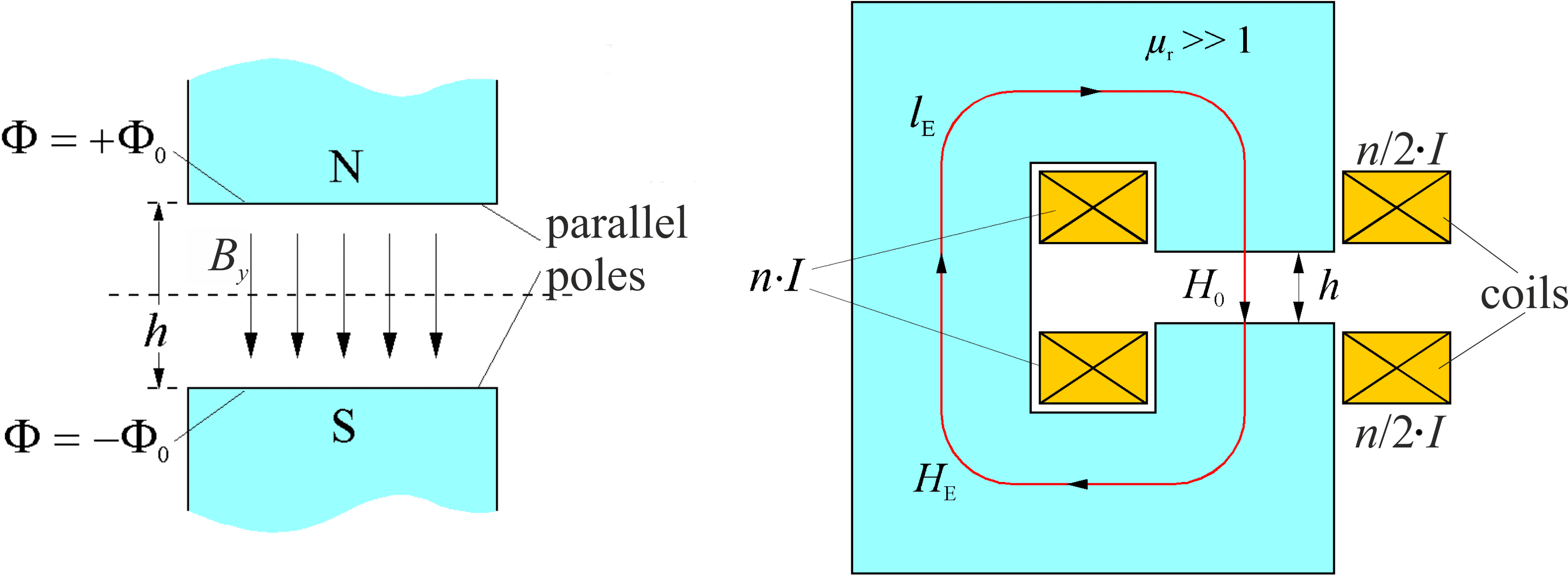}
\caption{Pole profile and cross section of a dipole magnet creating a vertical homogeneous $B$ field.}
\label{fig:dipole-magnet}
\end{center}
\end{figure}

%_______________________________________________________________________________
%---------------------------------------------------------------------------------------------------------------
In order to deflect a charged particle beam, a constant homogeneous magnetic field pointing perpendicular to the design orbit is required.
In flat horizontal accelerators the constant magnetic field has to point parallel to the vertical direction, which is the case for the first upright multipole component:
%------------------
\begin{equation}
\vec B(x,y) = B_0 \cdot \hat{e}_y = \text{const.} ~~~~ \leftrightarrow ~~~~ \Phi(x,y) = -B_0 \cdot y~.
\label{eq:B-dipole-field-potential}  
\end{equation}
%------------------
Two equipotential surfaces $\Phi = \pm \Phi_0$ are thus required defining the poles' profiles to be flat and parallel.
We therewith get a dipole magnet which is often build as C-magnet for which the C-shaped iron core act as return yoke and the magnetic field is exited by two coils wound around the poles (Fig. \ref{fig:dipole-magnet}).

%---------------------------------------------------------------------------------
%--------------------------------------------------------------------------------

The generated dipole field can be calculated from Ampère's law computing the line integral along the integration path indicated in Fig.~\ref{fig:dipole-magnet}
%------------------
\begin{equation}
n I =\oint \vec H \cdot \mathrm{d} \vec s =  \int_{\text{gap}} \vec H_0 \cdot \mathrm{d} \vec s + \int_{\text{yoke}} \vec H_{\text{iron}} \cdot \mathrm{d} \vec s~,
\label{eq:H-integral-dipole}  
\end{equation}
%------------------
where $n$ is the number of turns of the exitation coils ($\frac{n}{2}$ turns for the upper and $\frac{n}{2}$ turns for the lower coil) and $I$ the current which flows through the coils. 
Since $\mu_r \gg \mu_0$ and $\vec B_\text{iron} =\vec B_0$, we have $\vert H_\text{iron} \vert \ll \vert H_0 \vert$ and the integral along the path in the return yoke does not contribute substancially.
We therefore get in good approximation the simple relation
%------------------
\begin{equation}
B_0 = \mu_0 \frac{n I}{h}~,
\label{eq:amp-winding-dipole}  
\end{equation}
%------------------
indicating the required {\em ampere-turns} $n I$ to generate a magnetic field $B_0$ in a gap of height $h$.

\noindent\underline{Example:}\\
With a gap height of 5\,cm $n I = 40000$ Ampère-turns are needed to produce a bending field of 1\,Tesla.

According to Section \ref{sec:cartesian-coordinates} the {\em dipole strength} $\kappa$ can be derived from a normalization applying a~factor $\frac{q}{p}$.
It indicates the curvature of the orbit which is equivalent to the inverse bending radius and has the unit of inverse meters:
%------------------
\begin{equation}
\boxed{
\kappa = \frac{1}{\rho} = \frac{q}{p}B_0 = \frac{q \mu_0}{p}\frac{n I}{h}, ~~~~~ [\kappa] = \text{m}^{-1}
}~.
\label{eq:dipole-strength}  
\end{equation}
%------------------
It is obvious that the sign of the curvature $\kappa$ (and therewith also the sign of the bending radius $\rho$) can be positive or negative, depending on wether the particle is been bent outwards or inwards.
Using our right-handed coordinate system $(x,y,s)$ we obtain the following relation between bending and curvature:
\begin{itemize}
\item if a particle is deflected in the direction of $-\hat{e}_x$ the curvature is positive ($\kappa > 0$),
\item if a particle is deflected in the direction of $\hat{e}_x$ the curvature is negative ($\kappa < 0$).
\end{itemize}
Of course this is of no importance when only indicating the absolute ``strengh'' of a dipole magnet, but explicit care has to be taken when looking at the action of a dipole magnet on off-momentum particles (see Section~\ref{sec:off-momentum-particles}).

\begin{figure}[ht]
\begin{center}
\includegraphics[width=15cm]{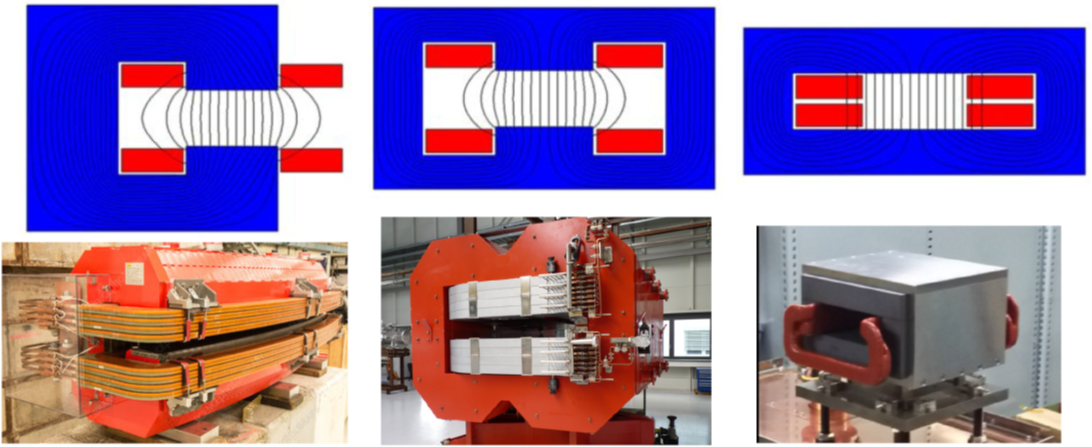}
\caption{Different types of iron dominated dipole magnets. Left: C dipole magnet; the picture shows a bending magnet of the SESAME storage ring. Middle: H dipole magnet; the picture shows a prototype of the normal conducting 11 Super-FRS  dipole magnet. Right: window-frame dipole magnet; the picture shows the injection kicker for the KEK photon factor advanced ring.}
\label{fig:pictures-dipole-magnets}
\end{center}
\end{figure}

Typical types of iron dominated dipole magnets are presented in Fig.~\ref{fig:pictures-dipole-magnets}.
For all of them the field shaping is accomplished by two parallel magnetic poles made of ferromagnetic material (e.g.\ iron).
%---------------------------------------------------------------------------------
%--------------------------------------------------------------------------------

%_______________________________________________________________________________
%---------------------------------------------------------------------------------------------------------------
\subsection{Quadrupole magnets and beam focusing}
\label{sec:quadrupole-magnets}
\begin{figure}[ht]
\begin{center}
\includegraphics[width=12cm]{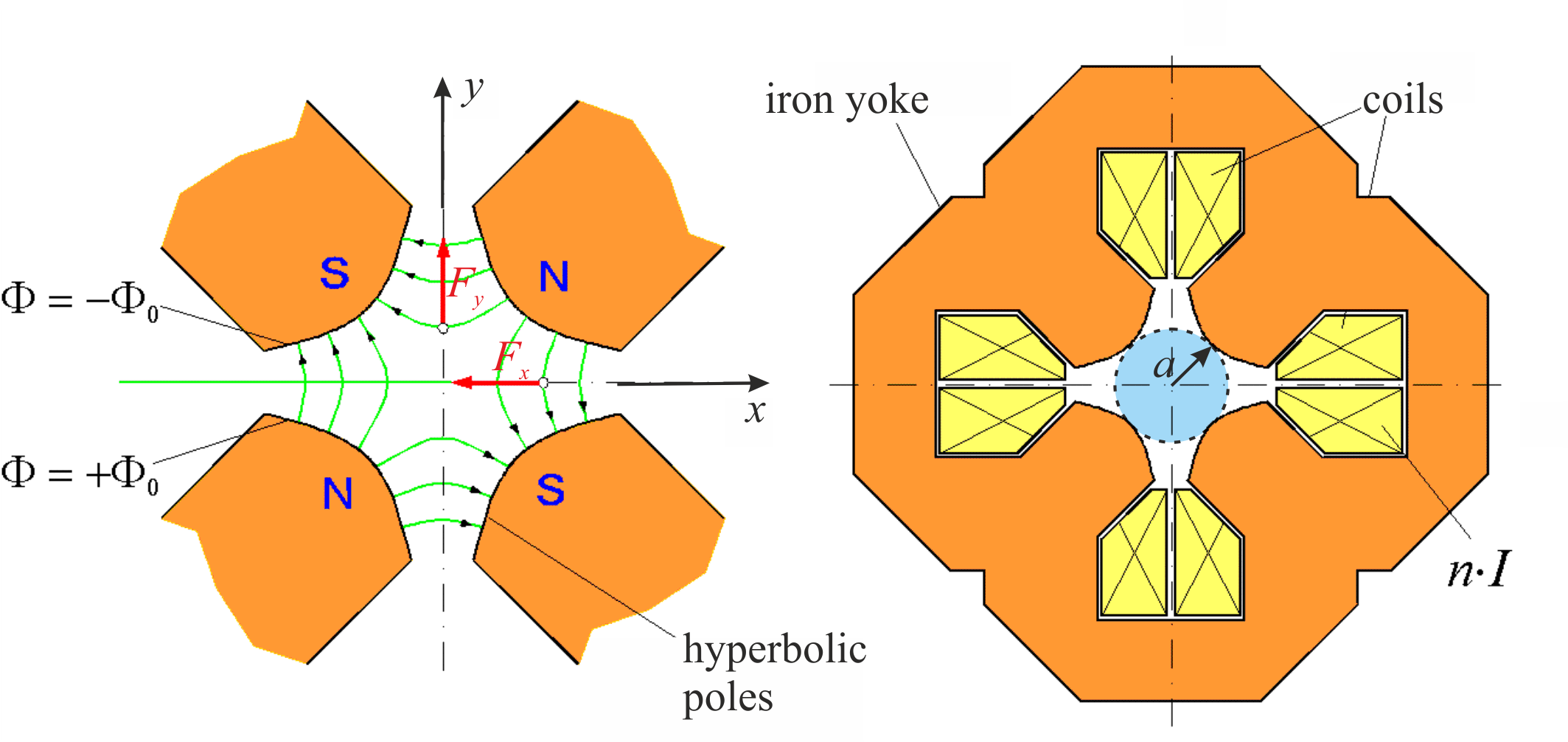}
\caption{Cross section of a quadrupole magnet creating a $B$ field which scales linearly with $x$ and $y$.}
\label{fig:quadrupole-magnet}
\end{center}
\end{figure}

%_______________________________________________________________________________
%---------------------------------------------------------------------------------------------------------------
Beam focusing requires restoring forces which scale linearly with the distance from the center defining the optical axis of the focusing device.
Since the Lorentz force is proportional to the magnetic field, beam focusing can be achieved with transverse magnetic fields scaling linearly with the transverse offset (x,y).
Whereas focusing in the horizontal plane requires a vertical field $B_y$ proportional to $x$, focusing in the vertical plane requires a horizontal field $B_x$ proportional to $y$.
Such fields are given by the second upright magnetic multipole, the quadrupole:
%------------------
\begin{equation}
B_y = -g\cdot x, ~~~ B_x = -g \cdot y ~~~~\text{with}~~~~g = -\frac{\partial B_y}{\partial x} =-\frac{\partial B_x}{\partial y} = \text{const.}
\label{eq:B-quadrupole-field-def}  
\end{equation}
%------------------
The corresponding potential can be taken from Eq.~\ref{eq:cartesian-potential} and reads using the gradient $g$ as proportionality factor
%------------------
\begin{equation}
\Phi(x,y) = g \cdot x \cdot y
\label{eq:B-quadrupole-potential-def}~.
\end{equation}
%------------------
Four hyperbolic equipotential surfaces are thus required defining the poles' profiles and position for positive and negative values for $\Phi_0$ and $x$ according to
%------------------
\begin{equation}
y(x) = \pm \frac{\Phi_0}{gx}~.
\label{eq:B-quadrupole-poles}  
\end{equation}
%------------------
We therewith get a quadrupole magnet which consists of four hyperbolic poles connected by a return yoke.
The magnetic field is exited by four coils wound around the poles (cf.\ Fig.~\ref{fig:quadrupole-magnet}).

%---------------------------------------------------------------------------------
\begin{figure}[ht]
\begin{center}
\includegraphics[width=6cm]{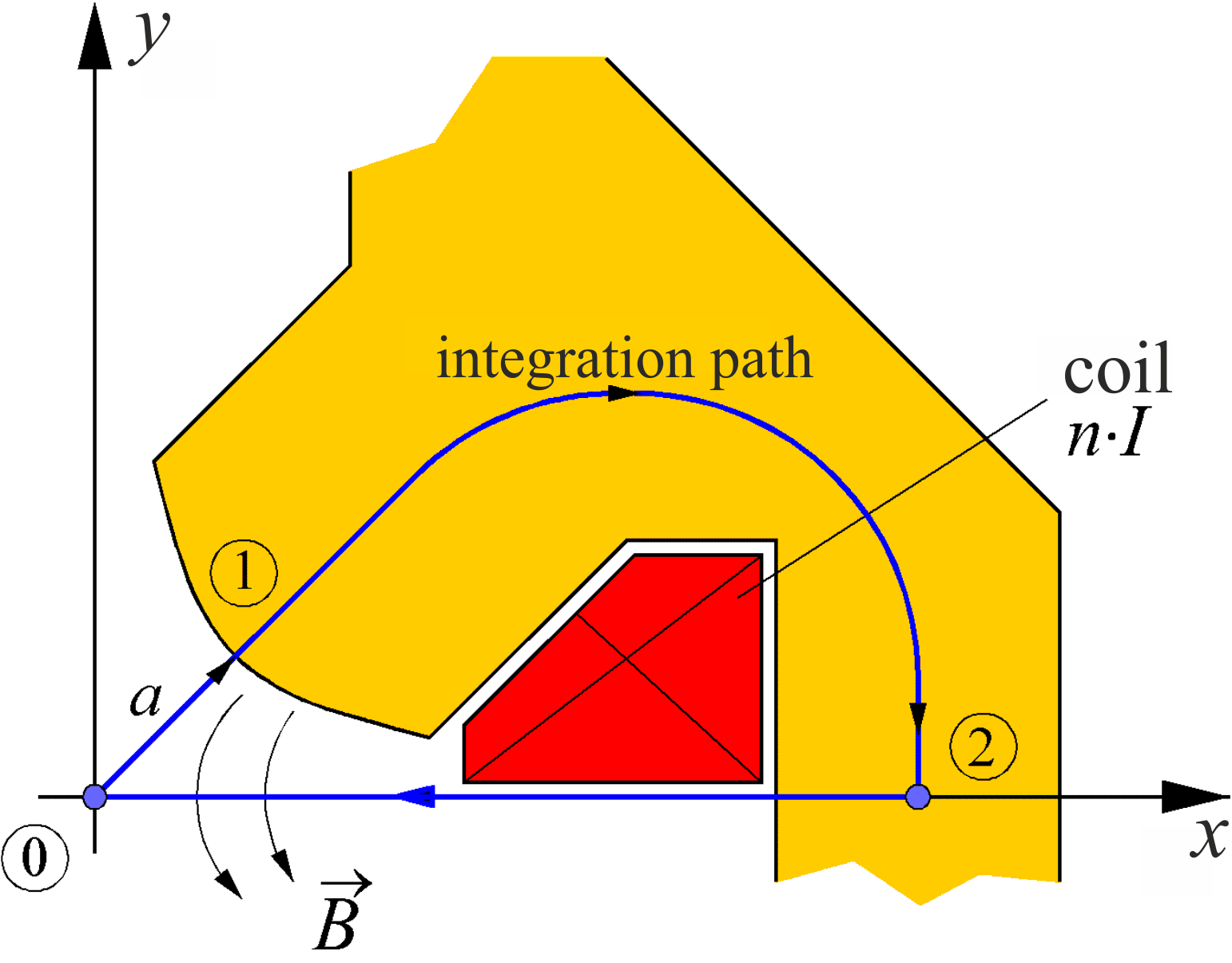}
\caption{Integration path used to compute the quadrupole gradient as a function of the Ampère-turns.}
\label{fig:quadrupole-magnet-path}
\end{center}
\end{figure}

%--------------------------------------------------------------------------------
\noindent $\Phi_0$ can be expressed by the smallest distance of the poles from the center which is called the {\em aperture} $a$
%------------------
\begin{equation}
\Phi_0 = \Phi(\frac{a}{\sqrt{2}},\frac{a}{\sqrt{2}}) = \frac{1}{2}g a^2~.
\label{eq:B-quadrupole-aperture}  
\end{equation}
%------------------
Hence, the pole geometry of a quadrupole is exclusively determined by its aperture $a$
%------------------
\begin{equation}
y(x) = \pm \frac{a^2}{2x}~.
\label{eq:B-quadrupole-poles-aperture}  
\end{equation}
%------------------
The generated quadrupole field can be calculated from Ampère's law computing the line integral along the~integration path indicated in Fig.~\ref{fig:quadrupole-magnet-path}

%---------------------------------------------------------------------------------
%--------------------------------------------------------------------------------
%------------------
\begin{equation}
n I =\oint \vec H \cdot \mathrm{d} \vec s =  \int_0^1 \vec H_0 \cdot \mathrm{d} \vec s + \int_1^2 \vec H_{\text{iron}} \cdot \mathrm{d} \vec s + \int_2^0 \vec H \cdot \mathrm{d} \vec s ~.
\label{eq:H-integral-quadrupole}  
\end{equation}
%------------------
Again $\vert H_\text{iron} \vert \ll \vert H_0 \vert$ and the second integral does not contribute substancially.
The third integral vanishes since $\vec H \perp \mathrm{d} \vec s$ along the integration path chosen.
We therefore get in good approximation
%------------------
\begin{equation}
n I = \frac{1}{\mu_0} \int_0^a g r \mathrm{d} r ~~~~ \rightarrow ~~~~~ g = \frac{2 \mu_0 n I}{a^2}~,
\label{eq:amp-winding-quadrupole}  
\end{equation}
%------------------
indicating the required ampere-turns $n I$ to generate a gradient $g$ for a quadrupole magnet with a given aperture $a$.

The restoring force acting on the charged particles traversing the quadrupole magnet can be derived from inserting the quadrupole field in the Lorentz force.
Again, we will use our right-handed coordinate system $(x,y,s)$ and obtain
%------------------
\begin{equation}
\vec F = q \cdot \left( \vec v \times \vec B \right) = qvg \cdot (x \hat{e}_x - y \hat{e}_y)~.
\label{eq:force-quadrupole}
\end{equation}
%------------------
\begin{center}
\textbf{A quadrupole magnet is therefore focusing only in one plane and defocusing in the other,\\
depending on the sign of the gradient {\em g} and the charge {\em q}!}
\end{center}

According to Section \ref{sec:cartesian-coordinates} the~{\em quadrupole strength} $k$ can be derived from a normalization with the~factor $\frac{q}{p}$.
It indicates the focusing strength of a quadrupole and has the unit of inverse square meters:
%------------------
\begin{equation}
\boxed{
k = \frac{q}{p}g = \frac{q \mu_0}{p}\frac{n I}{a^2}, ~~~~~ [k] = \text{m}^{-2}
}~.
\label{eq:quadrupole-strength}  
\end{equation}
%------------------
The sign of the quadrupole strength $k$ determines if a particle beam is focused or defocused.
It is common practice to name a quadrupole magnet a focusing quadrupole if it is focusing in the horizontal plane.
Such a quadrupole will act defocusing in the vertical plane.
According to Eqs.~(\ref{eq:force-quadrupole})--(\ref{eq:quadrupole-strength}) horizontal focusing is achieved for negative $k$ and vertical focusing is achieved for positive $k$:
\begin{itemize}
\item a focusing quadrupole magnet is horizontal focusing and vertical defocusing and has $k < 0$,
\item a defocusing quadrupole magnet is horizontal defocusing and vertical focusing and has $k > 0$.
\end{itemize}

In order to gain a more illustrative understanding of the quadrupole strength we will look at a very ``thin'' quadrupole, applying the approximation of a thin lens.
Thin in this context means that the length $L$ of the quadrupole magnet is much smaller than its focal length $f$.
In this approximation we can neglect that the beam size is already changing in the quadrupole field.

%---------------------------------------------------------------------------------
\begin{figure}[ht]
\begin{center}
\includegraphics[width=7cm]{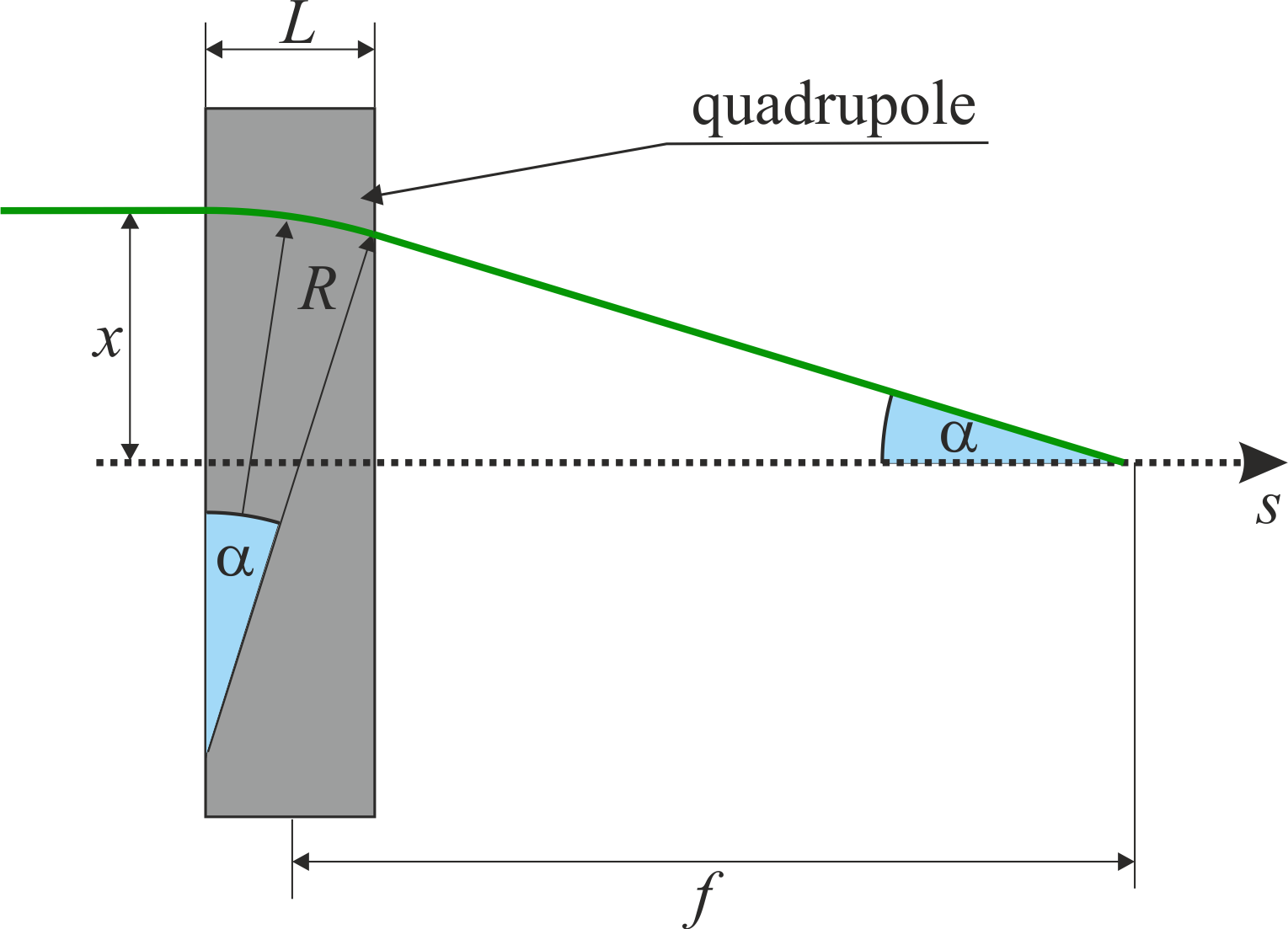}
\caption{Beam deflection in a ``thin'' quadrupole magnet with $L \ll f.$}
\label{fig:quad-focal-length}
\end{center}
\end{figure}
%--------------------------------------------------------------------------------

\noindent From the triangles in Fig.~\ref{fig:quad-focal-length} we get
%------------------
\begin{equation}
\tan \alpha = \frac{x}{f} = \frac{L}{R}=L\frac{q}{p}B_y=- \frac{q}{p}g x L = - x k L~.
\label{eq:triangles-quadrupole}
\end{equation}
%------------------
We therewith obtain the following useful relations for the horizontal ($f_x$) and vertical ($f_y$) focusing lengths of a thin quadrupole magnet acting as an ``ideal'' lens:
%------------------
\begin{equation}
\boxed{
\frac{1}{f_x} = - k L, ~~~~ \frac{1}{f_y} = k L ~~~~~ \text{for} ~~~~~ L \ll \vert f_x \vert
}~.
\label{eq:focal-length-quadrupole}
\end{equation}
%------------------

%_______________________________________________________________________________
%---------------------------------------------------------------------------------------------------------------
\subsection{Sextupole magnets}
\label{sec:sextupole-magnets}
%_______________________________________________________________________________
%---------------------------------------------------------------------------------------------------------------
We can extend our treatment to higher multipoles and will exemplarily discuss the multipole of next higher order, the sextupole.
Again, the magnetic field and potential can be taken from Eqs.~(\ref{eq:cartesian-magfield})--(\ref{eq:cartesian-potential}).
Using $g'$ as proportional factor the magnetic field is
%------------------
\begin{equation}
\vec B = \frac{1}{2}g^\prime \left( x^2-y^2 \right) \hat{e}_y + g^\prime x y \hat{e}_x ~~~~ \text{with} ~~~~
           g^\prime = \frac{\partial^2 B_y}{\partial x^2} = \text{const.}~,
\label{eq:B-setupole-field-def}  
\end{equation}
%------------------
and the corresponding potential reads
%------------------
\begin{equation}
\Phi(x,y) = \frac{1}{6}g^\prime \left( y^3 - 3 x^2 y \right)~.
\label{eq:B-sextupole-potential-def}  
\end{equation}
%------------------
Six equipotential surfaces are thus required defining the poles' profiles and position according to
%------------------
\begin{equation}
x(y) = \pm \sqrt{\frac{y^2}{3} \pm \frac{2 \Phi_0}{g^\prime y}}~.
\label{eq:B-setupole-poles}  
\end{equation}
%------------------
%---------------------------------------------------------------------------------
\begin{figure}[ht]
\begin{center}
\includegraphics[width=12cm]{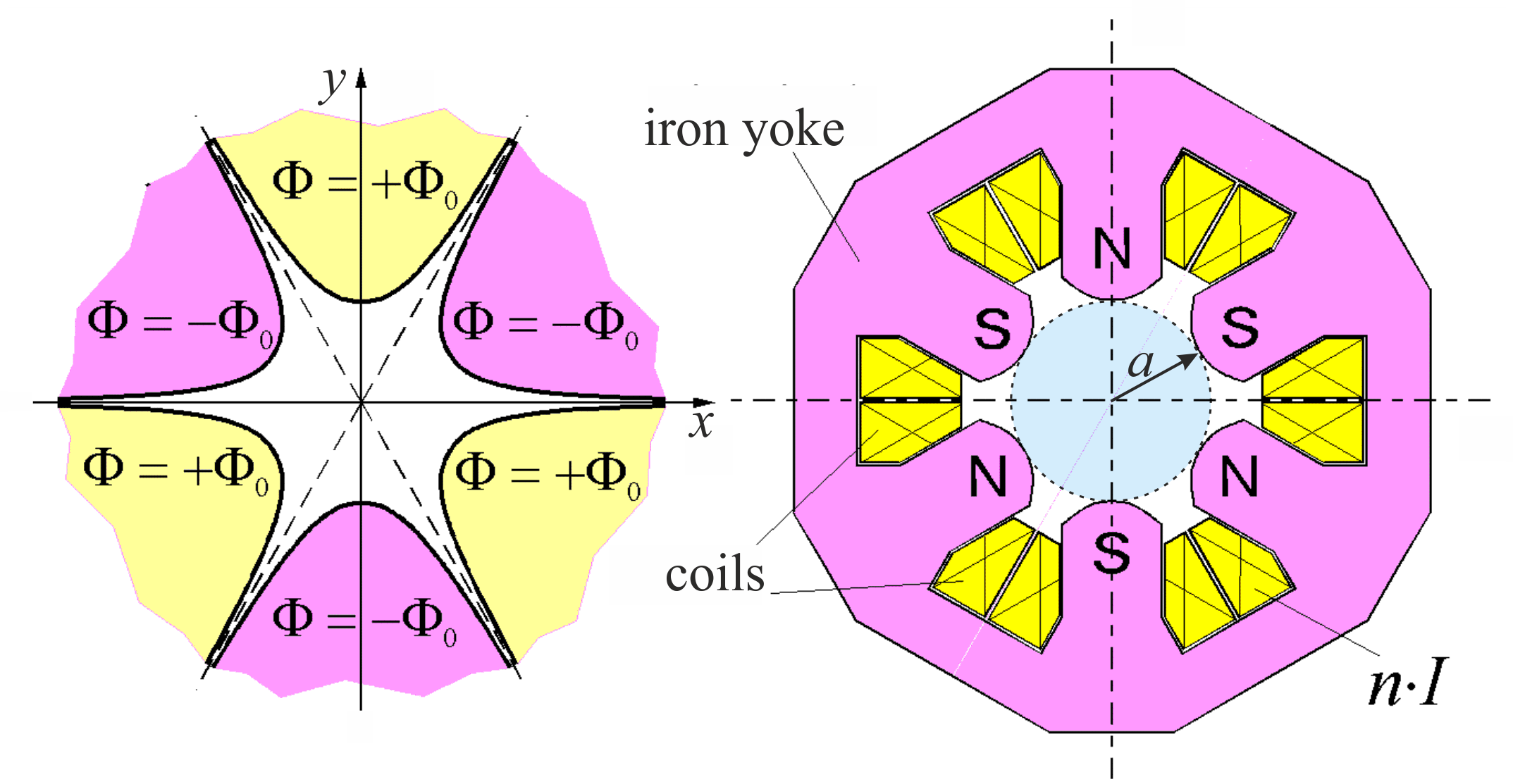}
\caption{Pole profile and cross section of a setupole magnet.}
\label{fig:sextupole-magnet}
\end{center}
\end{figure}
%--------------------------------------------------------------------------------
We therewith get a setupole magnet which consists of six poles connected by a return yoke.
The magnetic field is exited by six coils wound around the poles (cf.\ Fig.~\ref{fig:sextupole-magnet}).

\begin{figure}[ht]
\begin{center}
\includegraphics[width=4cm]{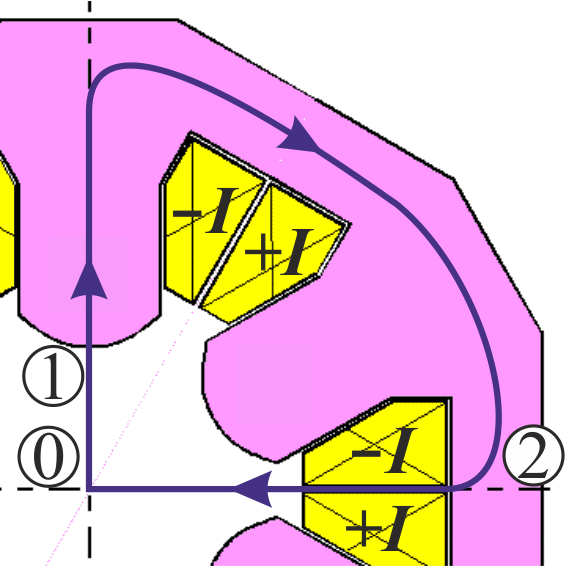}
\caption{Integration path used to compute the sextupole gradient g' as a function of the Ampère-turns.}
\label{fig:sextupole-magnet-path}
\end{center}
\end{figure}

\noindent Again $\Phi_0$ can be expressed by the aperture $a$
%------------------
\begin{equation}
\Phi_0 = \Phi(0,a) = \frac{1}{6}g' a^3~,
\label{eq:B-setupole-aperture}  
\end{equation}
%------------------
and the geometry of the poles is exclusively determined by the aperture $a$
%------------------
\begin{equation}
x(y) = \pm \sqrt{\frac{y^2}{3} \pm \frac{a^3}{3y}}~.
\label{eq:B-sextupole-poles-aperture}  
\end{equation}
%------------------
The $g'$ parameter may be related to the current of the coils by evaluating the line integral along the~integration path indicated in Fig.~\ref{fig:sextupole-magnet-path}
%------------------
\begin{equation}
n I =\oint \vec H \cdot \mathrm{d} \vec s =  \int_0^1 \vec H_0 \cdot \mathrm{d} \vec s + \int_1^2 \vec H_{\text{iron}} \cdot \mathrm{d} \vec s + \int_2^0 \vec H \cdot \mathrm{d} \vec s~.
\label{eq:H-integral-sextupole}  
\end{equation}
%------------------
%---------------------------------------------------------------------------------
%--------------------------------------------------------------------------------
Again $\vert H_\text{iron} \vert \ll \vert H_0 \vert$ and the second integral does not contribute substancially.
The third integral vanishes since $\vec H \perp d \vec s$ along the integration path chosen.
We therefore get in good approximation
%------------------
\begin{equation}
n I = \frac{1}{\mu_0} \int_0^a \frac{g^\prime}{2} y^2 \mathrm{d} y ~~~~ \rightarrow ~~~~~ g^\prime = \frac{6 \mu_0 n I}{a^3}~,
\label{eq:amp-winding-sextupole}  
\end{equation}
%------------------
indicating the required ampere-turns $n I$ to generate a gradient $g'$ for a sextupole magnet with a given aperture $a$.

The {\em sextupole strength} $m$ is derived from a normalization with the factor $\frac{q}{p}$ and has the unit inverse qubic meters:
%------------------
\begin{equation}
\boxed{
m = \frac{q}{p}g^\prime = \frac{6 q \mu_0}{p}\frac{n I}{a^3}, ~~~~~ [m] = \text{m}^{-3}
}~.
\label{eq:sextupole-strength}  
\end{equation}
%------------------
A simple understanding of the action of a sextupole will be given later in Section \ref{sec:chromaticity}.
In any case we will expect a coupling of the particles' motion in the horizontal and vertical plane from the $y$-dependence of the vertical and the $x$-dependence of the horizontal magnetic field.

%_______________________________________________________________________________
%---------------------------------------------------------------------------------------------------------------
\subsection{Effective field length}
\label{sec:field-length}
%_______________________________________________________________________________
%---------------------------------------------------------------------------------------------------------------
So far, we assumed that all magnets will produce only transverse magnetic fields components which do not vary inside the magnet and immediatly drop to zero outside the magnet.
However, in reality there will be fringe fields at the end of the magnets which will have a longitudinal field component and gradually decrease  to zero.
For the following we will neglect the longitudinal field components since they only marginally act on the beam.
In order to use a field distribution not varying with $s$, we define an {\em effective field length} $l_\text{eff}$ via
%------------------
\begin{equation}
\int_{-\infty}^\infty \vec B(x,y,s) \cdot \mathrm{d}s = \vec B_0(x,y)\cdot l_\text{eff}~,
\label{eq:effective-field-length}  
\end{equation}
%------------------
which enables us to use a pure transverse magnetic field $\vec B_0(x,y)$ which will not vary inside the magnet and will drop to zero outside the effective field length.
%---------------------------------------------------------------------------------
\begin{figure}[ht]
\begin{center}
\includegraphics[width=8cm]{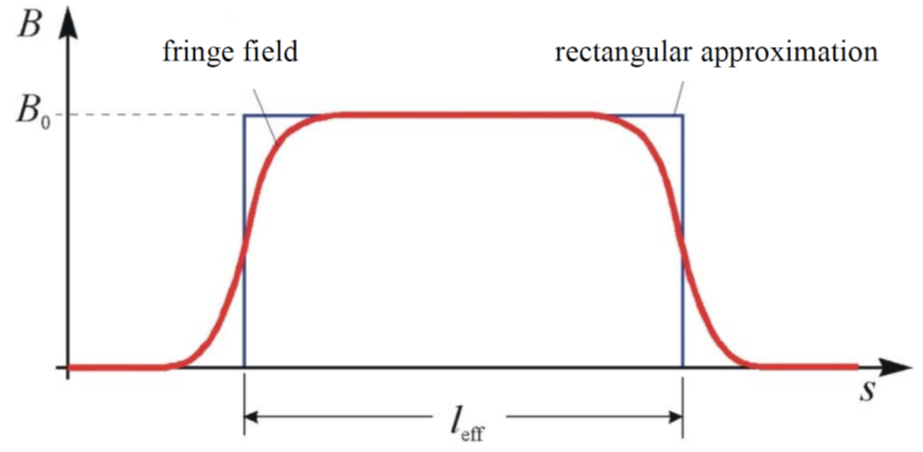}
\caption{Rectangular approximation of the evolution of the transverse field component along a magnet. Outside the effective field length, the field will immediatly drop to zero.}
\label{fig:effective-field-length}
\end{center}
\end{figure}
%--------------------------------------------------------------------------------
This approximation is illustrated in Fig.~\ref{fig:effective-field-length}.
Please note that the effective field length will not equal the iron length of a given magnet.

%===========================================================================
%===========================================================================
\section{Equations of motion in the transverse planes}
\label{sec:hill-equation}
%===========================================================================
%===========================================================================
The following sections will deal with the effect of transverse magnetic fields on the dynamics of charged particles utilizing the multipole decomposition in cartesian coordinates which has been dervied in Section~\ref{sec:cartesian-coordinates}.
We will concentrate on the transverse beam dynamics and therefore decouple the motion of the beam particles in the plane transverse to the orbit from their longitudinal motion.
The goal of this section is to derive the equations of motion in a transverse plane evolving along the orbit. 
We will utilize a special coordinate system which is following the design orbit.

%_______________________________________________________________________________
%---------------------------------------------------------------------------------------------------------------
\subsection{Coordinate system following the reference orbit}
\label{sec:coordinate-system}
\begin{figure}[ht]
\begin{center}
\includegraphics[width=14cm]{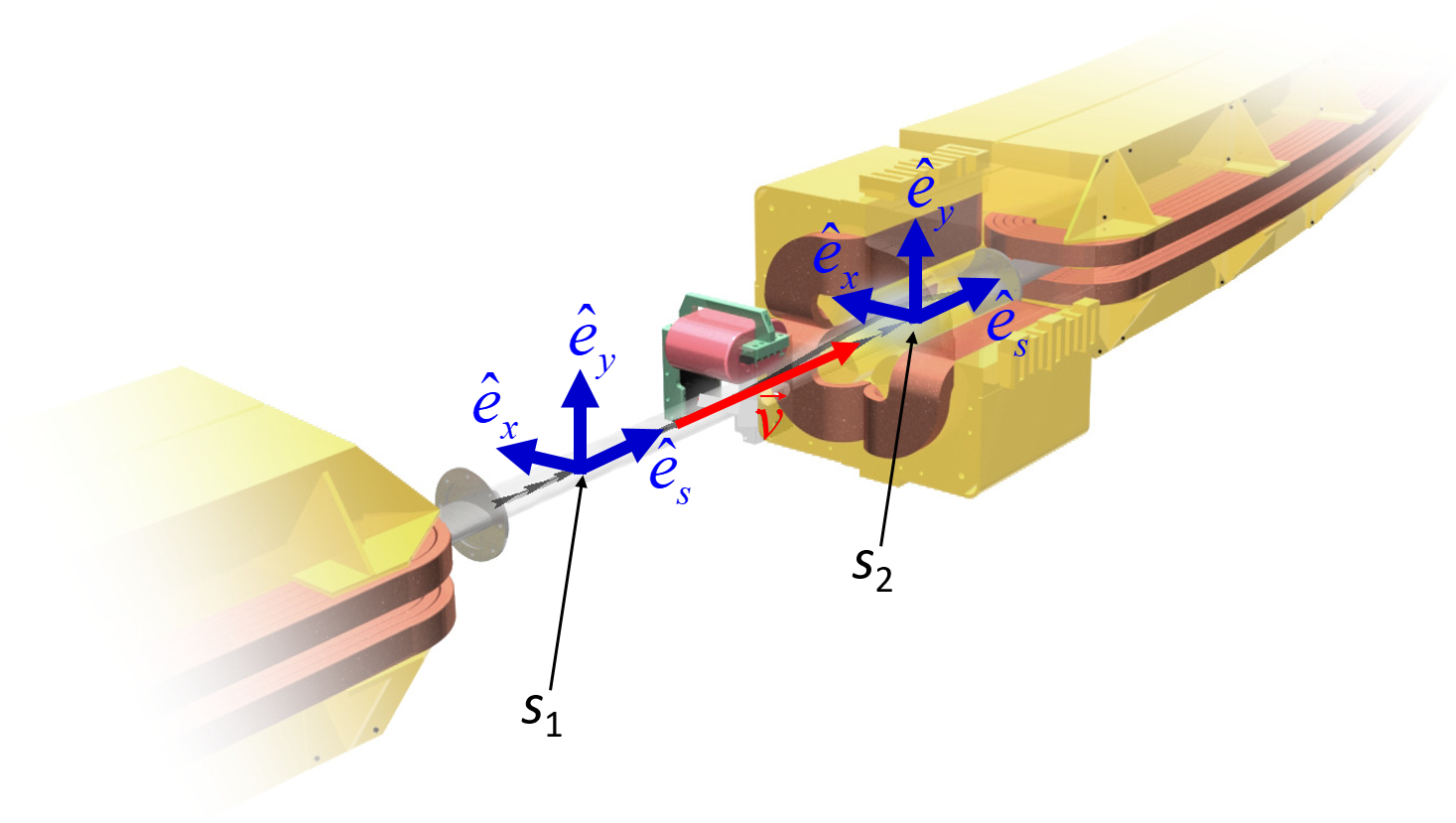}
\caption{Right-handed coordinate system ($x,y,s$) following the reference orbit. The unit vectors $\hat{e}_x$ and $\hat{e}_s$ define the orbit plane, $\hat{e}_y$ is perpendicular to it and pointing upwards.}
\label{fig:frame-following-orbit}
\end{center}
\end{figure}

%_______________________________________________________________________________
%---------------------------------------------------------------------------------------------------------------
Since we are interested in the evolution of the particles' small transverse deviations from the reference orbit, a fixed cartesian reference frame would not be well suited for the description of the transverse dynamics.
Instead, we will use an orthogonal right-handed reference frame ($x,y,s$) which is attached with its origin to the reference orbit.
We will restrict ourself to a flat reference orbit lying in the ($x,s$) plane and define the coordinate unit vectors $\hat{e}_x$, $\hat{e}_y$, $\hat{e}_s$ in the following way:
\begin{itemize}
\item $\hat{e}_s$ is tangential to the reference orbit and pointing in the direction of the reference particle's speed,
\item $\hat{e}_y$ is perpendicular to the plane defined by the flat reference orbit and is pointing upwards,
\item $\hat{e}_x =\hat{e}_y \times \hat{e}_s$ is in the plane defined by the reference orbit.
\end{itemize}
The longitudinal coordinate $s$ (sometimes named the {\em arc length parameter}) is measured along the design orbit starting from a position on the reference orbit which is often chosen as the start of a beam transfer line or a symmetry point in a circular accelerator.
Figure~\ref{fig:frame-following-orbit} illustrates the coordinate system following the reference orbit.
%---------------------------------------------------------------------------------
%--------------------------------------------------------------------------------

\noindent \underline{Remarks:}
{\em
It is important to notice that this frame is not co-moving with the particles (as misleadingly stated in many textbooks) but fixed to the laboratory at any time!
Otherwise, if using an accelerated co-moving frame complicated Lorentz transformations would be necessary.
However, the orientation of our frame following the orbit is defined by the curved reference orbit and thus changing.
We therefore name it a frame ``following the reference orbit''.
Sometimes it is referred to as ``Frenet-Serret'' coordinates, which is again not fully conform with our definitions.
In Frenet-Serret coordinates the ``normal'' and ``binormal'' unit vectors $\hat{n}$ and $\hat{b}$ are defined as
%------------------
\begin{equation}
\hat{n} = \frac{\frac{\partial \hat{e}_s}{\partial s}}{\vert \frac{\partial \hat{e}_s}{\partial s} \vert}, ~~~~~~~~
\hat{b} = \hat{e}_s \times \hat{n}~.
\label{eq:frenet-serret}  
\end{equation}
%------------------
Thus $\hat{n} = \pm \hat{e}_x$ and $\hat{b} = \pm \hat{e}_y$ depending on the sign of the curvature $\kappa = 1/ \rho$ which is always positive in Frenet-Serret coordinates according to its definition via $\frac{\partial \hat{e}_s}{\partial s}= \kappa \cdot \hat{n}$.
Although Frenet-Serret coordinates seem to be well defined in storage rings where all main dipole magnets deflect the beam in the same direction and thus allowing to define the coordinate vectors such that one obtains an always positive curvature, their usage will inevitably lead to confusion when being applied to beam transfer lines where the beam might be bent in opposite directions.
}

%_______________________________________________________________________________
%---------------------------------------------------------------------------------------------------------------
\subsection{Geometric optics}
\label{sec:geometric-optics}
\begin{figure}[ht]
\begin{center}
\includegraphics[width=12cm]{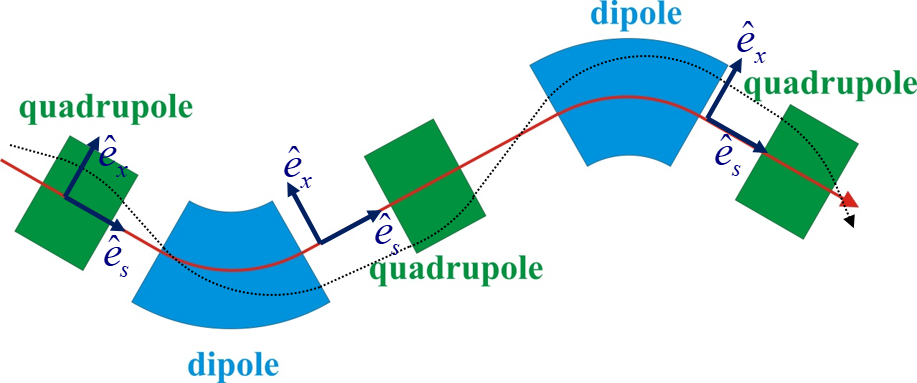}
\caption{Top view on the reference path (red curve) and the trajectory of a charged particle with horizontal transverse displacements $x(s)$ (dashed curve) moving through a sequence of different accelerator magnets.}
\label{fig:reference-path}
\end{center}
\end{figure}

\begin{figure}[ht]
\begin{center}
\includegraphics[width=16cm]{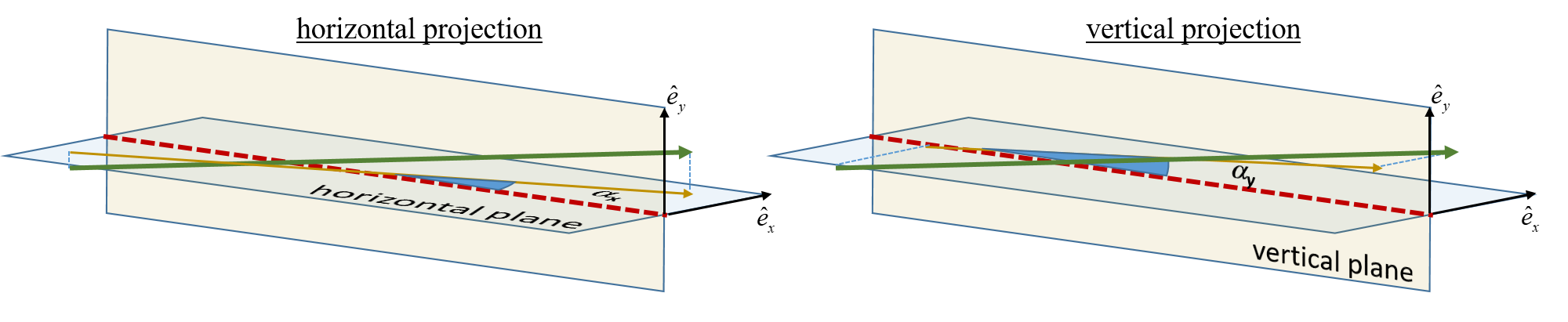}
\caption{Transverse position and angular displacements in paraxial optics can be derived from the projection of the~particle's orbit on the horizontal and vertical planes.}
\label{fig:paraxial-optics}
\end{center}
\end{figure}

%_______________________________________________________________________________
%---------------------------------------------------------------------------------------------------------------
We now want to start with a simplified treatment of the transverse beam dynamics utilizing an approach which is typically applied in geometric light optics.
Here, the transverse position of a light ray is characterized by its transverse displacements from the optical axis.
We can adopt this approach to charged particles moving through magnetic bending and focusing fields by first defining a reference path which corresponds to the optical axis in light optics.
The reference path indicates the trajectory of an ``ideal'' particle (the {\em reference particle}) moving along the design orbit and can thus be identified with the design orbit.
Using our coordinate system following the design orbit, we can define transverse displacements $x(s)$ and $y(s)$ from the design orbit at a given longitudinal position $s$ and therewith characterize the~transverse position ($x,y$) of an individual beam particle.
This is illustrated in Fig.~\ref{fig:reference-path} for the horizontal plane.
%---------------------------------------------------------------------------------
%--------------------------------------------------------------------------------
The change of the transverse displacements are defined by the derivatives
%------------------
\begin{equation}
x^{\prime} = \frac{\mathrm{d} x}{\mathrm{d} s} = \tan \alpha_x~, ~~~~~~~~~
y^{\prime} = \frac{\mathrm{d} y}{\mathrm{d} s} = \tan \alpha_y~,
\label{eq:pointing-definition}  
\end{equation}
%------------------
and are often called the {\em angular displacements} since they are related to the angles $\alpha_x, \alpha_y$ of the projection of the particle's trajectory on the ($x,s$) or ($y,s$) planes with the reference orbit as illustrated in Fig.~\ref{fig:paraxial-optics}.
%---------------------------------------------------------------------------------
%--------------------------------------------------------------------------------

In case of small displacements $x,y \ll \rho$ and $x^\prime,y^\prime \ll1$ which corresponds to the so-called {\em paraxial optics},  $\alpha_{x,y} = \tan \alpha_{x,y}$ holds in good approximation.
We can then characterize each particle by its horizontal and vertical position-dependent vectors
%------------------
\begin{equation}
\vec x = \begin{pmatrix} x \\x^\prime \end{pmatrix}, ~~~~~~~~~
\vec y = \begin{pmatrix} y \\\ y^\prime \end{pmatrix}~.
\label{eq:horizontal-position-vectors}  
\end{equation}
%------------------
The planes $(x,x^\prime)$, $(y, y^\prime)$ are called the transverse {\em trace spaces}.

We will now concentrate on first two magnetic multpoles and compute the changes of the initial vectors $\vec x_0$, $\vec y_0$ caused by a beam line ``element'' like a drift, dipole magnet or quadrupole magnet.
%---------------------------------------------------------------------------------
\begin{figure}[ht]
\begin{center}
\includegraphics[width=10cm]{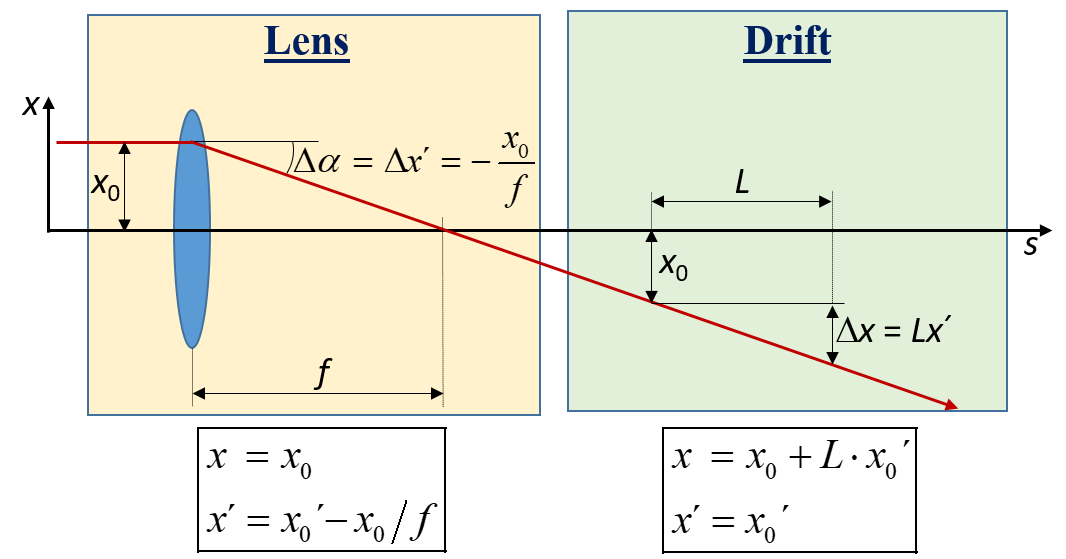}
\caption{Change of the transverse position and angular displacments in a drift and a focusing lens when applying the appromiation of paraxial optics.}
\label{fig:drift-lens-paraxial}
\end{center}
\end{figure}
%--------------------------------------------------------------------------------

In a {\bf drift section}, the angular displacements remain unchanged
%------------------
\begin{equation}
x^\prime = x^\prime_0~, ~~~~~~~~~
y^\prime = y^\prime_0~,
\label{eq:drift-x}  
\end{equation}
%------------------
whereas position displacements will change linearly with increasing drift length $L$
%------------------
\begin{equation}
x = x_0 + Lx^\prime~, ~~~~~~~~~
y = y_0 + L y^\prime~.
\label{eq:drift-xprime}  
\end{equation}
%------------------

Since a {\bf dipole magnet} will deflect each particle on a circular orbit with radius $\rho$ independent of its initial transverse displacements, it will to first approximation act like a drift in free space.

When a charged particle traverses a {\bf thin quadrupole magnet} of length $L \ll f$ it will only get a~kick changing the angular displacements by an angle $\Delta \alpha$ depending on the position displacement and the quadrupole's focal length
%------------------
\begin{equation}
x^\prime = x^\prime_0 - \frac{x^\prime_0}{f_x}~, ~~~~~~~~~
y^\prime = y^\prime_0 - \frac{y^\prime_0}{f_y}~, ~~~~~~~~~
f_x = - f_y~,
\label{eq:quad-xprime}  
\end{equation}
%------------------
whereas its position displacements will remain unchanged, thus
%------------------
\begin{equation}
x = x_0~, ~~~~~~~~~
y = y_0~.
\label{eq:quad-x}  
\end{equation}
%------------------
Hence, the action of a drift space and the first two magnetic multipoles can be described by linear maps, where the trace space vectors $\vec x$ and $\vec y$ are transformed by multiplying a matrix $\mathbf{M}_x$ or $\mathbf{M}_y$
%------------------
\begin{equation}
\vec x = \mathbf{M}_x \vec x~, ~~~~~~~~~
\vec y = \mathbf{M}_y \vec y~,
\label{eq:lin-map}  
\end{equation}
%------------------
which is called the horizontal or vertical {\em transfer matrix} of the corresponding beam line element.
From Eqs.~(\ref{eq:drift-x})--(\ref{eq:quad-x}) we obtain the following transfer matrix for a drift or dipole of length $L$
%------------------
\begin{equation}
\mathbf{M}_\text{drift,x} = \mathbf{M}_\text{drift,y} =
\mathbf{M}_\text{dip,x}= \mathbf{M}_\text{dip,y} =
\begin{pmatrix} 1 & L \\ 0 & 1 \\ \end{pmatrix}~,
\label{eq:maps-geom-drift}  
\end{equation}
%------------------
and for a thin quadrupole
%------------------
\begin{equation}
\mathbf{M}_\text{quad,x} = \begin{pmatrix} 1 & 0 \\ -\frac{1}{f_x} & 1 \\ \end{pmatrix}, ~~~~~~~~~
\mathbf{M}_\text{quad,y} = \begin{pmatrix} 1 & 0 \\ -\frac{1}{f_y} & 1 \\ \end{pmatrix}
\label{eq:maps-geom-quad}  ~.
\end{equation}
%------------------
Please note that these matrices represent only the lowest order approximation of the linear maps for the~first two magnetic multipoles (treating dipoles like a drift and restricting to thin quadrupoles)!

Applying this formalism to a real beam transfer line, the transformation of the trace space vectors of a particle can be calculated by simple matrix multiplication.
For instance, we obtain for the change of horizontal vector $\vec X_0$ at the entrance of our example beam line in Fig.~\ref{fig:reference-path}
%------------------
\begin{equation}
\vec x = \mathbf{M}_\text{drift,6} \cdot \mathbf{M}_\text{quad,3} \cdot \mathbf{M}_\text{drift,5} \cdot
\mathbf{M}_\text{dip,2} \cdot \mathbf{M}_\text{drift,4} \cdot \mathbf{M}_\text{quad,2} \cdot
\mathbf{M}_\text{drift,3} \cdot \mathbf{M}_\text{dip,1} \cdot \mathbf{M}_\text{drift,2} \cdot
\mathbf{M}_\text{quad,2} \cdot \mathbf{M}_\text{drift,1} \cdot \vec x_0~,
\label{eq:matrix-mult-beamline}  
\end{equation}
%------------------
where the matrices of the elements have to be ordered from right to left, thus starting with the last element (drift,6 - the last drift section) and ending with the first element (drift,1 - the first drift section).
The complete beam transfer line can thus be represented by a single matrix, called the {\em transfer matrix} $\mathbf{M}$, which is computed by multiplying the matrices $\mathbf{M}_i $ of the individual elements $i$:
%------------------
\begin{equation}
\mathbf{M} = \prod_i \mathbf{M}_i~.
\label{eq:matrix-mult}  
\end{equation}
%------------------
If one is interested in the displacements at an intermediate position $s$ in the beam line, this multiplication then only comprises the matrices of the elements located before this position:
%------------------
\begin{equation}
\mathbf{M}(s) = \prod_{i=1}^{N(s)} \mathbf{M}_i ~~~ \rightarrow ~~~ \vec x(s) = \mathbf{M}(s) \cdot \vec x_0~.
\label{eq:orbit-mult}  
\end{equation}
%------------------
This approach allows the piece-wise calculation of the particle's full trajectory.

%_______________________________________________________________________________
%---------------------------------------------------------------------------------------------------------------
\subsection{Equations of motion}
\label{sec:equations-of-motions}
%_______________________________________________________________________________
%---------------------------------------------------------------------------------------------------------------
A more correct treatment of the transverse beam dynamics requires solving the equations of motion.
For this, we first need to derive these equations.
In order to continue using linear maps, we will restrict our treatment to an approximation considering only linear terms.

We start introducing the transverse position vector of a moving particle at longitudinal position $s$:
%------------------
\begin{equation}
\vec R(s) = \left [ \rho + x(s) \right ] \cdot \ \hat{e}_x(s) + y(s) \cdot \hat{e}_y(s)~.
\label{eq:transv-position-vector}  
\end{equation}
%------------------
%---------------------------------------------------------------------------------
\begin{figure}[ht]
\begin{center}
\includegraphics[width=14cm]{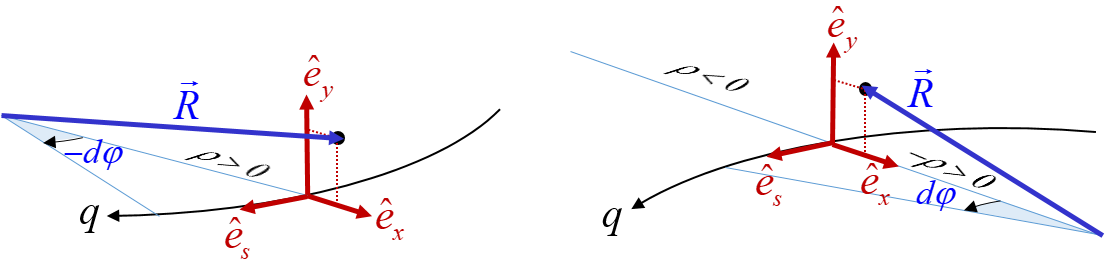}
\caption{Coordinates of a charged particle moving on a curved orbit. Left: local orbit with positive curvature $\kappa = \frac{1}{\rho} > 0$. Right: local orbit with negative curvature $\kappa = \frac{1}{\rho} < 0$.}
\label{fig:polar-coordinates}
\end{center}
\end{figure}
%--------------------------------------------------------------------------------
Please note that the unit coordinate vectors of our coordinate system following the orbit will change their orientation and will thus be dependent on $s$.
In the following, we will concentrate on a local orbit with positive curvature (the case $\rho < 0$ can be treated fully analog).
We use polar coordinates for the~horizontal plane and get
%------------------
\begin{equation}
\hat{e}_x = \begin{pmatrix}  \cos \varphi \\ \sin \varphi \\ \end{pmatrix}, ~~~~~~~~~
\hat{e}_\varphi = -\hat{e}_s = \begin{pmatrix} -\sin \varphi \\ \cos \varphi \\ \end{pmatrix}~.
\label{eq:polar-unit-vectors}  
\end{equation}
%------------------
An increase d$s$ along the reference orbit will thus lead to a rotation of $\hat{e}_x$ and $\hat{e}_s$ by 
%------------------
\begin{equation}
\mathrm{d} \varphi = - \frac{1}{\rho} \mathrm{d}s~,
\label{eq:change-polar-unit-vectors}  
\end{equation}
%------------------
and we get for the derivative of the unit vectors with respect to $s$
%------------------
\begin{equation}
\frac{\mathrm{d}}{\mathrm{d}s} \hat{e}_x = \hat{e}_x^\prime = + \frac{1}{\rho}\hat{e}_s~, ~~~~~~~~~
\frac{\mathrm{d}}{\mathrm{d}s} \hat{e}_s = \hat{e}_s^\prime = - \frac{1}{\rho}\hat{e}_x~.
\label{eq:derivative-polar-unit-vectors}  
\end{equation}
%------------------
From here on we will use ``prime'' for derivatives with respect to the longitudinal coordinate $s$ and ``dot'' for derivatives with repesct to time $t$.
For a particle moving with speed $v_s$ along the reference orbit these derivatives are linked via
%------------------
\begin{equation}
\frac{\mathrm{d}}{\mathrm{d}t} = \frac{\mathrm{d}}{\mathrm{d}s} \frac{\mathrm{d}s}{\mathrm{d}t} =
v_s \cdot \frac{\mathrm{d}}{\mathrm{d}s}~.
\label{eq:derivatives-prime-dot-ref-orbit}  
\end{equation}
%------------------
For the following, we make four fundamental assumptions simplifying our derivations:
\begin{itemize}
\item All particles are moving with individual, but constant longitudinal velocities $\dot s = v_s$ = const. The~longitudinal velocity is much larger than the transverse velocity components $v_x$, $v_z$.
\item The curvature of the orbit is varying ``slowly''. All derivatives $\rho^\prime$, $\rho^{\prime\prime}$, ... are therefore neglected.
\item The design orbit is in the $(x, s)$ plane. The vertical coordinate vector is independent of $s$, $\hat{e}_y^\prime = 0$.
\item The transverse displacements are small compared to the bending radius of the orbit (paraxial optics). There exist no coupling of the transverse to the longitudinal motion. We will therefore neglect all the longitudinal components and put $R_s = R_s^\prime = R_s^{\prime\prime} = 0$.
\end{itemize}
Setting $r(s) = \rho + x(s)$ and stop explicitly indicating the $s$ dependence, we simply write
%------------------
\begin{equation}
\vec R = r \cdot \hat{e}_x + y \cdot \hat{e}_y~,
\label{eq:transv-position-vector-r}  
\end{equation}
%------------------
and obtain for the derivatives
%------------------
\begin{eqnarray}
\vec R^\prime &=& x^\prime \cdot \hat{e}_x + \frac{r}{\rho} \cdot \hat{e}_s + y^\prime \cdot \hat{e}_y ~,\\
\vec R^{\prime\prime} &=& \left ( x^{\prime\prime} - \frac{r}{\rho^2} \right ) \cdot \hat{e}_x + 
y^{\prime\prime} \cdot \hat{e}_y + 2 \frac{x^\prime}{\rho^2} \cdot \hat{e}_s~,
\label{eq:transv-position-vector-derivs}  
\end{eqnarray}
where all terms along $\hat{e}_s$ will be ignored according to our fundamental assumptions.
The change of the~transverse position vector $\vec R$ is caused by the external Lorentz force leading to a momentum change
%------------------
\begin{equation}
\frac{\mathrm{d}\vec p}{\mathrm{d}t} = \gamma_r m_0 \frac{\mathrm{d}^2 \vec R}{\mathrm{d}t^2} =
 q \left ( \vec v \times \vec B \right )~.
\label{eq:Lorentz-force-coordinates}  
\end{equation}
%------------------
Since in this equation the second derivative of the position vector with respect to time appears, we have to transform ``prime'' to ``dot''.
The transformation for the reference particle given by Eq.~(\ref{eq:derivatives-prime-dot-ref-orbit}) has to be generalized for particles moving with a horizontal displacement $x$ from the reference orbit with constant $v_s$.
%---------------------------------------------------------------------------------
\begin{figure}[ht]
\begin{center}
\includegraphics[width=8cm]{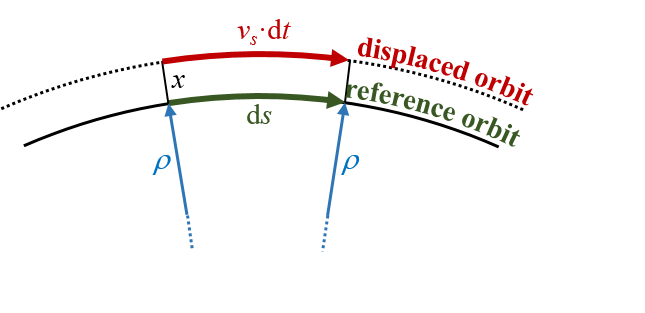}
\caption{Path length element $\mathrm{d}s$ for a particle moving with constant velocity $v_s$ and horizontal offset $x$ at radius $r = \rho + x$.}
\label{fig:path-length-difference}
\end{center}
\end{figure}
%--------------------------------------------------------------------------------
From geometrical considerations we obtain (cf.\ Fig.~\ref{fig:path-length-difference})
%------------------
\begin{equation}
\mathrm{d}s = v_s \mathrm{d}t \left ( \frac{\rho}{\rho+x} \right ) =
\left ( \frac{\rho}{r} \right ) v_s \mathrm{d}t ~~~~ \rightarrow ~~~~
\frac{\mathrm{d}}{\mathrm{d}t} = v_s \left ( \frac{\rho}{r} \right ) \frac{\mathrm{d}}{\mathrm{d}s}~,
\label{eq:derivatives-prime-dot}  
\end{equation}
%------------------
and therewith get
%------------------
\begin{equation}
\ddot{\vec R} = \left ( v_s \frac{\rho}{r} \right )^2 \vec R^{\prime\prime} =
\left ( v_s \frac{\rho}{r} \right )^2 \cdot
\left [ \left ( x^{\prime\prime} - \frac{r}{\rho^2} \right ) \cdot \hat{e}_x + 
y^{\prime\prime} \cdot \hat{e}_y \right ]~.
\label{eq:transv-position-vector-temp-deriv}  
\end{equation}
%------------------
Since $v_x \ll v_s$ and $v_y \ll v_s$ the Lorentz force can be simplified to
%------------------
\begin{equation}
q \left ( \vec v \times \vec B \right ) =
q \left [- v_s B_y \hat{e}_x + v_s B_x \hat{e}_y + \left ( B_y v_x - B_x v_y \right ) \hat{e}_s \right ] \approx
q v_s \left ( - B_y \hat{e}_x + B_x \hat{e}_y \right )~.
\label{eq:Lorentz-force-linearized}  
\end{equation}
%------------------
Combining Eqs.~(\ref{eq:Lorentz-force-coordinates}), (\ref{eq:transv-position-vector-temp-deriv}) and (\ref{eq:Lorentz-force-linearized}), and using $p=\gamma_r m_0 v_s$ we obtain the equations of motion with respect $s$:
%------------------
\begin{equation}
\left ( x^{\prime\prime} - \frac{x + \rho}{\rho^2} \right ) \cdot \hat{e}_x + y^{\prime\prime} \cdot \hat{e}_y =
\frac{q}{p} \left ( \frac{x + \rho}{\rho} \right )^2 \left ( -B_y \hat{e}_x + B_x \hat{e}_y \right )~.
\label{eq:transv-equation-B}  
\end{equation}
%------------------
We restrict the magnetic field components to the first two normal magnetic multipoles 
%------------------
\begin{equation}
B_x = - \frac{p_0}{q} ky~, ~~~~~~~~~~ B_y = \frac{p_0}{q} \left ( \frac{1}{\rho} - kx \right )~.
\label{eq:B-decomposition}  
\end{equation}
%------------------
where the multipole strengths $\frac{1}{\rho}$ and $k$ are normalized to the momentum $p_0$ of the reference particle, the {\em reference momentum}.
Since Eq.~(\ref{eq:transv-equation-B}) deals with the actual momentum $p$ of a non-reference particle, we have to find a simple approximation of the fraction $\frac{p_0}{p}$.
We therefore assume a small momentum deviation $\Delta p = p - p_0$ and approximate the reciprocal value $\frac{1}{p}$ by its first order Taylor expansion
%------------------
\begin{equation}
\frac{1}{p} \approx \frac{1}{p_0} + \Delta p \left. \frac{\partial \left ( 1/p \right )}{\partial p} \right|_{p=p_0} =
\frac{1}{p_0} \left( 1 - \frac{\Delta p}{p_0} \right )~.
\label{eq:p-Taylor-expansion}  
\end{equation}
%------------------
Inserting this we obtain
%------------------
\begin{eqnarray}
x^{\prime\prime} &=& \frac{x}{\rho^2} + \frac{1}{\rho} - \left ( 1 - \frac{\Delta p}{p_0} \right )
\left ( 1 + \frac{x}{\rho} \right )^2 \left ( \frac{1}{\rho} - kx \right )~, \\
y^{\prime\prime} &=& - \left ( 1 - \frac{\Delta p}{p_0} \right ) \left ( 1 + \frac{x}{\rho} \right )^2 ky~.
\end{eqnarray}
%------------------
Keeping only the lowest order and neglecting all non-linear terms finally lead to the linearized equations of motion
%------------------
\begin{equation}
\boxed{
\begin{split}
x^{\prime\prime}(s) + \left ( \frac{1}{\rho^2(s)} - k(s) \right ) \cdot x(s) &= \frac{1}{\rho} \frac{\Delta p}{p_0} \\
y^{\prime\prime}(s) + k(s) \cdot y(s) &= 0
\end{split}
}~.
\label{eq:linearized-equations-of-motion} 
\end{equation}
%------------------

It is once again worth mentioning the list of approximations which have been used to derive these equations:
\begin{itemize}
\item {\bf paraxial optics} where $x, y \ll \rho$ and $v_s = \text{const.}$,
\item {\bf flat accelerators} for which the design orbit lies in a plane,
\item {\bf linear field changes} represented by the first two magnetic multipoles (dipoles and quadrupoles),
\item {\bf upright magnets} ensuring that the motion in the horizontal and vertical plane is decoupled,
\item {\bf quasi monochromatic beam} with only small deviations $\Delta p$ from the reference momentum $p_0$,
\item {\bf only linear terms} in $x, y, \Delta p$ thus restricting to linear beam optics.
\end{itemize}
One could think that these restrictions would significantly limit the usability of the linearized equations of motion.
In fact it turns out that this is not the case.
We will discover that the application of Eqs.~(\ref{eq:linearized-equations-of-motion}) in the framework of the {\em transverse linear beam dynamics} will lead to valuable and deep insights into beam dynamics in particle accelerators.
A treatment of more sophisticated non-linear problems is, however, not possible using Eqs.~(\ref{eq:linearized-equations-of-motion}).
This requires a different approach using the Hamilton formalism and non-linear maps, which is out of the scope of this introductory course.

%===========================================================================
%===========================================================================
\section{Matrix formalism}
\label{sec:matrix-formalism}
%===========================================================================
%===========================================================================
In the linearized equations of motion, the transverse dynamics of charged particles in an accelerator (or a~beam transfer line) is fully determined by the location of the installed dipole and quadrupole magnets (the~{\em lattice}), their individual multipole strengths $\frac{1}{\rho}$ and $k$, and the particles' relative momentum deviations $\frac{\Delta p}{p_0}$.
Since the multipole strengths will change from magnetic element to magnetic element and thus with longitudinal position $s$, an analytical solution of the linearized equations of motion can only be derived for special and simple cases.
However, when concentrating on a beam line element for which the multipole strengths remain constant, the equations of motion turn to linear differential equations with constant coefficients which can be easily solved analytically.
Thus, one powerful method to handle the problem of the $s$ dependence of the coefficients is to segment the orbit in segments with constant multipole strengths, solve Eq.~(\ref{eq:linearized-equations-of-motion}) for each section seperately, and connect the solutions for the different sections to a continuous function.
This is possible since the equations of motion are linear, allowing to represent the solutions for the different sections by linear maps (matrices).
As already presented in Section \ref{sec:geometric-optics}, these matrices can be applied to the position vectors $\vec x$ and $\vec y$ transforming the particle's initial coordinates before the element to its final coordinates after the element.
This procedure can be repeated element by element by taking the final coordinates after the previous element as initial coordinates before the next one and thus leading to a sequential application of the matrices of the individual elements as in Eq.~(\ref{eq:orbit-mult}), from which the transfer matrix $\mathbf{M}(s_0,s)$ of a choosen larger section $s_0 \rightarrow s$ of a complicated magnetic lattice can be calculated:
%------------------
\begin{equation}
\mathbf{M}(s_0,s) = \prod_{i=m(s_0)}^{n(s)} \mathbf{M}_i ~~~ \rightarrow ~~~ 
\vec x(s) = \mathbf{M}(s_0,s) \cdot \vec x(s_0)~.
\label{eq:orbit-mult-s}  
\end{equation}
%------------------
Hence, the problem reduces to derive the transfer matrices of the individual beam line elements with constant multipole strengths, from which the transfer matrix of a given lattice can be calculated according to Eq.~(\ref{eq:orbit-mult-s}).
These are namely the free drift, the dipole magnets and the quadrupole magnets (we will not consider so-called {\em combined function magnets} with non-vanishing dipole and quadrupole strengths). 
Since we are in the first instance not interested in the influence of a momentum deviation, we will derive the matrices setting $\frac{\Delta p}{p_0} = 0$ (monochromatic beam).
The impact of a non-vanishing momentum deviation will be presented in the last section on dynamics with off momentum particles.

%_______________________________________________________________________________
%---------------------------------------------------------------------------------------------------------------
\subsection{4x4 vectors and transverse trace spaces}
\label{sec:4x4-vectors-and-transverse-trace-spaces}
%_______________________________________________________________________________
%---------------------------------------------------------------------------------------------------------------
We will characterize a particle's state by a 4x4 vector built from the transverse displacements from the~design orbit (our relative coordinates):
%------------------
\begin{equation}
\begin{pmatrix}  x \\ x^\prime \\ y \\ y^\prime \end{pmatrix} =
\begin{pmatrix} \text{horizontal displacement} \\ \text{horizontal angular displacement} \\
\text{vertical displacement} \\ \text{vertical angular displacement} \end{pmatrix}~.
\label{eq:4x4-vector}  
\end{equation}
%------------------
The plane $(x, x^\prime )$ forms the {\em horizontal trace space}, the plane $(y, y^\prime )$ forms the {\em vertical trace space}.
Both are often called {\em phase space}, which is, strictly speaking, not correct since the phase space is represented by the canonical conjugated coordinates $(x, p_x)$ or $(y, p_y)$ respectively.
In case of monochromatic beams with constant momentum (e.g.\ no post acceleration), the angular displacements are linked to the transverse momenta by a constant factor $\beta_r \gamma_r m_0 c$ (cf.\ Section~\ref{sec:dynamics-acceleration})
%------------------
\begin{equation}
p_x = \beta_r \gamma_r m_0 c \cdot x^\prime~, ~~~~~~~~~ p_y = \beta_r \gamma_r m_0 c \cdot y^\prime~,
\end{equation}
%------------------
and the trace space can be identified with the phase space.

We will use the matrix formalism applying 4x4 matrices to the 4x4 vectors to compute particle trajectories.
Since we restricted our treatment to upright magnets, there will be no coupling of the~transverse planes.
Thus, the 4x4 matrices of all magnetic variants will simplify to
%------------------
\begin{equation}
\mathbf{M} = \begin{pmatrix}
r_{11} & r_{12} & 0 & 0 \\ r_{21} & r_{22} & 0 & 0 \\ 0 & 0 & r_{33} & r_{44} \\ 0 & 0 & r_{43} & r_{44} \end{pmatrix}~,
\end{equation}
%------------------
with maximum 8 non-vanishing elements $r_{ik}$.

%_______________________________________________________________________________
%---------------------------------------------------------------------------------------------------------------
\subsection{Drift space}
\label{sec:drift-space}
%_______________________________________________________________________________
%---------------------------------------------------------------------------------------------------------------
Since $k = \frac{1}{\rho} = 0$, the equations of motion for a free drift simplify to 
%------------------
\begin{equation}
\begin{split}
x^{\prime\prime}(s) = 0~, \\
y^{\prime\prime}(s) = 0~.
\end{split}
\end{equation}
%------------------
We thus get for known initial values $x_0, x^\prime_0, y_0, y^\prime_0$ and a drift length $L$
%------------------
\begin{equation}
\begin{array} {ll}
x(L) = x_0 + L \cdot x^\prime_0 \\ y(L) = y_0 + L \cdot y^\prime_0
\end{array}
~~~~~~~ \text{and} ~~~~~~~
\begin{array} {ll}
x^\prime(L) = x^\prime_0 \\ y^\prime(L) = y^\prime_0
\end{array}~,
\end{equation}
%------------------
yielding the transfer matrix of a drift with length $L$:
%------------------
\begin{equation}
\mathbf{M}_\text{drift} = \begin{pmatrix} 1 & L & 0 & 0 \\ 0 & 1 & 0 & 0 \\ 0 & 0 & 1 & L \\ 0 & 0 & 0 & 1 \end{pmatrix}~.
\label{eq:drift-matrix}  
\end{equation}
%------------------

%_______________________________________________________________________________
%---------------------------------------------------------------------------------------------------------------
\subsection{Dipole magnet}
\label{sec:dipole-magnet}
%_______________________________________________________________________________
%---------------------------------------------------------------------------------------------------------------
We will first consider a sector dipole magnet for which the end faces are perpendicular to the reference orbit.
%---------------------------------------------------------------------------------
\begin{figure}[ht]
\begin{center}
\includegraphics[width=8cm]{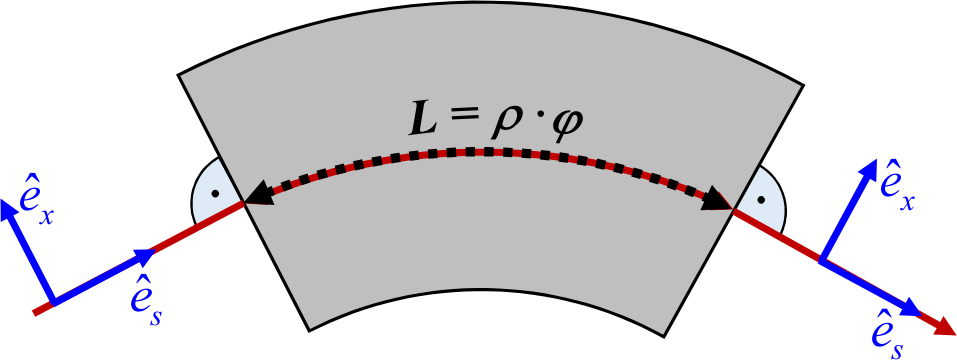}
\caption{Left: Sector dipole magnet for which both magnet end faces are perpendicular to the reference orbit. The~length of the curved orbit is the arc length $L$.}
\label{fig:sector-dipole}
\end{center}
\end{figure}
%--------------------------------------------------------------------------------
When assuming that the magnetic field drops instantaneously to zero outside the dipole and remains constant in the dipole, the reference orbit is bent with constant radius $\rho$ between the two end faces. The length $L$ of the curved reference orbit (the ``arc length'') is determined by the deflection angle $\varphi$ and the bending radius $\rho$ and equals
%------------------
\begin{equation}
L = \rho \cdot \varphi~.
\end{equation}
%------------------
Since $k = 0$ for a dipole magnet, the equations of motion reduce to
%------------------
\begin{equation}
{
\begin{split}
x^{\prime\prime}(s) + \frac{x}{\rho^2} &= 0~, \\
y^{\prime\prime}(s) &= 0~,
\end{split}
}
\end{equation}
%------------------
and are solved for known initial values $x_0, x^\prime_0, y_0, y^\prime_0$ by
%------------------
\begin{equation}
{
\begin{split}
x(L) &= x_0 \cdot \cos \varphi + \frac{x^\prime_0}{\rho} \cdot \sin \varphi~, \\
y(L) &= y_0 + L \cdot y^\prime_0~,
\end{split}
}
\end{equation}
%------------------
thus yielding the transfer matrix of a sector dipole magnet of arc length $L = \rho \varphi$
%------------------
\begin{equation}
\mathbf{M}_\text{dip} = \begin{pmatrix} \cos \varphi & \rho \sin \varphi & 0 & 0 \\
- \frac {1}{\rho} \sin \varphi & \cos\varphi & 0 & 0 \\
0 & 0 & 1 & \rho \varphi \\ 0 & 0 & 0 & 1 \end{pmatrix}~.
\label{eq:sector-dipole-matrix}  
\end{equation}
%------------------
A sector dipole is therefore focusing in the horizontal plane and acts like a drift in the vertical plane.

\begin{figure}[ht]
\begin{center}
\includegraphics[width=15cm]{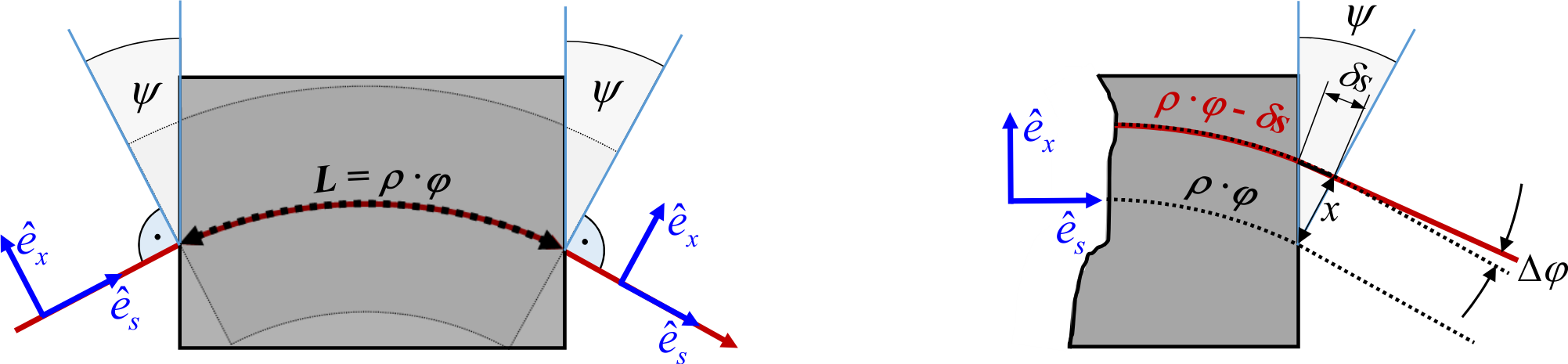}
\caption{Left: Rectangular dipole magnet for which both magnet end faces are parallel and passed by the reference orbit under an angle 90°+\,$\psi$ greater than 90°. The length of the curved reference orbit is the arc length $L$.
Right: A particle with offset $x$ passes the dipole face which is inclined in the $(x, s)$ plane. Due to the reduced path length $\delta s$ in the dipole it is less deflected by an angle $\Delta \varphi$.}
\label{fig:rectangular-dipole}
\end{center}
\end{figure}

Often, dipole magnets are built as rectangular magnets for which the end faces are parallel to each other and passed by the beam under an angle different from 90°.
Here, we will only consider the situation with a symmetric path, cf.\ Fig.~\ref{fig:rectangular-dipole}.
%---------------------------------------------------------------------------------
%--------------------------------------------------------------------------------

In the bending plane, the reference orbit has an angle $\Psi$ with the normal to the end surface.
Particles with a horizontal offset $x$ either have a longer or shorter path in the dipole field than the reference particle, the path  length difference is $\delta s = - x \cdot \tan \Psi$.
These particles will be bent more or less, resulting in an increase or decrease of total deflection angle by $\Delta \varphi = \frac{\delta s}{\rho} = - \frac{x}{\rho} \tan \Psi$ and thus leading to an~additional change of the angular displacement by $\Delta x^\prime = - \Delta \varphi$.
This additional bending is called the~horizontal ``edge effect'' and can be described by an additional matrix $\mathbf{M}_{\text{edge},x}$ acting on $x$ and $x^\prime$ in the~horizontal trace space
%------------------
\begin{equation}
\mathbf{M}_{\text{edge},x} = \begin{pmatrix} 1 & 0 \\ \frac{\tan \Psi}{\rho} & 1 \end{pmatrix}~,
\label{eq:edge-matrix-x}  
\end{equation}
%------------------
following and/or preluding the matrix of a sector dipole.
It acts focusing or defocusing on the beam, depending on the sign of $\Psi$ and is called {\em edge focusing}.
For a symmetric path in a rectangular magnet, $\Psi$ is positive and we obtain an additional horizontal defocusing.

The situation in the vertical plane is different.
A correct treatment of the vertical edge effect requires to consider that the magnetic field cannot instantaneously drop off behind the end faces, thus leading to so called ``fringe fields'' having a non-vanishing component $B_z \neq 0$ perpendicular to the end face which decreases with increasing distance $z$ to the end face.
%---------------------------------------------------------------------------------
\begin{figure}[ht]
\begin{center}
\includegraphics[width=15cm]{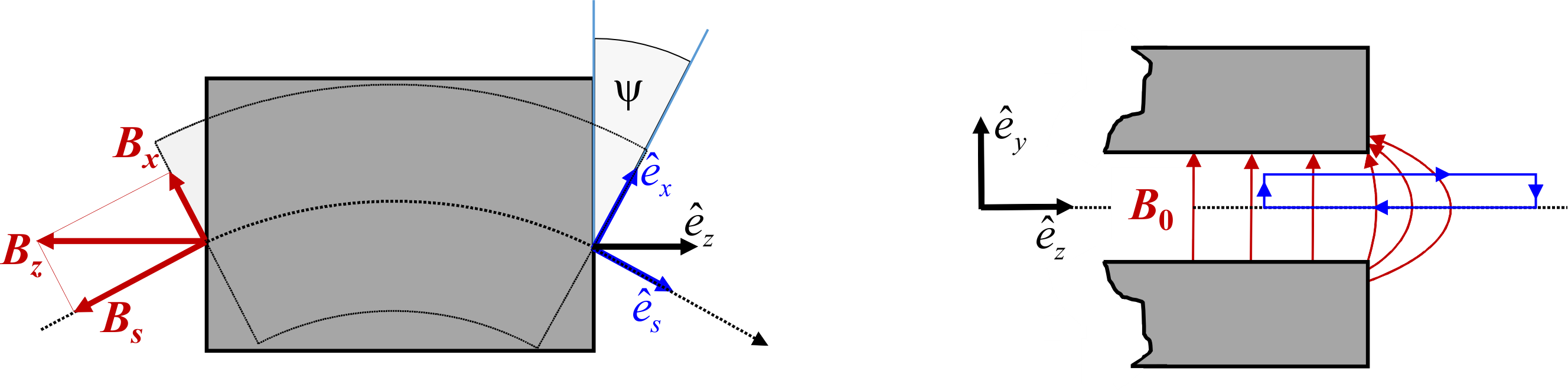}
\caption{Left: Decomposition of the $z$ component of the fringe field into components parallel and perpendicular to the particle's orbit.
Right: The chosen integration path for calculating the $z$ component of the fringe field.}
\label{fig:edgefield-path}
\end{center}
\end{figure}
%--------------------------------------------------------------------------------

The integral effect of this fringe field component can be estimated by applying Ampère's law and chosing a closed integration path in the plane perpendicular to the end face, cf.\ Fig.~\ref{fig:edgefield-path}
%------------------
\begin{equation}
0 = \oint \vec B \cdot \mathrm{d}\vec s = \left. \int_0^y B_y \mathrm{d}y \right|_\text{inside} +
\int_\text{inside}^\text{outside} B_z(y) \mathrm{d}z + \left. \int_y^0 B_y \mathrm{d}y \right|_\text{outside} +
\int^\text{inside}_\text{outside} B_z(y=0) \mathrm{d}z~.
\end{equation}
%------------------
Since neither the horizontal path at the reference orbit nor the vertical path far away outside the magnet make a contribution,
we simply get
%------------------
\begin{equation}
B_0 \cdot y = - \int_\text{inside}^\text{outside} B_z(y) \mathrm{d}z~,
\label{eq:vert-fringe}
\end{equation}
%------------------
where $B_0$ denotes the homogeneous vertical bending field inside the dipole magnet.
The field $B_z$ can be decomposed in a longitudinal and a transverse component (cf.\ Fig.~\ref{fig:edgefield-path}),
%------------------
\begin{equation}
B_x= - B_z \cdot \sin \Psi~, ~~~~~~~~~~ B_s = B_z \cdot \cos \Psi~,
\end{equation}
%------------------
of which only the transverse component will have an impact on the beam.
The resulting vertical bending angle can be calculated by integrating along $\mathrm{d}s = \frac{\mathrm{d}z}{\cos \Psi}$, yielding
%------------------
\begin{equation}
\Delta y^\prime = \frac{q}{p_0} \int B_x \mathrm{d}s = \frac{q}{p_0} \tan \Psi \int B_z \mathrm{d}z~.
\end{equation}
%------------------
Using Eq.~(\ref {eq:vert-fringe}) and $\frac{q}{p_0} = \frac{1}{B_0 \rho}$, we finally obtain
%------------------
\begin{equation}
\Delta y^\prime = - \frac{\tan \Psi}{\rho} y~.
\end{equation}
%------------------
Thus, the vertical edge effect can be described by an additional matrix $\mathbf{M}_{\text{edge},y}$ acting on $y$ and $y^\prime$ in the~vertical trace space
%------------------
\begin{equation}
\mathbf{M}_{\text{edge},y} = \begin{pmatrix} 1 & 0 \\ - \frac{\tan \Psi}{\rho} & 1 \end{pmatrix}~,
\label{eq:edge-matrix-y}  
\end{equation}
%------------------
again following and/or preluding the matrix of a sector dipole.
For a symmetric path in a rectangular magnet, we obtain an additional vertical focusing.

Both matrices $\mathbf{M}_{\text{edge},x}$ and $\mathbf{M}_{\text{edge},y}$ can be combined to a 4x4 matrix describing the edge effect when the beam enters / exits the magnetic field of a rectangular dipole magnet:
%------------------
\begin{equation}
\mathbf{M}_\text{edge} = \begin{pmatrix} 1 & 0 & 0 & 0\\ \frac{\tan \Psi}{\rho} & 1 & 0 & 0 \\
0 & 0 & 1 & 0\\ 0 & 0 &- \frac{\tan \Psi}{\rho} & 1 \end{pmatrix}~.
\label{eq:edge-matrix}  
\end{equation}
%------------------
Since for a symmetric path $\Psi = \frac{\varphi}{2}$, we obtain for the transfer matrix of a rectangular dipole magnet
%------------------
\begin{equation}
\mathbf{M}_\text{rect} =\mathbf{M}_\text{edge} \cdot \mathbf{M}_\text{dip} \cdot \mathbf{M}_\text{edge} =
\begin{pmatrix} 1 & \rho \sin \varphi & 0 & 0 \\ 0 & 1 & 0 & 0 \\
0 & 0 & 1 - \frac{\rho \varphi}{f} & \rho \varphi \\ 0 & 0 & \frac{\rho \varphi}{f^2} - \frac{2}{f} & 1 - \frac{\rho \varphi}{f}
\end{pmatrix}~,
\label{eq:rect-dipole-matrix}  
\end{equation}
%------------------
where we have put $\frac{1}{f}=\frac{1}{\rho}\tan\left ( \frac{\varphi}{2} \right )$.
Using $\tan \left (\frac{ \varphi}{2} \right ) = \frac{1 - \cos \varphi}{\sin \varphi}$ and $\rho \sin \varphi \approx \rho\varphi$ we approximately get
%------------------
\begin{equation}
\mathbf{M}_\text{rect} \approx \begin{pmatrix} 1 & \rho\varphi & 0 & 0 \\ 0 & 1 & 0 & 0 \\
0 & 0 & \cos \varphi & \rho\varphi \\ 0 & 0 & -\frac{\sin^2 \varphi}{\rho\varphi} & \cos \varphi
\end{pmatrix}~.
\label{eq:rect-dipole-matrix-approx}  
\end{equation}
%------------------
A rectangular magnet thus shows exactly the opposite behaviour of a sector dipole manget: it is focusing in the vertical plane and acts like a drift in the horizontal plane.

%_______________________________________________________________________________
%---------------------------------------------------------------------------------------------------------------
\subsection{Quadrupole magnet}
\label{sec:quadrupole-magnet}
%_______________________________________________________________________________
%---------------------------------------------------------------------------------------------------------------
For a pure quadrupole magnet the bending term vanishes, $\frac{1}{\rho} = 0$.
The equations of motion reduce to 
%------------------
\begin{equation}
{
\begin{split}
x^{\prime\prime}(s) - k(s)  \cdot x(s) = 0~, \\
y^{\prime\prime}(s) + k(s) \cdot y(s) = 0~.
\end{split}
}
\end{equation}
%------------------
Their solutions depend on the sign of the quadrupole strength $k$.
For  $k < 0$ we get the solution for a~quadrupole magnet QF which is horizontal focusing and vertical defocusing
%------------------
\begin{equation}
{
\begin{split}
x(s) &= a \cdot \cos \left ( \sqrt{\vert k \vert}s \right ) + b \cdot \sin \left ( \sqrt{\vert k \vert}s \right )~, \\
y(s) &= c \cdot \cosh \left ( \sqrt{\vert k \vert}s \right ) + d \cdot \sinh \left ( \sqrt{\vert k \vert}s \right )~.
\end{split}
}
\end{equation}
%------------------
The integration constants $a, b, c, d$ can be derived from the initial values $x_0, x^\prime_0, y_0, y^\prime_0$ at $s = 0$:
%------------------
\begin{equation}
{
\begin{split}
x(s=0) = a = x_0~, ~~~~~~~~ x^\prime(s=0) = \sqrt{\vert k \vert} b = x^\prime_0~,\\
y(s=0) = c = y_0~, ~~~~~~~~ y^\prime(s=0) = \sqrt{\vert k \vert} d = y^\prime_0~.
\end{split}
}
\end{equation}
%------------------
Substituting and building the first derivative, we obtain the transfer matrix $\mathbf{M}_\text{QF}$ of a horizontal focusing quadrupole QF of length $L$ which for $L \rightarrow 0$ reduces to the first order approximation presented in Section~\ref{sec:geometric-optics} (cf.\ Eq.~(\ref{eq:maps-geom-quad}))
%------------------
\begin{equation}
\mathbf{M}_\text{QF} = \begin{pmatrix} 
\cos \Omega & \frac{1}{\sqrt{\vert k \vert}} \sin \Omega & 0 & 0 \\
-\sqrt{\vert k \vert} \sin \Omega & \cos \Omega & 0 & 0 \\
0 & 0 & \cosh \Omega & \frac{1}{\sqrt{\vert k \vert}} \sinh \Omega \\
0 & 0 & \sqrt{\vert k \vert} \sinh \Omega & \cosh \Omega 
\end{pmatrix}
\overset{L \rightarrow 0}{~~~\approx~~~}
\begin{pmatrix}
1 & 0 & 0 & 0 \\ -\frac{1}{f} & 1 & 0 & 0 & \\ 0 & 0 & 1 & 0 \\ 0 & 0 & \frac{1}{f} & 1
\end{pmatrix}~,
\label{eq:QF-matrix-approx}  
\end{equation}
%------------------
where we have set $\Omega = \sqrt{\vert k \vert}\cdot L$ and $\frac{1}{f} = -k L$.

The case $k > 0$ can be treated completely analog.
It deals with a horizontal defocusing quadrupole QD represented by the transfer matrix $\mathbf{M}_\text{QD}$
%------------------
\begin{equation}
\mathbf{M}_\text{QD} = \begin{pmatrix} 
\cosh \Omega & \frac{1}{\sqrt{\vert k \vert}} \sinh \Omega & 0 & 0 \\
\sqrt{\vert k \vert} \sinh \Omega & \cosh \Omega & 0 & 0 \\
0 & 0 & \cos \Omega & \frac{1}{\sqrt{\vert k \vert}} \sin \Omega \\
0 & 0 & -\sqrt{\vert k \vert} \sin \Omega & \cos \Omega 
\end{pmatrix}
\overset{L \rightarrow 0}{~~~\approx~~~}
\begin{pmatrix}
1 & 0 & 0 & 0 \\ \frac{1}{f} & 1 & 0 & 0 & \\ 0 & 0 & 1 & 0 \\ 0 & 0 & -\frac{1}{f} & 1
\end{pmatrix}~.
\label{eq:QD-matrix-approx}  
\end{equation}
%------------------

%_______________________________________________________________________________
%---------------------------------------------------------------------------------------------------------------
\subsection{General properties of the transfer matrices}
\label{sec:matrix-properties}
%_______________________________________________________________________________
%---------------------------------------------------------------------------------------------------------------
Each 4x4 transfer matrix consists of two 2x2 matrices $\mathbf{M}_x$, $\mathbf{M}_y$ parametrizing the transformation of the~coordinates in the horizontal and the vertical trace space:
%------------------
\begin{equation}
\mathbf{M} = \begin{pmatrix} \mathbf{M}_x & \begin{pmatrix} 0 & 0 \\ 0 & 0 \end{pmatrix} \\
\begin{pmatrix} 0 & 0 \\ 0 & 0 \end{pmatrix} & \mathbf{M}_y \end{pmatrix}~.
\label{eq:matrix-decomposition}
\end{equation}
%------------------
The elements of the 2x2 matrices can be expressed by the two independent fundamental solutions $C(s)$ and $S(s)$ of the equations of motion and their derivatives $C^\prime(s)$ and $S^\prime(s)$.
This representation of the 2x2 transfer matrix acting either on $(x, x^\prime)$ or $(y, y^\prime)$ is known as the {\em fundamental matrix} or the ``CS-matrix''
%------------------
\begin{equation}
\mathbf{M} = \begin{pmatrix} C & S \\ C^\prime & S^\prime \end{pmatrix}~,
\label{eq:CS-matrix}
\end{equation}
%------------------
where we have skipped the index $x$ or $y$ for reasons of simplicity.
Please note that only in case of constant multipole strengths, $C(s)$ and $S(s)$ are simple Cosine and Sine functions.
In all cases, the fundamental solutions do not only depend on $s$ but as well on the initial position $s_0$.
In particular we have for $s = s_0$
%------------------
\begin{equation}
\mathbf{M}(s_0, s_0) = \begin{pmatrix} 1 & 0 \\ 0 & 1 \end{pmatrix} = \mathbf{I}~.
\label{eq:id-matrix}
\end{equation}
%------------------
So, the determinant in this case is $\det\left( \mathbf{M}(s_0,s_0) \right) = 1$, which in general is represented by the so-called {\em Wronskian}
%------------------
\begin{equation}
\det\left( \mathbf{M} \right) = \det \begin{pmatrix} C & S\\ C^\prime & S^\prime \end{pmatrix} = C S^\prime - S C^\prime~.
\label{eq:Wronskian}
\end{equation}
%----------------
The derivative of the Wronskian vanishes
%------------------
\begin{equation}
\frac{\mathrm{d}}{\mathrm{d}s} \det\left( \mathbf{M} \right) =
C^\prime S^\prime +C S^{\prime\prime} - S^\prime C^\prime - S C^{\prime\prime} =
C S^{\prime\prime} - S C^{\prime\prime} = 0~,
\label{eq:Wronskian-deriv}
\end{equation}
%----------------
since $C$ and $S$ are solutions of the equations of motion and thereby satisfy $C^{\prime\prime} = -KC$ and $S^{\prime\prime} = -KS$.
Thus, the Wronskian is constant.
Using Eq.~(\ref{eq:id-matrix}) we obtain the very important result that the determinant of the transfer matrix always equals one:
%------------------
\begin{equation}
\det\left( \mathbf{M} \right) = \det \begin{pmatrix} C & S\\ C^\prime & S^\prime \end{pmatrix} = 
C S^\prime - S C^\prime = 1~.
\label{eq:det-transfer-matrix}
\end{equation}
%----------------

%_______________________________________________________________________________
%---------------------------------------------------------------------------------------------------------------
\subsection{Particle trajectories}
\label{sec:particle-trajectories}
\begin{figure}[ht]
\begin{center}
\includegraphics[width=12cm]{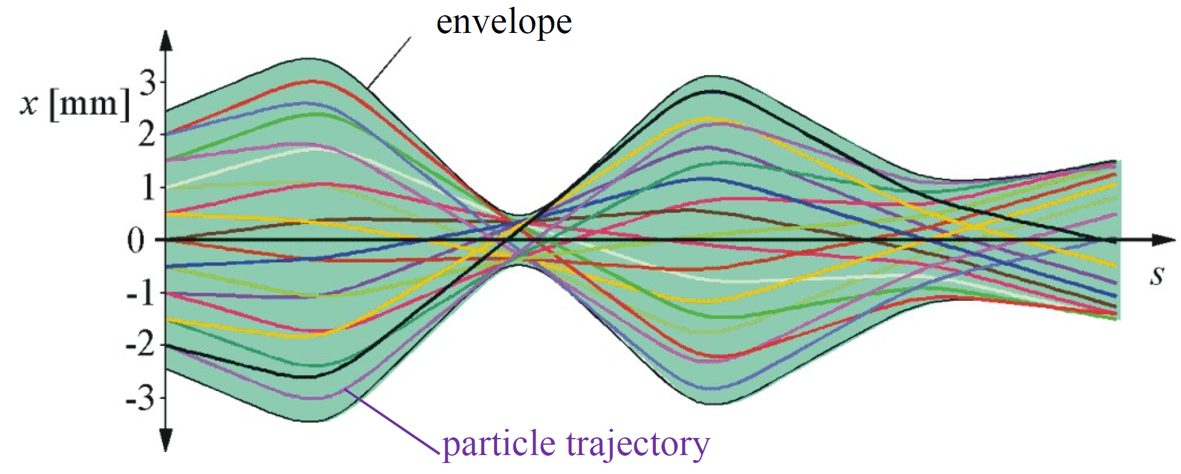}
\caption{Evolution of the horizontal envelope of a real beam formed by very many particles with individual trajectories.}
\label{fig:beam-envelope}
\end{center}
\end{figure}

%_______________________________________________________________________________
%---------------------------------------------------------------------------------------------------------------
With the derived transfer matrices particle trajectories may be calculated for any given arbitrary beam transport line by applying the matrix formalism consisting of a sequential application of the transfer matrices according to Eq.~(\ref{eq:orbit-mult-s}).
However, a real beam consists of very many particles and we are mainly interested in the evolution of macroscopic beam parameters like beam size and beam divergence along a~beam transport line.
One could think that the envelope of a real beam might be represented by a~single trajectory of a ``characteristic particle'' with maximum transverse displacement from the~reference orbit.
Unfortunately, this is not the case since all particles are performing individual oscillations around the~reference orbit (cf.\ Fig.~\ref{fig:beam-envelope}).
%---------------------------------------------------------------------------------
%--------------------------------------------------------------------------------

Thus, we have to extract the beam size from the whole set of particles forming an ``ensemble''. 
In a straight-forward approach, the beam envelope could be calculated by determining the maximum of all individual transverse displacements at a given longitudinal position $s$ and repeating this procedure for different $s$.
Of course, this brute-force approach will be very time-consuming and in addition not well suited for gaining more insight in the transverse dynamics of real beams.
We are rather interested in a~more elegant formalism in which we can take use of the derived transfer matrices.
It turns out that such a formalism exists.
It is based on transforming the distribution of particles in the trace space, which will be defined in the next section.

%===========================================================================
%===========================================================================
\section{Linear beam optics in transfer lines}
\label{sec:linear-beam-optics}
%===========================================================================
%===========================================================================
%_______________________________________________________________________________
%---------------------------------------------------------------------------------------------------------------
\subsection{Particle beams and trace space}
\label{sec:particle-beams-and-trace-space}
\begin{figure}[ht]
\begin{center}
\includegraphics[width=6cm]{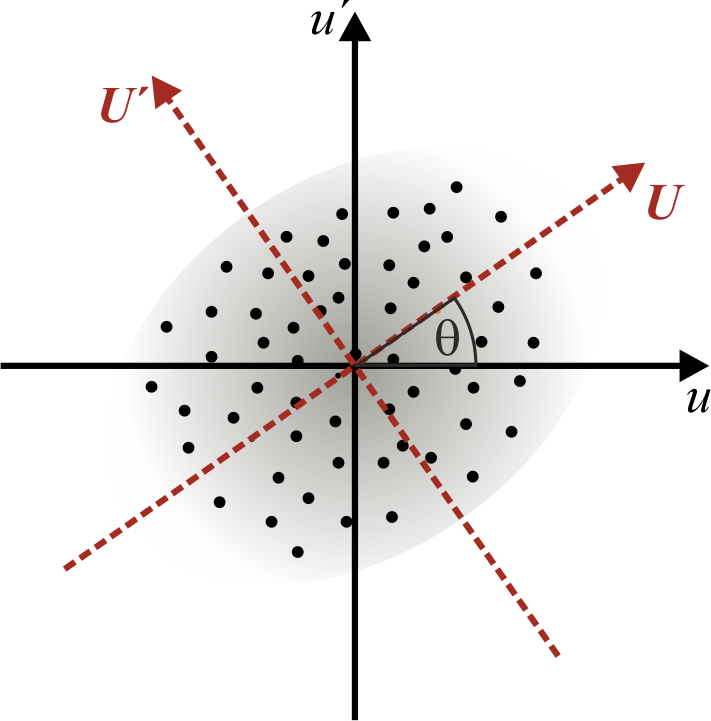}
\caption{Distribution of points in trace space $(u, u^\prime)$ at a fixed longitudinal position $s$. Each point represents a~single particle. Since the distribution of points shows a correlation of $u_i$ and $u^\prime_i$, further calculations are carried out in a~rotated frame $(U, U^\prime)$ with vanishing correlation of $U_i$ and $U^\prime_i$.}
\label{fig:trace-space-distribution}
\end{center}
\end{figure}

%_______________________________________________________________________________
%---------------------------------------------------------------------------------------------------------------
Each particle of a beam is characterized by its transverse displacements $x, x^\prime, y, y^\prime$.
These displacements are dependent on the longitudinal position $s$ and thus vary along the beam transfer line.
If we take a~snapshot at a fixed $s$ (which often is called a {\em Poincaré section}), we will obtain a distribution of points in the 4-dimensional transverse trace space, where each point represents a single particle.
The beam's transverse properties at position $s$ are therefore completely characterized by the distribution of points in the transverse trace space $(x, x^\prime, y, y^\prime)$.
Since in linear beam dynamics the motions in the horizontal and vertical planes are decoupled, we can treat the 2-dimensional horizontal and vertical trace spaces seperatly.
In the following, we will therefore concentrate on one of the two 2-dimensional trace spaces $(x, x^\prime)$ and $(y, y^\prime)$ using the coordinate $u$ which can be either $x$ or $y$.
In this universal notation the 2-dimensional trace space is then formed by $(u, u^\prime)$ and can be either the horizontal or the vertical one.
Figure~\ref{fig:trace-space-distribution} represents a typical distribution of particles in such a trace space at a fixed $s$.
%---------------------------------------------------------------------------------
%--------------------------------------------------------------------------------

In the following, we will assume that the beam is centered on the reference orbit.
Since $u$ and $u^\prime$ indicate the displacements from the reference orbit, this implies that their mean values will vanish:
%------------------
\begin{equation}
\langle u \rangle = \frac{1}{N} \sum_{i=1}^N u_i =0~, ~~~~~~~ \langle u^\prime \rangle = \frac{1}{N} \sum_{i=1}^N u_i^\prime =0~.
\end{equation}
%------------------
The transverse distributions of the beam can thus be characterized by the variances 
%------------------
\begin{equation}
\sigma_u^2 = \langle u^2 \rangle = \frac{1}{N} \sum_{i=1}^N u_i^2~, ~~~~~~~
\sigma_{u^\prime}^2 = \langle {u^\prime}^2 \rangle = \frac{1}{N} \sum_{i=1}^N {u_i^\prime}^2~.
\end{equation}
%------------------

%_______________________________________________________________________________
%---------------------------------------------------------------------------------------------------------------
\subsection{Beam emittance}
\label{sec:beam-emittance}
%_______________________________________________________________________________
%---------------------------------------------------------------------------------------------------------------
We still restrict our treatment to monochromatic beams, where the momenta $\vert \vec p_i \vert$ of all particles are constant and equal to the reference momentum $p_0$.
Therefore, the transverse angular displacements $u_i^\prime$ differ from the transverse momenta $p_{u_i}$ by only a constant factor $\beta_r \gamma_r m_0 c$ and the famous theorem of Liouville can be applied as well to the trace space.
It states that the phase space distribution function, describing the density of possible states around a phase space point, is invariant under conservative forces.
Applying this to a freely chosen subset of particles (representing possible ``states'' of our system) we can conclude that the phase space area occupied by any subset of particles is constant.
Thus, it shall also apply for the beam as a whole stating, that the phase space area covered by the beam remains constant.
Since for constant momenta the areas in phase space and trace space differ by only a constant factor, the same applies for the trace space.

This gives rise to the definition of the {\em beam emittance} $\epsilon_u$ as the area covered by the beam divided by $\pi$:
%------------------
\begin{equation}
\boxed
{
\epsilon_u = \frac{\text{trace space area}}{\pi}
}~.
\label{eq:emittance-definition}
\end{equation}
%------------------
But how do we define the phase space area covered by the beam?
A first possibility is to request that this area comprises all the particles.
This would immediately lead to difficulties if the beam has some irregular tails which then would completely dominate the emittance.
Instead, it is more advisable to define the emittance with a fraction of the beam only.
The most ``clean'' definition is the one using the~variances $\sigma_u^2$ and $\sigma_{u^\prime}^2$.
Since, as can be seen in Fig.~\ref{fig:trace-space-distribution}, $u_i$ and $u^\prime_i$ are usually correlated, we will change to a rotated coordinate system in which the correlation vanishes.
We therefore define a system $(U, U^\prime)$ which is rotated by an angle $\theta$ by the following coordinate transformations:
%------------------
\begin{equation}
{
\begin{split}
U_i &= u_i \cdot \cos \theta + u_i^\prime \cdot \sin \theta~, \\
U_i^\prime &= -u_i \cdot \sin \theta + u_i^\prime \cdot \cos \theta~,
\label{eq:coordinate-rotation}
\end{split}
}
\end{equation}
%------------------
and obtain for the variances in the rotated system
%------------------
\begin{equation}
{
\begin{split}
\sigma_U^2 = \frac{1}{N} \sum_{i=1}^N U_i^2 = 
\langle u^2 \rangle \cos^2 \theta + \langle {u^\prime}^2 \rangle \sin^2 \theta + \langle u u^\prime \rangle \sin (2 \theta)~, \\
\sigma_{U^\prime}^2 = \frac{1}{N} \sum_{i=1}^N {U_i^\prime}^2 = 
\langle u^2 \rangle \sin^2 \theta + \langle {u^\prime}^2 \rangle \cos^2 \theta - \langle u u^\prime \rangle \sin (2 \theta)~. \\
\end{split}
}
\label{eq:variances-uup}
\end{equation}
%------------------
The variances are minimized / maximized with respect to the angle $\theta$ when
%------------------
\begin{equation}
\frac{\partial \sigma_U^2}{\partial \theta} = \frac{\partial \sigma_{U^\prime}^2}{\partial \theta} = 0~,
\label{eq:rotation-angle-uncorr}
\end{equation}
%------------------
yielding
%------------------
\begin{equation}
\tan (2 \theta) = \frac {2 \langle u u^\prime \rangle}{\langle u^2 \rangle - \langle {u^\prime}^2 \rangle}~,
\label{eq:min-theta}
\end{equation}
%------------------
and leading to a vanishing correlation of $U$ and $U^\prime$.
We now build
%------------------
\begin{equation}
\sigma_U^2 + \sigma_{U^\prime}^2 = \langle u^2 \rangle + \langle {u^\prime}^2 \rangle~,
~~~~~~ \text{and} ~~~~~~
\sigma_U^2 - \sigma_{U^\prime}^2 = \frac{2 \langle u u^\prime \rangle}{\sin (2 \theta)}~,
\end{equation}
%------------------
and obtain
%------------------
\begin{equation}
{
\begin{split}
\sigma_U^2 = \frac{1}{2} \left ( \langle u^2 \rangle + \langle {u^\prime}^2 \rangle + \frac{2 \langle u u^\prime \rangle}{\sin (2 \theta)} \right )~, \\
\sigma_{U^\prime}^2 = \frac{1}{2} \left ( \langle u^2 \rangle + \langle {u^\prime}^2 \rangle - \frac{2 \langle u u^\prime \rangle}{\sin (2 \theta)} \right )~. \\
\end{split}
}
\label{eq:variances-rotated}
\end{equation}
%------------------
The emittance $\epsilon_u$ is equal to the square root of the product of the variances in the rotated system which can be expressed by $u$ and $u^\prime$ using Eqs.~(\ref{eq:variances-uup}) and (\ref{eq:min-theta})
%------------------
\begin{equation}
\boxed
{
\epsilon_u = \sigma_U \cdot \sigma_{U^\prime} = \sqrt{\langle u^2  \rangle \langle {u^\prime}^2 \rangle - \langle u u^\prime \rangle^2}
}~.
\label{eq:eps-variances}
\end{equation}
%------------------

It is important to once again note that this is a statistical definition of the emittance based on the~variances of the distribution of points in the 2-dimensional trace space!
More general, the emittance can be defined by the integral
%------------------
\begin{equation}
\epsilon_u = \frac{1}{\pi} \iint \mathrm{d} u \mathrm{d}u^\prime~.
\end{equation}
%------------------

Due to Liouville's theorem, the emittance remains constant under conservative forces.
It thus describes an intrinsic property of the particle beam and therefore often serves as a parameter indicating the beam's ``quality''.
According to its definition by $\epsilon_u = \sigma_U \cdot \sigma_{U^\prime}$, the product $\pi \epsilon_u$ represents the area of an {\em envelope ellipse} with the two semi-axes $\sigma_U$ and $\sigma_U^\prime$ which covers a certain fraction of the beam depending on the beam's shape.
For example, for a beam with Gaussian distribution this fraction amounts to $39.4\%$.

The parametrization of the ellipse by the rotated coordinates can be derived from the well-known ellipse equation
%------------------
\begin{equation}
\frac{U^2}{\sigma_U^2} + \frac{{U^\prime}^2}{\sigma_{U^\prime}^2} = 1~,
\end{equation}
%------------------
yielding together with Eq.~(\ref{eq:eps-variances})
%------------------
\begin{equation}
U^2 \sigma_{U^\prime}^2 + {U^\prime}^2 \sigma_U^2 = \epsilon_u^2~.
\end{equation}
%------------------
Squaring Eq.~(\ref{eq:coordinate-rotation}) and combining it with Eqs.~(\ref{eq:min-theta})--(\ref{eq:variances-rotated}), the envelope ellipse can be parameterized by the original trace space coordinates $u$ and $u^\prime$:
%------------------
\begin{equation}
\epsilon_u^2 = u^2 \sigma_{u^\prime}^2 - 2 u u^\prime \langle u u^\prime \rangle + {u^\prime}^2 \sigma_u^2~.
\end{equation}
%------------------
When using the correlation coefficient $r$ which is defined by
%------------------
\begin{equation}
r = \frac{\langle u u^\prime \rangle}{\sqrt{\langle u^2 \rangle \langle {u^\prime}^2 \rangle}}~,
\end{equation}
%------------------
the ellipse equation can be rewritten as follows:
%------------------
\begin{equation}
\epsilon_u^2 = u^2 \sigma_{u^\prime}^2 - 2 u u^\prime r \sigma_u \sigma_{u^\prime} + {u^\prime}^2 \sigma_u^2~.
\label{eq:ellipse-equation-variances}
\end{equation}
%------------------

%_______________________________________________________________________________
%---------------------------------------------------------------------------------------------------------------
\subsection{Optical functions}
\label{sec:optical-functions}
%_______________________________________________________________________________
%---------------------------------------------------------------------------------------------------------------
So far, we have found a new and very important invariant, the beam emittance $\epsilon_u$ indicating the area in trace space covered by the beam and thus characterizing an intrinsic property of the beam.
Of course, this does not mean that the variances $\sigma_u$ and $\sigma_{u^\prime}$ and the correlation coefficient $r$ will stay constant as well.
Since $\sigma_u$ represents the r.m.s.\ width (or r.m.s.\ beam size) and $\sigma_{u^\prime}$ represents the r.m.s.\ divergence of the beam, both (as well as $r$) will be affected by the magnet optics (the magnetic lattice) and will vary with $s$.
In order to decouple the impact of the magnet optics from the intrinsic properties of the beam, it is useful to make a normalization to $\epsilon_u$
%------------------
\begin{eqnarray}
\label{eq:def-beta}
\sigma^2_u(s) = \langle u^2(s) \rangle = \epsilon_u \cdot \beta_u(s)~, \\
\label{eq:def-gamma}
\sigma^2_{u^\prime}(s) = \langle {u^\prime}^2(s) \rangle = \epsilon_u \cdot \gamma_u(s)~, \\
\label{eq:def-alpha}
r \sigma_u \sigma_{u^\prime} = \langle u u^\prime \rangle = -\epsilon_u \cdot \alpha_u(s)~,
\end{eqnarray}
%------------------
which leads to the definition of three new functions $\alpha_u(s), \beta_u(s), \gamma_u(s)$.
These functions are called the~{\em optical functions} (which later will be named Twiss parameters as well).
Using them, the equation for the~envelope ellipse reads in the ``conventional'' form
%------------------
\begin{equation}
\boxed
{
\epsilon_u = \gamma_u u^2 + 2 \alpha_u u u^\prime + \beta_u {u^\prime}^2
}~.
\label{eq:ellipse-equation}
\end{equation}
%------------------
It is worth to mention that all the above derived equations appear in identical form for the horizontal and vertical plane.
In the following, we will therefore skip the index $u$ for reasons of simplicity.
Please note, that this does not imply that the emittances and corresponding optical functions are equal in both planes.
They are in fact not!

\begin{figure}[ht]
\begin{center}
\includegraphics[width=10cm]{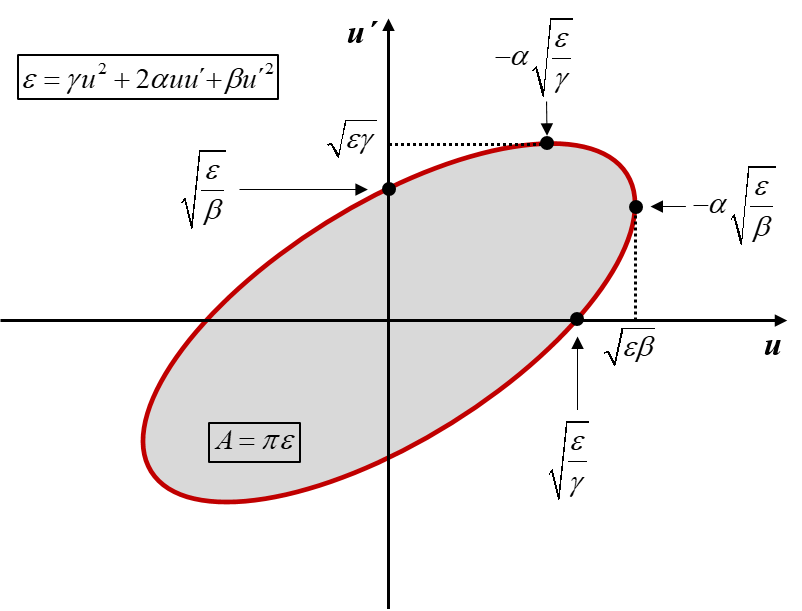}
\caption{Trace space ellipse of the beam with some characteristic coordinates.}
\label{fig:ellipse-parameters}
\end{center}
\end{figure}

\noindent The meaning of the optical functions can be summarized as follows:
\begin{itemize}
\item $\sqrt{\beta}$ represents the r.m.s. beam envelope per unit emittance,
\item $\sqrt{\gamma}$ represents the r.m.s. beam divergence per unit emittance,
\item $\sqrt{\alpha}$ is proportional to the correlation of $u$ and $u^\prime$.
\end{itemize}
Figure~\ref{fig:ellipse-parameters} visualizes the meaning of the optical functions and some more characteristic coordinates.
%---------------------------------------------------------------------------------
%--------------------------------------------------------------------------------

Still, the problem remains how to determine the evolution of the optical functions for a given magnetic lattice.
Using Eqs.~(\ref{eq:def-beta}) -- (\ref{eq:def-gamma}), only the initial values $\beta_0$, $\alpha_0$ and $\gamma_0$ can be extracted from the trace space distribution of the beam at the beginning of a beam transfer line, specifying the incoming beam.
Their evolution is exclusivly determined by the magnetic lattice, described by the sequence of transfer matrices.
To progress here we first have to gain some further insight in $\beta$, $\gamma$ and $\alpha$ and how they are linked together.
This can be obtained from a more detailed study of the equations of motion.

%_______________________________________________________________________________
%---------------------------------------------------------------------------------------------------------------
\subsection{Twiss parameters}
\label{sec:twiss-parametersl}
%_______________________________________________________________________________
%---------------------------------------------------------------------------------------------------------------
For the following, we rewrite the equations of motion (Eq.~(\ref{eq:linearized-equations-of-motion})) in a universal form using our general coordinate $u$ and a new parameter $K$:
%------------------
\begin{equation}
u^{\prime\prime}(s) + K(s) \cdot u(s) = 0~.
\label{eq:equations-of-motion-generalized}
\end{equation}
%------------------
The definition of the magnet parameter $K(s)$ depends on the plane being considered: 
%------------------
\begin{equation}
K(s) = \left \{
\begin{array}{l l}
\frac{1}{\rho^2(s)} - k(s) ~~~~\text{horizontal plane} \\
~~~~~ k(s) ~~~~~~~~~~~~\text{vertical plane}
\end{array}
\right . ~.
\label{eq:K-generalized}
\end{equation}
%------------------
For $K(s) = \text{const.} > 0$ Eq.~(\ref{eq:equations-of-motion-generalized}) would describe a harmonic oscillator with constant oscillation frequency $\sqrt{K}$ and amplitude $A$.
Since $K(s)$ is now varying we expect that the amplitude as well as the~oscillation frequency will vary and thus be explicitly dependent on $s$.
We therefore make the Ansatz
%------------------
\begin{equation}
u(s) = A \cdot w(s) \cdot \cos \left ( \mu(s) + \varphi_0 \right )~.
\label{eq:ansatz}
\end{equation}
%------------------
The argument $\mu(s)$ is called {\em phase advance} and is monotonously increasing:
%------------------
\begin{equation}
\mu(s) > 0~, ~~~~~~ \mu^\prime(s) > 0~, ~~~~~~ \mu(0) = \mu_0 = 0~.
\end{equation}
%------------------
The constant parameters $A$ and $\varphi_0$ are defined by the individual particle.
Since only the product of $A$ and an amplitude function $w(s)$ enter, both are defined except for a scaling factor which in principle can be chosen freely.
As will be discovered later, it will be advantageous to define this scaling factor such that
%------------------
\begin{equation}
w_0 = w(0) = \sqrt{\frac{1}{\mu^\prime_0}} ~~~~~~ \rightarrow ~~~~~ w_0^2 \mu^\prime_0 = 1~.
\label{eq:scaling}
\end{equation}
%------------------
Using the Ansatz the equation of motion reads
%------------------
\begin{equation}
\left ( w^{\prime\prime} - w {\mu^\prime}^2 + K w \right ) \cos \left ( \mu + \varphi_0 \right ) -
\left ( 2 w^\prime \mu^\prime + w \mu^{\prime\prime} \right ) \sin \left ( \mu + \varphi_0 \right ) = 0~.
\end{equation}
%------------------
Since we request this equation to be fulfilled for any initial phase $\varphi_0$ the sine and cosine terms have to vanish seperately:
%------------------
\begin{eqnarray}
\label{eq:equation1-ansatz}
w^{\prime\prime} - w {\mu^\prime}^2 + K w =0~, \\
\label{eq:equation2-ansatz}
2 w^\prime \mu^\prime + w \mu^{\prime\prime} = 0~.
\end{eqnarray}
%------------------
Integration of Eq.~(\ref{eq:equation2-ansatz}) yields
%------------------
\begin{equation}
\int_0^s \frac{\mu^{\prime\prime}}{\mu^\prime} \mathrm{d}s = -2 \int_0^s \frac{w^\prime}{w} \mathrm{d}s
~~~~~~\rightarrow ~~~~~~ \frac{\mu^\prime (s)}{\mu^\prime_0} = \left ( \frac{w_0}{w(s)} \right )^2~,
\end{equation}
%------------------
and after a subsequent integration using Eq.~(\ref{eq:scaling}) we get
%------------------
\begin{equation}
\mu(s) = w_0^2 \mu_0^\prime \int_0^s \frac{\mathrm{d}\tilde s}{w^2(\tilde s)} =
\int_0^s \frac{\mathrm{d}\tilde s}{w^2(\tilde s)}~.
\label{eq:phase-adv-int-w}
\end{equation}
%------------------
We now define a new function $\beta(s)$, the {\em beta function}, which at present has nothing in common with the~function defined in Eq.~(\ref{eq:def-beta}) (but will turn out later to be identical) by
%------------------
\begin{equation}
\beta(s) = w^2(s)~.
\label{eq:def-beta-Twiss}
\end{equation}
%------------------
Using Eq.~(\ref{eq:phase-adv-int-w}), the phase advance can be computed from the beta function via
%------------------
\begin{equation}
\boxed
{
\mu(s) = \int_0^s \frac{\mathrm{d}\tilde s}{\beta(\tilde s)}
}~.
\label{eq:phase-adv-int}
\end{equation}
%------------------
The transverse position displacement then reads
%------------------
\begin{equation}
u(s) = A \sqrt{\beta(s)} \cos \left ( \mu(s) + \varphi_0 \right )~,
\label{eq:ansatz-beta}
\end{equation}
%------------------
and represents an oscillation around the reference orbit with varying amplitude $A \sqrt{\beta}$.
Building the~derivative of Eq. (\ref{eq:ansatz-beta}) we get for the angular displacement
%------------------
\begin{equation}
u^\prime(s) =
- \frac{A}{\sqrt{\beta(s)}} \left [ \alpha(s) \cos \left ( \mu(s) + \varphi \right ) + \sin \left ( \mu(s) + \varphi \right ) \right ]~,
\label{eq:uprime}
\end{equation}
%------------------
where we again have used a new function $\alpha(s)$ defined by
%------------------
\begin{equation}
\boxed
{
\alpha(s) = -\frac{\beta^\prime(s)}{2}
}~.
\label{eq:def-alpha-Twiss}
\end{equation}
%------------------
Equation~(\ref{eq:ansatz-beta}) can be transformed to
%------------------
\begin{equation}
\cos^2 \left (\mu(s) + \varphi_0 \right ) = \frac{u^2(s)}{A^2 \beta(s)}~,
\end{equation}
%------------------
which can be used in combination with Eq.~(\ref{eq:uprime}) to obtain
%------------------
\begin{equation}
\sin^2 \left (\mu(s) + \varphi_0 \right ) =
\left ( \frac{\sqrt{\beta(s)}}{A} u^\prime(s) + \frac{\alpha(s)}{A \sqrt{\beta(s)}} u(s) \right )^2~.
\end{equation}
%------------------
Using $\cos^2 + \sin^2 = 1$ we derive
%------------------
\begin{equation}
\frac{u^2(s)}{\beta(s)} + \left ( \frac{\alpha(s)}{\sqrt{\beta(s)}} u(s) + \sqrt{\beta(s)} u^\prime(s) \right )^2 = A^2~.
\end{equation}
%------------------
We finally define a third function $\gamma(s)$ by
%------------------
\begin{equation}
\boxed
{
\gamma(s) = \frac{1 + \alpha^2(s)}{\beta(s)}
}~,
\label{eq:def-gamma-Twiss}
\end{equation}
%------------------
and obtain a relation
%------------------
\begin{equation}
A^2 = \gamma(s) u^2(s) + 2 \alpha(s) u(s) u^\prime(s) + \beta(s) {u^\prime}^2(s)~,
\label{eq:ellipse-equation-A}
\end{equation}
%------------------
which looks exactly like the envelope ellipse equation (\ref{eq:ellipse-equation}) but with $A^2$ instead of $\epsilon$.
Are the newly defined functions $\alpha(s)$, $\beta(s)$ and $\gamma(s)$ identical to those defined in Eqs.~(\ref{eq:def-beta})  -- (\ref{eq:def-alpha})?
Since Eq.~(\ref{eq:ellipse-equation}) parametrizes the envelope ellipse of the beam, which was based on the second statistical moments $\langle u^2 \rangle$, $\langle {u^\prime}^2 \rangle$ and the correlation coefficient $\langle u u^\prime \rangle$, whereas Eq.~(\ref{eq:ellipse-equation-A}) indicates an ellipse for a single particle with individual $A^2$ and $\varphi_0$, we expect that this will be the case.

For a more rigorous proof we have to calculate the statistical moments by averaging over the~distribution of particles.
Each particle is represented by its individual $A_i$ and $\varphi_{0,i}$ which determine via Eq.~(\ref{eq:ansatz-beta}) its transverse coordinates $(u_i, u_i^\prime)$.
We will evaluate the second statistical moments at a fixed $s$ position and drop the $s$ dependence by e.g.\ writing $u_i, \mu$ instead of $u_i(s), \mu(s)$.
Please note that the~beta function and the phase advance are determined by Eq.~(\ref{eq:equations-of-motion-generalized}), thus only depend on the position $s$, and remain the same for all particles!
Again, we will assume a centered beam and no correlation of amplitude and phase:
%------------------
\begin{eqnarray}
\label{eq:assumption-centered}
0 &=& \langle u_i \rangle~, \\
\label{eq:assumption-uncorr}
0 &=& \left\langle \cos \left ( \mu + \varphi_{0,i} \right ) \right\rangle = 
\left\langle \sin \left ( \mu + \varphi_{0,i} \right ) \right\rangle =
\left\langle \cos \left ( \mu + \varphi_{0,i} \right ) \sin \left ( \mu + \varphi_{0,i} \right ) \right\rangle~.
\end{eqnarray}
%------------------
For the quadratic terms we get from averaging over all phases $\varphi_{0,i}$
%------------------
\begin{equation}
\left\langle \cos^2 \left ( \mu + \varphi_{0,i} \right ) \right\rangle =
\left\langle \sin^2 \left ( \mu + \varphi_{0,i} \right ) \right\rangle = \frac{1}{2}~.
\end{equation}
%------------------
We finally obtain for the second statistical moments and the emittance 
%------------------
\begin{eqnarray}
\sigma_u^2 = \left\langle u^2 \right\rangle &=&
\left\langle A_i^2 \beta \cos^2 (...) \right\rangle = \frac{1}{2} \left\langle A_i^2  \right\rangle \beta~, \\
\sigma_{u^\prime}^2 = \left\langle {u^\prime}^2 \right\rangle &=&
\left\langle \frac{A_i^2}{\beta} \left \{ \alpha^2 \cos^2 (...) + \sin^2 (...)+ 2 \alpha \cos (...) \sin (...) \right \} \right\rangle
= \frac{1}{2} \left\langle A_i^2 \right\rangle \gamma~, \\
\left\langle u u^\prime \right\rangle &=& \left\langle -A_i^2 \left \{\alpha \cos^2 (...) + \cos (...) \sin (...) \right \}
\right\rangle = -\frac{1}{2} \left\langle A_i^2 \right\rangle \alpha~, \\
\epsilon &=& \sqrt{ \left\langle u^2 \right\rangle \left\langle{u^\prime}^2 \right\rangle - \left\langle u u^\prime \right\rangle^2} = \frac{1}{2} \left\langle A_i^2 \right\rangle \sqrt{\beta\gamma - \alpha^2} =
\frac{1}{2} \left\langle A_i^2 \right\rangle~,
\end{eqnarray}
%------------------
and indeed find, that the definition of the optical functions comply with the settings made in Eqs.~(\ref{eq:def-beta-Twiss}), (\ref{eq:def-alpha-Twiss}) and (\ref{eq:def-gamma-Twiss})!
The 4 parameters $\beta$, $\alpha$, $\gamma$ and $\mu$ are called the {\em Twiss parameters}.
The first three of them (the optical functions) determine the orientation and shape of the beam ellipse in trace space (not its area, which is specified by the emittance only!), wheras the last specifies the phase advance of the~particles' oscillations around the reference path according to Eq.~(\ref{eq:ansatz-beta}).
These oscillations are named {\em betatron oscillations} since they were first investigated for a betatron.

We obtained another remarkable result:
The beam emittance can be determined by averaging over the squared amplitude factors $A_i^2$ of the individual particles.
$A_i^2$ remains constant for a given particle $i$ and $\pi A_i^2$ specifies the area of the particle's individual ellipse in trace space.
$A^2$ is called the {\em Courant-Snyder invariant}, a tribute to E.~D.~Courant and H.~S.~Synder who pioneered the theory of the strong focusing synchrotron in the mid 50's (cf.\ \cite{Courant-AG1,Courant-AG2}).

Each particle will stay on its own ellipse whose shape and orientation is determined by the optical functions only.
Particles with same $A^2$ will be on the same ellipse but at different locations, determined by their individual phases $\varphi$.
Particles with different $A^2$ will be on ellipses with different areas but same orientation and shape, since these are determined by the Twiss parameters only.
%---------------------------------------------------------------------------------
\begin{figure}[ht]
\begin{center}
\includegraphics[width=14cm]{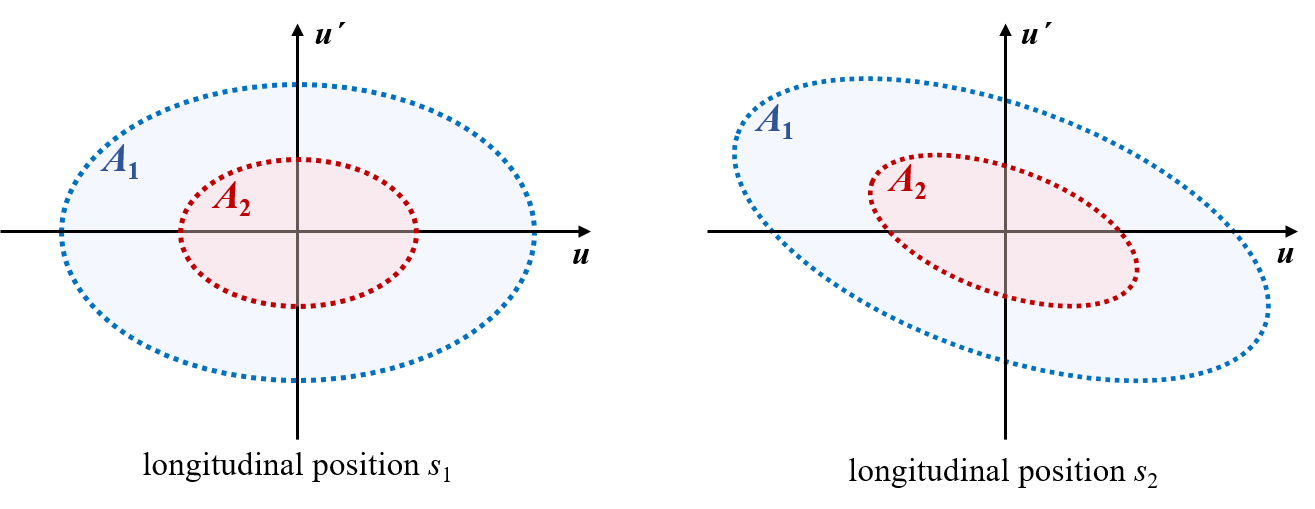}
\caption{Trace space points of particles with two different Courant-Snyder invariants $A_1$ and $A_2$ and different phases $\varphi$. Left: Distribution at longitudinal position $s_1$ where $\alpha = 0$. Right: Distribution at longitudinal position $s_2$ where $\alpha \neq 0$.}
\label{fig:A-ellipses}
\end{center}
\end{figure}
%--------------------------------------------------------------------------------

Figure~\ref{fig:A-ellipses} illustrates this situation.
All particles represented by dots of the same color have identical $A^2$, but different $\varphi$.
The shape and orientation of the ellipses depends on $s$ and will change from $s_1$ to $s_2$, their areas will remain constant.

%_______________________________________________________________________________
%---------------------------------------------------------------------------------------------------------------
\subsection{Transformation of beta functions}
\label{sec:transformation-of-beta-functions}
%_______________________________________________________________________________
%---------------------------------------------------------------------------------------------------------------
Let's consider a single particle with coordinates $u_0$ and $u_0^\prime$ at position $s_0$ travelling through a magneto-optic system.
At position $s$ its coordinates will have changed to $u$ and $u^\prime$, but its Courant-Snyder invariant $A^2$ will remain the same.
Thus, Eq.~(\ref{eq:ellipse-equation-A}) holds for both positions, linking the coordinates and the Twiss parameters at $s_0$ and $s$:
%------------------
\begin{equation}
\gamma_0 u_0^2 + 2 \alpha_0 u_0 u_0^\prime + \beta_0 {u_0^\prime}^2 = A^2 =
\gamma u^2 + 2 \alpha u u^\prime + \beta {u^\prime}^2~.
\label{eq:ellipse-u-u0}
\end{equation}
%------------------
On the other hand, the coordinate transformation from $s_0$ to $s$ is defined by applying the transfer matrix $\mathbf{M}(s_0, s)$ to the position vector $\vec u$
%------------------
\begin{equation}
\vec u = \mathbf{M}(s_0, s) \vec u_0~,
\end{equation}
%------------------
which reads in its fundamental form (cf.\ Eq.~(\ref{eq:CS-matrix}))
%------------------
\begin{equation}
\begin{pmatrix} u \\ u^\prime \end{pmatrix} = \begin{pmatrix} C & S \\ C^\prime & S^\prime \end{pmatrix}\cdot \begin{pmatrix} u_0 \\ u^\prime_0 \end{pmatrix}~.
\end{equation}
%------------------
Building the inverse, we obtain with Eq.~(\ref{eq:det-transfer-matrix})
%------------------
\begin{equation}
\begin{pmatrix} u_0 \\ u^\prime_0 \end{pmatrix} = 
\frac{1}{C S^\prime - S C^\prime} \begin{pmatrix} S^\prime & -S \\ -C^\prime & C \end{pmatrix}\cdot
\begin{pmatrix} u \\ u^\prime \end{pmatrix} =
\begin{pmatrix} S^\prime u - S u^\prime \\ -C^\prime u + C u^\prime \end{pmatrix}~.
\end{equation}
%------------------
We therewith get for the quadratic and mixed terms
%------------------
\begin{eqnarray}
u_0^2 &=& {S^\prime}^2 u^2 - 2 S S^\prime u u^\prime + S^2 {u^\prime}^2~, \\
{u_0^\prime}^2 &=& {C^\prime}^2 u^2 - 2 C C^\prime u u^\prime + C^2 {u^\prime}^2~, \\
u_0 u_0^\prime &=& -C^\prime S^\prime u^2 + \left ( C^\prime S + C S^\prime \right ) u u^\prime - C S {u^\prime}^2~,
\end{eqnarray}
%------------------
which we insert in Eq.~(\ref{eq:ellipse-u-u0}) yielding
%------------------
\begin{equation}
{
\begin{split}
\underbrace{
\left ( {S^\prime}^2 \gamma_0 - 2 S^\prime C^\prime \alpha_0 + {C^\prime}^2 \beta_0 \right )
}_{\substack{= \gamma}}
 u^2 +
\underbrace{
2 \left ( -S S^\prime \gamma_0 + \left ( S^\prime C + S C^\prime \right ) \alpha_0 - C C^\prime \beta_0 \right )
}_{\substack{= \alpha}}
u u^\prime \\ +
\underbrace{
\left ({S^\prime}^2 \gamma_0 - 2 S C \alpha_0 + C^2 \beta_0 \right )
}_{\substack{= \beta}}
{u^\prime}^2 = A^2 ~.~~~~~~~~~~~~~~~~~~~~~~~~~~~~~~
\end{split}
}
\end{equation}
%------------------
This reveals the transformation of the Twiss parameters in matrix formulation
%------------------
\begin{equation}
\boxed{
\begin{pmatrix} \beta \\ \alpha \\ \gamma \end{pmatrix} = 
\begin{pmatrix} C^2 & -2 S C & S^2 \\ -C C^\prime & S^\prime C + S C^\prime & -S S^\prime \\
{C^\prime}^2 & -2 S^\prime C^\prime & {S^\prime}^2 \end{pmatrix}\cdot
\begin{pmatrix} \beta_0 \\ \alpha_0 \\ \gamma_0 \end{pmatrix}
}~.
\label{eq:Beta-transformation-CS}
\end{equation}
%------------------
By means of this matrix equation one can transform the optical functions piecewise through the lattice and determine their evolution $\beta(s)$, $\alpha(s)$, and $\gamma(s)$.
Of course, this requires a cumbersome rearrangement of the elements $C, S, C^\prime, S^\prime$ of the transfer matrices in a new 3x3 matrix according to Eq.~(\ref{eq:Beta-transformation-CS}).

There exist another, even more useful procedure in which the transfer matrices can be used unchanged.
It is based on a new matrix $\mathbf{B}$, the {\em beta matrix}, which is defined by
%------------------
\begin{equation}
\mathbf{B} = \begin{pmatrix} \beta & -\alpha \\ -\alpha & \gamma \end{pmatrix}~,
\label{eq:Beta-matrix}
\end{equation}
%------------------
and can be related to the {\em beam matrix} $\mathbf{\Sigma}_\text{beam}$, which is the covariance matrix of the particle distribution, by
%------------------
\begin{equation}
\mathbf{\Sigma}_\text{beam} = \begin{pmatrix} \sigma_u^2 & \sigma_{u u^\prime} \\ \sigma_{u u^\prime} & \sigma_{u^\prime}^2 \end{pmatrix}=
\epsilon \cdot \begin{pmatrix} \beta & -\alpha \\ -\alpha & \gamma \end{pmatrix} = \epsilon \cdot \mathbf{B}~,
\label{eq:beam-matrix}
\end{equation}
%------------------
where $\sigma_{u u^\prime} = \langle u u^\prime \rangle$ and we have used Eqs.~(\ref{eq:def-beta}), (\ref{eq:def-gamma}), and (\ref{eq:def-alpha}).
According to Eq.~(\ref{eq:def-gamma-Twiss}), the determinant of the beta matrix is
%------------------
\begin{equation}
\det\left( \mathbf{B} \right) = \beta \gamma -\alpha^2 = 1~.
\end{equation}
%------------------
Using the inverse of the beta matrix
%------------------
\begin{equation}
\mathbf{B}^{-1} = \begin{pmatrix} \gamma & \alpha \\ \alpha & \beta \end{pmatrix}~,
\end{equation}
%------------------
Eq.~(\ref{eq:ellipse-u-u0}) can be written in rather compact form
%------------------
\begin{equation}
A^2 = {^T}{\vec u} \cdot \mathbf{B}^{-1} \cdot \vec u = {^T}{\vec u_0} \cdot \mathbf{B}_0^{-1} \cdot \vec u_0~.
\label{eq:Beta-transform}
\end{equation}
%------------------
Since the position vector can be transformed by applying the transfer matrix $\mathbf{M}$
%------------------
\begin{equation}
\vec u = \mathbf{M} \cdot \vec u_0~,
\end{equation}
%------------------
and accordingly
%------------------
\begin{equation}
{^T}{\vec u} = {^T}{\left ( \mathbf{M} \cdot \vec u_0 \right ) } = {^T}{\vec u_0} \cdot {^T}{\mathbf{M}}~,
\label{eq:vec-transform}
\end{equation}
%------------------
we can derive a simple transformation rule for the beta matrix.
By inserting $\mathbf{I} = \mathbf{M}^{-1} \cdot \mathbf{M}$ in Eq.~(\ref{eq:Beta-transform}), we obtain
%------------------
\begin{equation}
\begin{split}
A^2 &= {^T}{\vec u_0} \cdot  {^T}{\mathbf{M}} \cdot  {^T}{\mathbf{M}^{-1}} \cdot \mathbf{B}_0^{-1} \cdot \mathbf{M}^{-1} \cdot \mathbf{M} \cdot \vec u_0 \\
&= {^T}{\left ( \mathbf{M} \cdot \vec u_0 \right )} \cdot \left( {^T}{\mathbf{M}^{-1}} \cdot \mathbf{B}_0^{-1} \cdot \mathbf{M}^{-1} \right ) \cdot \left ( \mathbf{M} \cdot \vec u_0 \right ) \\
&= {^T}{\vec u} \cdot (
\underbrace{
 \mathbf{M} \cdot \mathbf{B}_0 \cdot {^T}{\mathbf{M}}
}_{\substack{= \mathbf{B}}}
)^{-1} \cdot \vec u~,
\end{split}
\end{equation}
%------------------
from which we can read off the transformation of the beta matrix:
%------------------
\begin{equation}
\boxed{
\mathbf{B} = \mathbf{M} \cdot \mathbf{B}_0 \cdot {^T}{\mathbf{M}}
}~.
\label{eq:beta-matrix-transformation}
\end{equation}
%------------------

\begin{figure}[ht]
\begin{center}
\includegraphics[width=8cm]{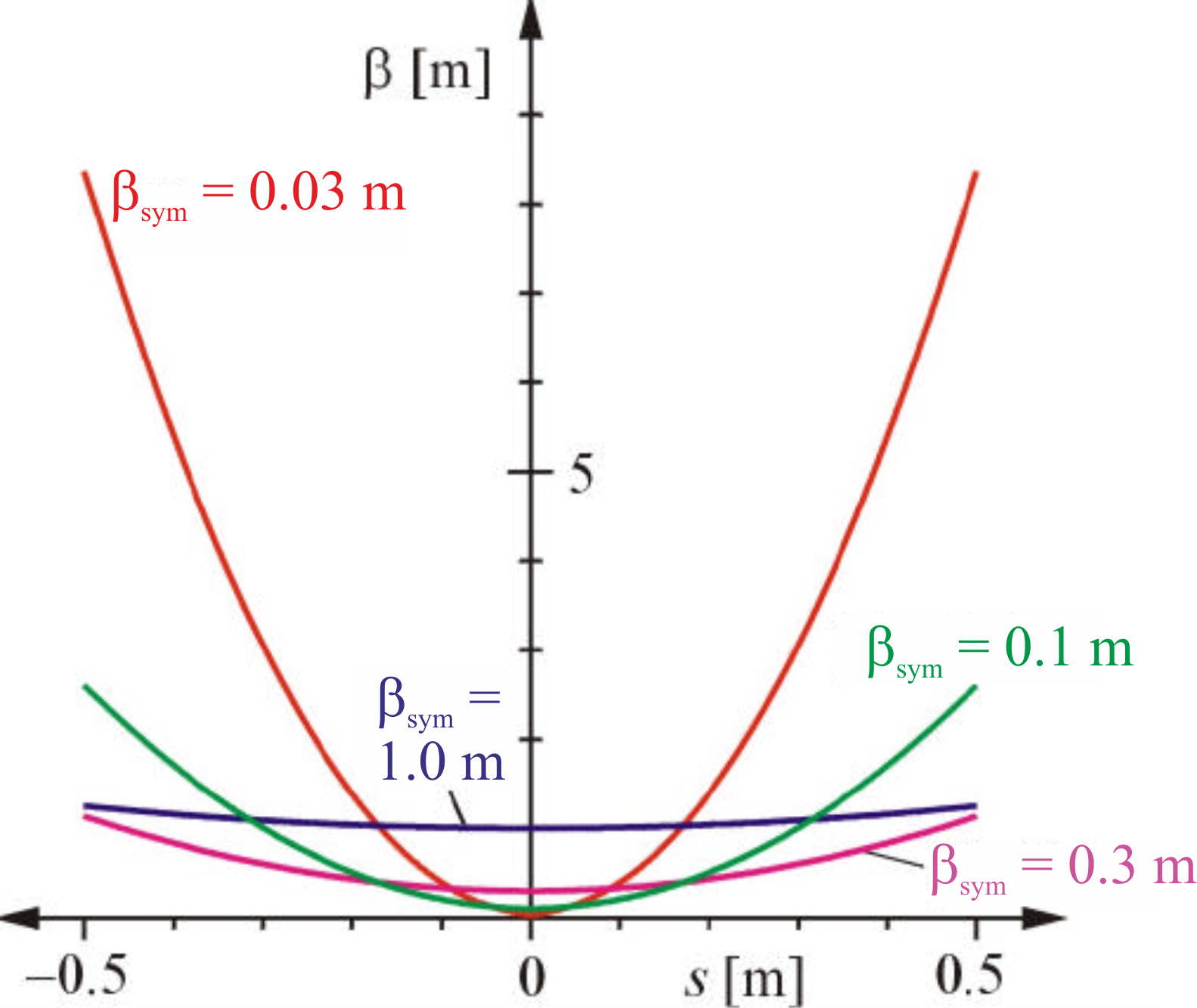}
\caption{Evolution of the beta function around a symmetry point $\alpha_\text{sym}=0$. Smaller values of $\beta_\text{sym}$ lead to a more pronounced increase of beta with increasing distance from the symmetry point.}
\label{fig:beta-functions-drift}
\end{center}
\end{figure}

As a simple example, we consider a free drift in a beam transfer line.
More specifically, we want to investigate the evolution of the beta function around a symmetry point at position $s = 0$ where $\alpha_\text{sym} = 0$.
We then have $\gamma_\text{sym} = \frac{1}{\beta_\text{sym}}$ and get
%------------------
\begin{equation}
\mathbf{B}(s) = \begin{pmatrix} 1 & s \\ 0 & 1 \end{pmatrix} \cdot
\begin{pmatrix} \beta_\text{sym} & s \\ 0 & \frac{1}{\beta_\text{sym}} \end{pmatrix} \cdot
\begin{pmatrix} 1 & 0 \\ s & 1 \end{pmatrix} =
\begin{pmatrix} \beta_\text{sym} + \frac{s^2}{\beta_\text{sym}} & \frac{s}{\beta_\text{sym}} \\ \frac{s}{\beta_\text{sym}} & \frac{1}{\beta_\text{sym}} \end{pmatrix}~,
\label{eq:beta-matrix-drift}
\end{equation}
%------------------
from which the evolution of the optical functions can be extracted:
%------------------
\begin{eqnarray}
\beta(s) &=& \beta_\text{sym} + \frac{s^2}{\beta_\text{sym}}~, \\
\alpha(s) &=& -\frac{s}{\beta_\text{sym}}~, \\
\gamma(s) &=& \frac{1}{\beta_\text{sym}}~.
\end{eqnarray}
%------------------
Figure~\ref{fig:beta-functions-drift} illustrates the evolution of the beta function for different values of $\beta_\text{sym}$.
%---------------------------------------------------------------------------------
%--------------------------------------------------------------------------------
The corresponding r.m.s.\ beam size scales with $\sigma = \sqrt{\epsilon \cdot \beta(s)}$, thus
%------------------
\begin{equation}
\sigma(s) = \sigma_0 \cdot \sqrt{1 + \left ( \frac{s}{\beta_\text{sym}} \right )^2}~,
\label{eq:sigma-evolution-drift}
\end{equation}
%------------------
whereas the r.m.s.\ beam divergence remains constant:
%------------------
\begin{equation}
\sigma^\prime(s) = \frac{\epsilon}{\sigma_0} = \text{const.}
\label{eq:sigmap-evolution-drift}
\end{equation}
%------------------
To obtain further insights, we will compare the particle beam with a Gaussian $\text{TEM}_{00}$ light beam, characterized by its waist radius $w(s)$ and the Rayleigh length $z_R$, in which $w^2$ is doubled.
From diffraction theory, we know (cf.~\cite{Born})
%------------------
\begin{equation}
w(s) = w_0 \cdot \sqrt{1 + \left ( \frac{s}{z_R} \right )^2}~.
\label{eq:w-evolution-drift}
\end{equation}
%------------------
The Rayleigh length is determined by the waist radius at the focal point $w_0$ and the wavelength $\lambda$ of the~light beam
%------------------
\begin{equation}
z_R = \frac{\pi w_0^2}{\lambda}~.
\label{eq:Rayleigh-length}
\end{equation}
%------------------
Comparing Eqs.~(\ref{eq:sigma-evolution-drift}), (\ref{eq:w-evolution-drift}), and (\ref{eq:Rayleigh-length}) we obtain
%------------------
\begin{equation}
\beta_\text{sym} ~~\widehat{=}~~ z_R = \frac{\pi w_0^2}{\lambda}~.
\label{eq:Rayleigh-beta}
\end{equation}
%------------------
Thus, in a distance $\beta_\text{sym}$ from the focal point, the variance $\sigma^2$ of the particle beam is doubled.
Since the~light beam's intensity distribution $I(x,y)$ is defined by
%------------------
\begin{equation}
I(x,y) = I_0 \cdot \left ( \frac{w_0}{w} \right )^2 \cdot e^{- \frac{2 \left( x^2 + y^2 \right)}{w^2}}~,
\end{equation}
%------------------
the variances of a round beam are linked to the squared waist radius by a factor $\frac{1}{4}$:
%------------------
\begin{equation}
\sigma_x^2 = \frac{\iint x^2 I(x,y) \mathrm{d}x\mathrm{d}y}{\iint I(x,y) \mathrm{d}x\mathrm{d}y} = 
\frac{\iint y^2 I(x,y) \mathrm{d}x\mathrm{d}y}{\iint I(x,y) \mathrm{d}x\mathrm{d}y} = \sigma_y^2 = \frac{w^2}{4}~.
\end{equation}
%------------------
Replacing $w_0 = 2 \sigma_0 = 2 \sqrt{\epsilon \beta}$ in Eq.~(\ref{eq:Rayleigh-beta}) we obtain the important relation
%------------------
\begin{equation}
4 \pi \epsilon ~~\widehat{=}~~ \lambda~.
\label{eq:emittance-lambda}
\end{equation}
%------------------
A particle beam with emittance $\epsilon$ thus evolves like a Gaussian $\text{TEM}_{00}$ light beam with wavelength $4 \pi \epsilon$.
%---------------------------------------------------------------------------------
\begin{figure}[ht]
\begin{center}
\includegraphics[width=13cm]{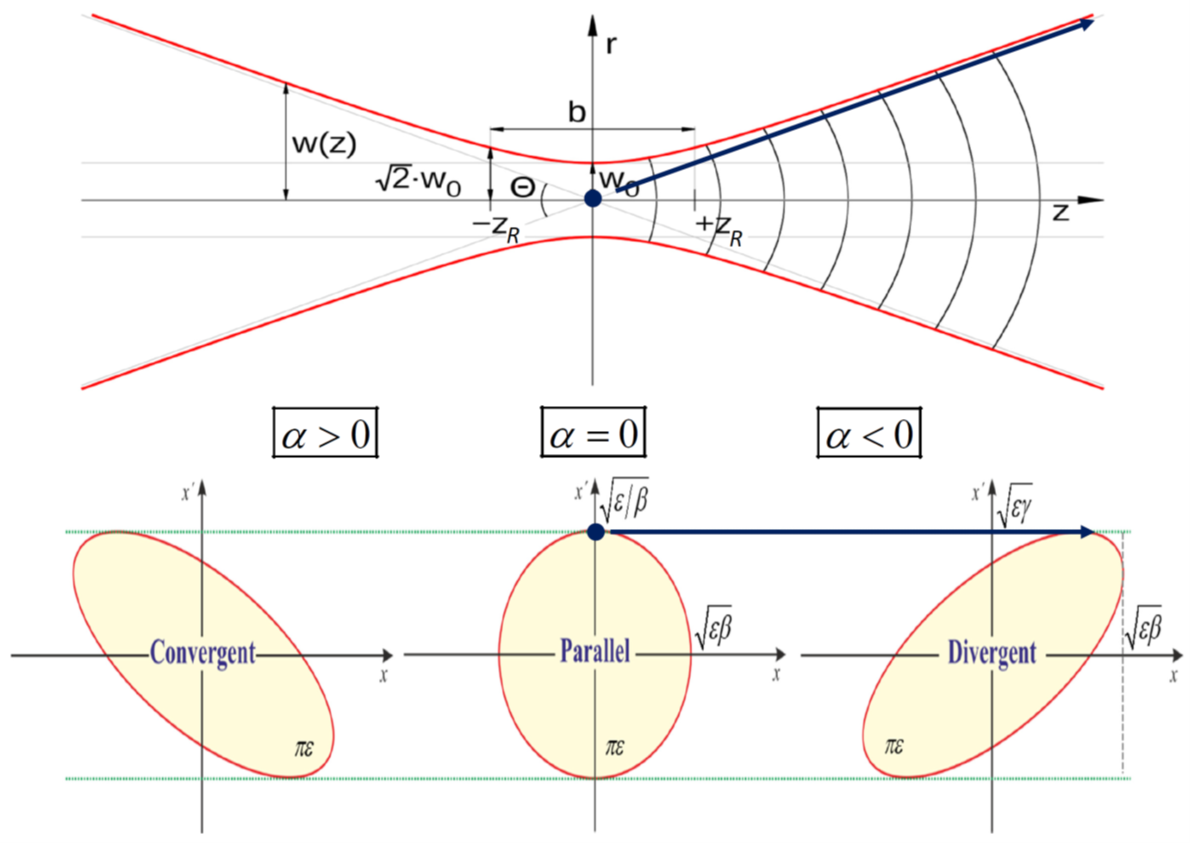}
\caption{Evolution of a beam in horizontal configuration space (upper figure) and horizontal trace space (lower figure).}
\label{fig:beam-evolution}
\end{center}
\end{figure}
%--------------------------------------------------------------------------------

Figure~\ref{fig:beam-evolution} illustrates the evolution of a beam in the~horizontal configuration space $(x, s)$ and the~horizontal trace space $(x, x^\prime)$ in case of a free drift.
Before the focal plane, the beam is convergent ($\alpha > 0$) and $x$ and $x^\prime$ have a negative correlation leading to a clockwise rotated and sheared ellipse.
In the focal plane $\alpha = 0$, the correlation vanishes and the ellipse is symmetric with respect to the $(x, x^\prime)$ axes.
Behind the focal plane, the beam is divergent ($\alpha < 0$) and $x$ and $x^\prime$ have a positive correlation leading to an anticlockwise rotated and sheared ellipse.
For large distances, the trajectory of the particle with maximum $x_0^\prime$ and $x_0 = 0$ in the focal plane defines the beam envelope.

The transformation matrix $\mathbf{M}$ can also be derived from the Twiss parameters.
If the initial values $\alpha_0$, $\beta_0$ and $\gamma_0$ at the beginning (at position $s_0$) and the final values $\alpha$, $\beta$ and $\gamma$ at the end (at position $s$) of a transfer line section are known, $\mathbf{M}$ can be expressed by these quantities.
In addition, and this is important, the phase advance $\mu$ needs to be known as well.
Then with
%------------------
\begin{equation}
u(s) =\sqrt{\epsilon \beta} \cos\left(\mu + \varphi_0 \right) = \sqrt{\epsilon \beta} \left( \cos \mu \cos \varphi_0 - \sin \mu \sin \varphi_0 \right)~,
\label{eq:pos-twiss}
\end{equation}
%------------------
its derivative
%------------------
\begin{equation}
u^\prime(s) =-\frac{\sqrt{\epsilon}}{\sqrt{\beta}} \left\{\alpha \left( \cos \mu \cos \varphi_0 - \sin \mu \sin \varphi_0 \right)
- \left( \sin \mu \cos \varphi_0 - \cos \mu \sin \varphi_0 \right) \right\}~,
\label{eq:angpos-twiss}
\end{equation}
%------------------
and the initial values $u(s_0) = u_0$, $u^\prime(s_0) = u_0^\prime$, $\mu(s_0) = 0$ which transform to
%------------------
\begin{eqnarray}
\cos \varphi_0 &=& \frac{u_0}{\sqrt{\epsilon \beta_0}}~, \\
\sin \varphi_0 &=& -\frac{1}{\sqrt{\epsilon}} \left( u_0^\prime \sqrt{\beta_0} + \frac{\alpha_0 u_0}{\sqrt{\beta_0}} \right)~,
\end{eqnarray}
%------------------
we obtain $\mathbf{M}$, only dependent on the initial and final Twiss parameters
%------------------
\begin{equation}
\boxed{
\mathbf{M}(s_0, s) = \begin{pmatrix} \frac{\sqrt{\beta}}{\sqrt{\beta_0}} \left( \cos \mu + \alpha_0 \sin \mu \right) &
\sqrt{\beta \beta_0} \sin \mu \\
\frac{\alpha_0 - \alpha}{\sqrt{\beta \beta_0}} \cos \mu - \frac{1 + \alpha \alpha_0}{\sqrt{\beta \beta_0}} \sin \mu &
\frac{\sqrt{\beta_0}}{\sqrt{\beta}} \left( \cos \mu - \alpha \sin \mu \right)
\end{pmatrix}
}~.
\label{eq:transfer-matrix-twiss}
\end{equation}
%------------------
This representation of the transfer matrix will turn out to be very useful when discussing circular accelerators.

%===========================================================================
%===========================================================================
\section{Circular accelerators}
\label{sec:circular-accelerators}
%===========================================================================
%===========================================================================
In the following, we will apply the derived formalism to circulator accelerators.
After looking into general requirements for orbit stability, we will investigate the impact of field errors on the beam dynamics.

%_______________________________________________________________________________
%---------------------------------------------------------------------------------------------------------------
\subsection{Weak focusing}
\label{sec:weak-focusing}
%_______________________________________________________________________________
%---------------------------------------------------------------------------------------------------------------
In the ``old'' days of the early synchrotrons, the concept of {\em weak focusing} has been applied to achieve confined beams in both horizontal planes.
It is based on having transverse focusing in both planes at the~same time.
Recalling the equations of motions which read
%------------------
\begin{eqnarray}
x^{\prime\prime}(s) + \left( \frac{1}{\rho^2} - k(s) \right) \cdot x(s) = 0~, \\
y^{\prime\prime}(s) + k(s) \cdot y(s) = 0~,
\end{eqnarray}
%------------------
we must demand that the factors before $x(s)$ and $y(s)$ always remain positive, which is equivalent to 
%------------------
\begin{equation}
0 < k(s) = -\frac{q}{p_0} \frac{\partial B_y}{\partial x} < \frac{1}{\rho^2}~,
\label{eq:weak-k}
\end{equation}
%------------------
where we have used the definition of $k$ according to Eqs.~(\ref{eq:B-quadrupole-field-def}) and (\ref{eq:quadrupole-strength}).
With $p_0 = q \rho B_y(0)$, where $B_y(0)$ defines the bending field at the desing orbit, we can define a new parameter, the dimensionsless {\em field index} $n$ by
%------------------
\begin{equation}
n = -  \frac{\rho}{B_y(0)} \frac{\partial B_y}{\partial x}~,
\label{eq:field-index}
\end{equation}
%------------------
which describes the change of the bending field with horizontal displacement $x$ from the reference orbit
%------------------
\begin{equation}
B_y(x) = B_y(0)\cdot \left( \frac{x + \rho}{\rho} \right)^{-n}~.
\label{eq:weak-gradient}
\end{equation}
%------------------
With Eq.~(\ref{eq:weak-k}) we get the well-known criterion of weak focusing 
%------------------
\begin{equation}
\boxed{
0 < n < 1
}~.
\label{eq:Steenbeck}
\end{equation}
%------------------
Thus an overall focusing in both planes can be acchieved by a constant field index.

\begin{figure}[ht]
\begin{center}
\includegraphics[width=13cm]{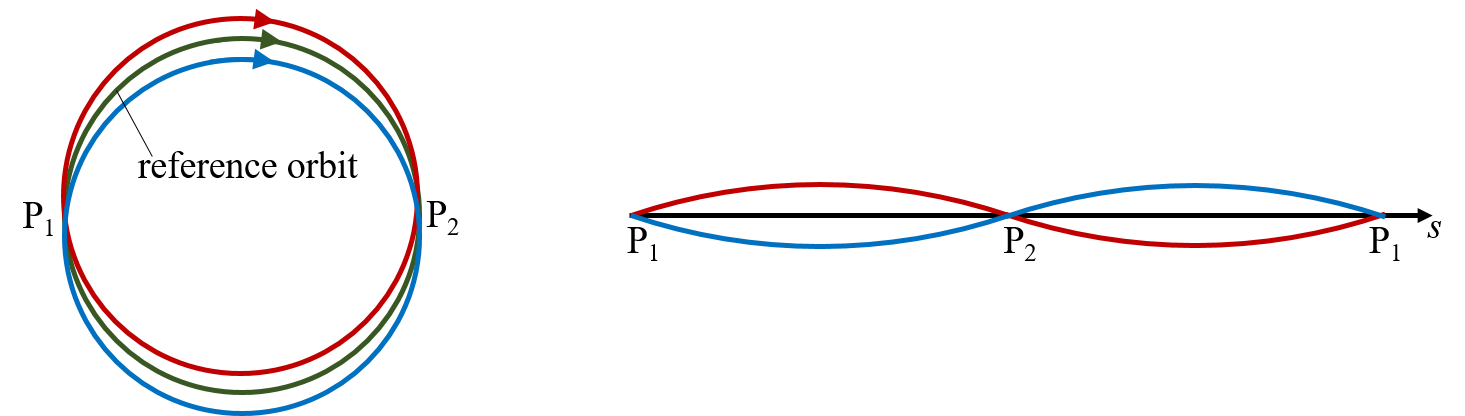}
\caption{Geometrical focusing in a homogeneous magnetic field. Left: all particles starting in point $P_1$ with slightly different angles will meet again in point $P_2$. Right: Particles are bent back to the reference orbit and thus experience a horizontal focusing.}
\label{fig:geometrical-focusing}
\end{center}
\end{figure}

Recalling that a quadrupole magnet with constant quadrupole strength $k$ only focuses in one plane but acts defocusing in the other, the question arises why a non-varying (with longitudinal coordinate $s$) ``moderate'' decrease of $B_y$ with $x$ according to Eq.~(\ref{eq:weak-gradient}) can lead to focusing in both transverse planes.
The reason lies in an additional focusing which is of pure geometrical nature and already gave rise to the horizontal weak focusing of a sector dipole magnet (cf.\ Eq.~(\ref{eq:sector-dipole-matrix})).
This {\em geometrical focusing} can best be explained considering a homogeneous bending field with $n=0$ in which particles with the~same momentum $p_0$ will move on circles with the same radii (cf.\ Fig.~\ref{fig:geometrical-focusing}).
%---------------------------------------------------------------------------------
%--------------------------------------------------------------------------------
All particles starting at a~point $P_1$  with slightly different angular displacements $x^\prime$ will meet again at point $P_2$.
With respect to the~reference orbit (as denoted by the transverse displacements $x$ and $y$ in our coordinate system following the reference orbit), this looks like focusing.
For $n < 1$ this geometrical focusing overcompensates the horizontal defocusing caused by the horizontal decrease of $B_y$.

A weak focusing machine is a circular accelerator made of identical dipole magnets with radially decreasing bending field strength fullfilling the weak focusing criterion Eq.~(\ref{eq:Steenbeck}).
The overall focusing of constant strength will cause an oscillation of the beam particles around the reference trajectory with constant spatial frequencies $\omega_x, \omega_y$ in the horizontal and vertical plane:
%------------------
\begin{eqnarray}
\omega_x &=& \sqrt{\frac{1}{\rho^2}-k} = \frac{\sqrt{1-n}}{\rho}~, \\
\omega_y &=& \sqrt{k} = \frac{\sqrt{n}}{\rho}~.
\end{eqnarray}
%------------------
The number $Q$ of oscillations per turn of length $L=2\pi\rho$ will then be
%------------------
\begin{eqnarray}
\label{eq:qx-weak}
Q_x &=& \rho\omega_x = \frac{1}{2\pi} \oint \frac{\mathrm{d}s}{\beta_x} = \sqrt{1-n} < 1~, \\
\label{eq:qy-weak}
Q_y &=& \rho\omega_y = \frac{1}{2\pi} \oint \frac{\mathrm{d}s}{\beta_y} = \sqrt{n} < 1~.
\end{eqnarray}
%------------------
Orbit stability is thus guaranted.
Why have such accelerators then been replaced by strong focusing machines applying alternating gradient focusing (as will we explained in the next section)?
The reason lies in the following: since the beta functions are constant, we get from Eqs.~(\ref{eq:qx-weak}) and (\ref{eq:qy-weak})
%------------------
\begin{equation}
\beta_x > \rho~, ~~~~~~~~~~~~~~ \beta_y > \rho~.
\end{equation}
%------------------
Thus the r.m.s.\ beam size defined by Eq.~(\ref{eq:def-beta}) will increase with increasing radius according to
%------------------
\begin{equation}
\sigma = \sqrt{\epsilon \beta} > \sqrt{\epsilon \rho}~,
\end{equation}
%------------------
and reach intolerable values for large orbit radii which are indispensably necessary for high energy circular accelerators like colliders and brilliant synchrotron radiation sources.

\noindent \underline{Remark:} {\em So far we have only considered monochromatic beams. Thus only the transverse displacements caused by betatron oscillations contributed to the r.m.s.\ beam size. As we will discover in section \ref{sec:off-momentum-particles}, there will be another contribution coming from the local dispersion and the momentum spread of the beam making the situation even worse.}

%_______________________________________________________________________________
%---------------------------------------------------------------------------------------------------------------
\subsection{Strong focusing}
\label{sec:atrong-focusing}
\begin{figure}[ht]
\begin{center}
\includegraphics[width=7cm]{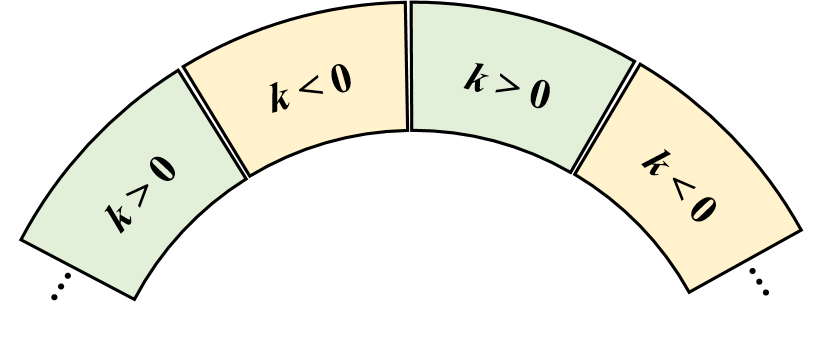}
\caption{Basic scheme of an alternating gradient strong focusing magnetic structure.}
\label{fig:ag-focusing}
\end{center}
\end{figure}

%_______________________________________________________________________________
%---------------------------------------------------------------------------------------------------------------
As we have seen in the last subsection, the problem of a weak focusing isomagnetic ring is that the net focusing is getting weaker and weaker with increasing orbit radius $\rho$, leading to a an r.m.s.\ beam size which increases linearly with $\sqrt{\rho}$.
This can be ``cured'' by applying a new concept: the {\em strong focusing} by alternating gradients.
Strong focusing simply means that $\vert n \vert \gg 1$, causing either horizontal defocusing in case of $n \gg 1$ or vertical defocusing in case of $n \ll 0$ in isomagnetic rings.
The basic idea of alternating gradient focusing is depicted in Fig.~\ref{fig:ag-focusing}.
%---------------------------------------------------------------------------------
%--------------------------------------------------------------------------------
Instead of using an isomagnetic ring, the ring is build from segments with large quadrupole strengths $k$ of opposite sign.
In both transverse planes, we then have a periodic sequence of strongly focusing and strongly defocusing sectors.
The key question is if this  leads to stable betatron oscillations in both planes, meaning that the oscillation amplitudes will remain finite even after a large number of turns in a circular accelerator.
In terms of the matrix formalism introduced in the last section, this translates to the requirement
%------------------
\begin{equation}
\lim \limits_{N \to \infty} \mathbf{M}^N = \text{finite}~,
\end{equation}
%------------------
where $\mathbf{M}$ denotes the transfer matrix for one periodic cell (or one turn).
This matrix has to be derived from the solutions of the linearized equations of motion, which read in the generalized form (cf.\ Eq.~(\ref{eq:equations-of-motion-generalized}))
%------------------
\begin{equation}
u^{\prime\prime}(s) + K(s) \cdot u(s) = 0~.
\label{eq:equations-of-motion-generalized-2}
\end{equation}
%------------------
Now, we have the situation that $K(s)$ is a periodic function since it is reproduced after each segment (or at least after one turn) of length $L$:
%------------------
\begin{equation}
K(s+L) = K(s)~.
\end{equation}
%------------------
In this case, Eq.~(\ref{eq:equations-of-motion-generalized-2}) is called {\em Hill's differential equation} named after the astronomer G.W.~Hill (cf.\ \cite{Hill}). ç
A powerful theorem known as {\em Floquet's theorem} states that if $K(s)$ is periodic, the amplitude function\textemdash and therewith the beta function $\beta(s)$\textemdash is periodic as well (cf.\ \cite{Floquet}).
With Eqs.~(\ref{eq:def-alpha-Twiss}) and (\ref{eq:def-gamma-Twiss}) this translates to so-called {\em periodic boundary condititions} for which
%------------------
\begin{eqnarray}
\label{eq:periodic-beta}
\beta(s+L) = \beta(s)~, \\
\label{eq:periodic-alpha}
\alpha(s+L) = \alpha(s)~, \\
\label{eq:periodic-gamma}
\gamma(s+L) = \gamma(s)~.
\end{eqnarray}
%------------------
Please note that this does not imply that the phase advance $\mu(s)$ an therewith the transverse displacements $x(s), y(s)$ are periodic as well.
This would be an exception and deal with a catastrophic situation, as we will discover later.
Applying the periodic boundary conditions to the transfer matrix of the periodic segment (or the one-turn matrix), we can simplify its Twiss parameters' dependent representation (cf.\ Eq.~(\ref{eq:transfer-matrix-twiss})) significantly since $\beta = \beta_0, \alpha = \alpha_0, \gamma = \gamma_0$:
%------------------
\begin{equation}
\boxed{
\mathbf{M} = \begin{pmatrix} \cos \mu + \alpha_0 \sin \mu & \beta_0 \sin \mu \\
-\gamma_0 \sin \mu & \cos \mu - \alpha_0 \sin \mu
\end{pmatrix}
}~.
\label{eq:Twiss-matrix}
\end{equation}
%------------------
This matrix was first derived by R.Q.~Twiss from general mathematics principles and is called the {\em Twiss matrix}.
In order to get an information on what is happening when $\mathbf{M}^N$ is built, we calculate the eigenvalues of $\mathbf{M}$ from 
%------------------
\begin{equation}
\det \left( \mathbf{M} - \lambda \cdot \mathbf{I} \right) = \lambda^2 - \mathrm{Tr}\left( \mathbf{M} \right)  \cdot\lambda + \det \mathbf{M} = 0~.
\end{equation}
%------------------
With $\mathrm{Tr}\left(\mathbf{M}\right) = 2 \cos\mu$ and $\det\mathbf{M}=1$ we obtain
%------------------
\begin{equation}
\lambda_{1,2} = \cos \mu \pm i \sin \mu = e^{\pm i \mu}~.
\end{equation}
%------------------
This reveals an important information: even for large values of $\mu$ we get eigenvalues $\vert \lambda_{1,2} \vert = 1$ in case of a real phase advance since then $\vert e^{\pm i \mu} \vert = 1$.
This is ensured when
%------------------
\begin{equation}
\vert\mathrm{Tr}\left(\mathbf{M}\right)\vert = \vert 2\cos\mu \vert \le 2~.
\label{eq:trace-M}
\end{equation}
%------------------
It should be noted that  the phase advance $\mu$ of a periodic structure and therewith the trace of its transfer matrix $\mathbf{M}$ are independent of the chosen starting point $s$ in the periodic lattice.
This can be seen as follows:\\
We can express the transfer matrix from position $s_1$ to position $s_2 + L$ as
%------------------
\begin{equation}
\mathbf{M}(s_1, s_2+L) = \mathbf{M}(s_2, s_2+L) \cdot \mathbf{M}(s_1,s_2)~,
\end{equation}
%------------------
or by
%------------------
\begin{equation}
\mathbf{M}(s_1, s_2+L) = \mathbf{M}(s_1+L, s_2+L) \cdot \mathbf{M}(s_1,s_1+L)~.
\end{equation}
%------------------
Since $\mathbf{M}(s_1+L,s_2+L) = \mathbf{M}(s_1,s_2)$, the matrix $\mathbf{M}(s_2,s_2+L)$ differs from the matrix $\mathbf{M}(s_1,s_1+L)$ only by a similarity transformation
%------------------
\begin{equation}
\mathbf{M}(s_2, s_2+L) = \mathbf{M}(s_1, s_2) \cdot \mathbf{M}(s_1,s_1+L) \cdot \mathbf{M}^{-1}(s1,s2)~,
\end{equation}
%------------------
which does not change the trace.
But take care: the transfer matrix $\mathbf{M}$ as a whole does depend on the~starting position $s$.
This is caused by the dependence of the optical functions on s.

\noindent We now apply a ``trick'' and rewrite the Twiss matrix using
%------------------
\begin{equation}
\mathbf{J} = \begin{pmatrix} \alpha & \beta \\ -\gamma & -\alpha \end{pmatrix}~,
\end{equation}
%------------------
and the 2x2 identity matrix $\mathbf{I}$ in the form
%------------------
\begin{equation}
\mathbf{M} = \mathbf{I}\cdot\cos\mu + \mathbf{J}\cdot\sin\mu~.
\end{equation}
%------------------
Since $\mathbf{J}^2= -\mathbf{I}$ we obtain
%------------------
\begin{equation}
\mathbf{M}^2 = \mathbf{I}\cdot\left( \cos^2\mu - \sin^2\mu\right) + 2\mathbf{J}\sin\mu\cos\mu = \mathbf{I}\cdot\cos(2\mu) + \mathbf{J}\cdot\sin(2\mu)~,
\end{equation}
%------------------
which can be extended using Moivre's formula to
%------------------
\begin{equation}
\mathbf{M}^N = \mathbf{I}\cdot\cos(N\mu) + \mathbf{J}\cdot\sin(N\mu)~.
\label{eq:Moivre}
\end{equation}
%------------------
Now coming back to Eq.~(\ref{eq:trace-M}).
With Eq.~(\ref{eq:Moivre}) we have
%------------------
\begin{equation}
\vert\mathrm{Tr}\left(\mathbf{M}^N\right)\vert = \vert 2\cos(N\mu) \vert \le2~.
\label{eq:trace-M^N}
\end{equation}
%------------------
So we found:\\[0.3cm]
\begin{center}
\textbf{
The matrix for an infinite number of turns in a periodic structure stays finite as long as the phase advance per period $\mu$ is a real number.
This can be easily checked by computing the trace of the~transfer matrix of the periodic structure which must satisfy 
}
\end{center}
%------------------
\begin{equation}
\boxed{
\vert \mathrm{Tr}\left( M \right) \vert \le 2
}~.
\label{eq:trace-condition}
\end{equation}
%------------------

%_______________________________________________________________________________
%---------------------------------------------------------------------------------------------------------------
\subsection{Regular FODO lattice}
\label{sec:FODO}
\begin{figure}[ht]
\begin{center}
\includegraphics[width=10cm]{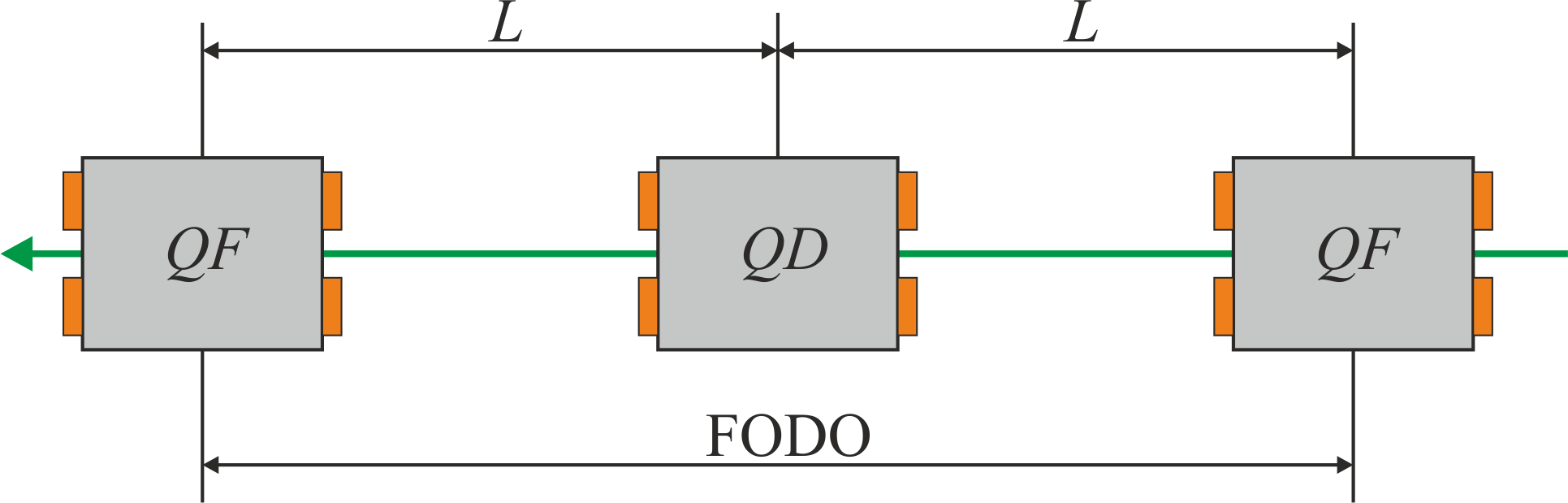}
\caption{One period of a FODO structure made from a periodic sequence of quadrupole magnets of alternating focusing.}
\label{fig:FODO}
\end{center}
\end{figure}

\begin{figure}[ht]
\begin{center}
\includegraphics[width=7cm]{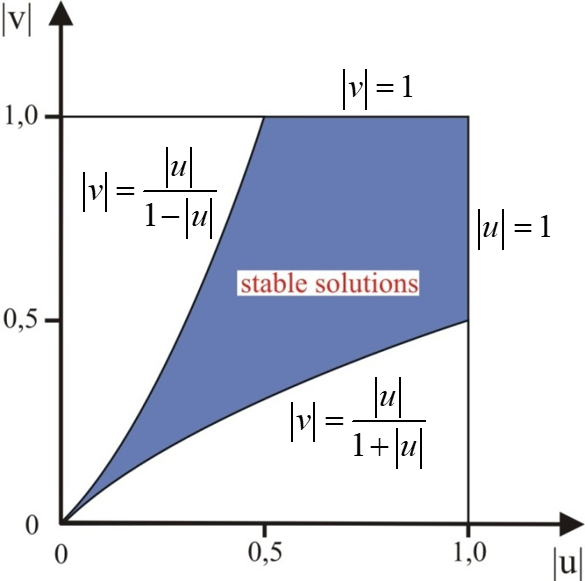}
\caption{Necktie diagram of stable solutions for a regular FODO lattice in thin lens approximation. The parameters $u$ and $v$ are linked to the focusing strengths $f_{QF}$ and $f_{QD}$ of the quadrupoles and the distance between the~quadrupoles $L$ by $u=\frac{L}{2f_{QF}}$ and $v=\frac{L}{2f_{QD}}$.}
\label{fig:necktie}
\end{center}
\end{figure}

%_______________________________________________________________________________
%---------------------------------------------------------------------------------------------------------------
The most popular strong focusing lattice is a regular (equidistant) sequence of focusing ($QF$) and defocusing ($QD$) quadrupole magnets with either free space or dipole magnets in between.
Since the~focusing is completely dominated by the quadrupole magnets, dipole magnets are treated as ``0'' focusing.
We thus have a sequence of F - 0 - D - 0 focusing.
Replacing ``0'' by ``O'' the optics is called a ``FODO'' lattice.
One period of such a structure is presented in Fig.~\ref{fig:FODO}.
%---------------------------------------------------------------------------------
%--------------------------------------------------------------------------------
The FODO geometry can be expressed symbolically by the sequence
%------------------
\begin{equation}
\underbrace{
\frac{1}{2} QF, O, \frac{1}{2} QD}
_{\substack{= \mathbf{M}_{-1/2}}}, 
\underbrace{
\frac{1}{2} QD, O, \frac{1}{2} QF}
_{\substack{= \mathbf{M}_{1/2}}}~.
\end{equation}
%------------------
We will treat the quadrupoles as thin magnets and set the focal lengths of half the quadrupoles to $f_1 = 2f_{QF} > 0, f_2 = 2f_{QD} < 0$ and the drift length to $L$.
Defining
%------------------
\begin{equation}
\frac{1}{f^\ast}=\frac{1}{f_1}+\frac{1}{f_2}-\frac{L}{f_1\cdot f_2}~,
\end{equation}
%------------------
the transformation matrix of half a FODO cell is
%------------------
\begin{equation}
\mathbf{M}_{1/2}=\begin{pmatrix}1&0\\-\frac{1}{f_2}&1\end{pmatrix}\cdot
\begin{pmatrix}1&L\\0&1\end{pmatrix}\cdot\begin{pmatrix}1&0\\-\frac{1}{f_1}&1\end{pmatrix}=
\begin{pmatrix}1-\frac{L}{f_1}&L\\ -\frac{1}{f^\ast}&1-\frac{L}{f_2}\end{pmatrix}~,
\end{equation}
%------------------
and
%------------------
\begin{equation}
\mathbf{M}_{-1/2}=\begin{pmatrix}1&0\\-\frac{1}{f_1}&1\end{pmatrix}\cdot
\begin{pmatrix}1&L\\0&1\end{pmatrix}\cdot\begin{pmatrix}1&0\\-\frac{1}{f_2}&1\end{pmatrix}=
\begin{pmatrix}1-\frac{L}{f_2}&L\\ -\frac{1}{f^\ast}&1-\frac{L}{f_1}\end{pmatrix}~.
\end{equation}
%------------------
Multiplying both matrices gives the transformation matrix of the FODO cell
%------------------
\begin{equation}
\mathbf{M}_\text{FODO}=\begin{pmatrix}1-2\frac{L}{f^\ast}&2L\left( 1-\frac{L}{f_2}\right) \\
-\frac{2}{f^\ast}\left(1-\frac{L}{f_1}\right)&1-\frac{2L}{f^\ast}\end{pmatrix}~.
\end{equation}
%------------------
Overall stability is guaranted if Eq.~(\ref{eq:trace-condition}) is fullfilled.
This gives
%------------------
\begin{equation}
\vert \mathrm{Tr}\left( \mathbf{M}_\text{FODO} \right) \vert = \vert 2-\frac{4 L}{f^\ast}\vert \le 2~,
\end{equation}
%------------------
which is equivalent to
%------------------
\begin{equation}
1 \ge \frac{L}{f^\ast} \ge 0~.
\end{equation}
%------------------
Defining two parameters $u = \frac{L}{f_1}$ ($QF$ has $f_1>0$) and $v = \frac{L}{f_2}$ ($QD$ has $f_2<0$) we get
%------------------
\begin{equation}
0 \le u+v-u\cdot v \le 1~,
\end{equation}
%------------------
from which we derive the boundaries of the stability region
%------------------
\begin{eqnarray}
\vert u\vert \le 1~, ~~~~ \vert v\vert \ge \frac{\vert u \vert}{1+\vert u\vert}~, \\
\vert v\vert \le 1~, ~~~~ \vert v\vert \le \frac{\vert u \vert}{1-\vert u\vert}~,
\end{eqnarray}
%------------------
and obtain the ``famous'' {\em necktie diagram} of stability for a thin lens approximation which is presented in Fig.~\ref{fig:necktie}.
%---------------------------------------------------------------------------------
%--------------------------------------------------------------------------------
In the simplest case of equal focusing strengths $f=f_{QF} = - f_{QD}$, we arrive at
%------------------
\begin{equation}
\frac{L}{2f} = \frac{L_\text{FODO}}{4f} \le 1~,
\end{equation}
%------------------
which corresponds to the minimum $4f$ imaging in light optics.

Beyond looking at the limits of stability, we can further optimize the focusing with respect to the~achievment of minimum beta functions and therewith beam sizes.
In doing so, we will again concentrate on the symmetric case $f = f_{QF} = - f_{QD}$.
The transfer matrix of a FODO cell then reduces to
%------------------
\begin{equation}
\mathbf{M}_\text{F/2-ODO-F/2}=\begin{pmatrix}1-\frac{L^2}{2f^2}&L\left( 2+\frac{L}{f}\right) \\
-\frac{L}{f^2}\left( 1-\frac{L}{2f}\right)&1-\frac{L^2}{2f^2}\end{pmatrix}~,
\end{equation}
%------------------
when starting in the center of a focusing quadrupole and
%------------------
\begin{equation}
\mathbf{M}_\text{D/2-OFO-D/2}=\begin{pmatrix}1-\frac{L^2}{2f^2}&L\left( 2-\frac{L}{f}\right) \\
-\frac{L}{f^2}\left( 1+\frac{L}{2f}\right)&1-\frac{L^2}{2f^2}\end{pmatrix}~,
\end{equation}
%------------------
when starting in the center of a defocusing quadrupole.
From the trace of the matrices, we get the phase advance
%------------------
\begin{equation}
\cos\mu = 1-\frac{L^2}{2f^2}~~~~~~\rightarrow~~~~~~\frac{L^2}{4f^2} = \frac{1}{2}(1-\cos\mu) = \sin^2\left(\frac{\mu}{2}\right)~,
\end{equation}
%------------------
whereas the beta function in the center of the quadrupole magnets can be derived from the matrix element $r_{12}=\beta\sin\mu$ revealing the maximum and minimum $\beta$ values
%------------------
\begin{eqnarray}
\beta_\text{max} = \frac{2L}{\sin\mu}\left( 1+\sin\left(\frac{\mu}{2}\right)\right)~, \\
\beta_\text{min} = \frac{2L}{\sin\mu}\left( 1-\sin\left(\frac{\mu}{2}\right)\right)~.
\end{eqnarray}
%------------------
The average beta function is
%------------------
\begin{equation}
\left\langle\beta\right\rangle = \frac{2L}{\sin\mu}~,
\end{equation}
%------------------
and is minimized for a phase advance of $\mu = \pi/2$.
For this often chosen case we have the striking result that the average beta function is the period length $2L$ of a FODO cell.

%_______________________________________________________________________________
%---------------------------------------------------------------------------------------------------------------
\subsection{Periodic beta functions}
\label{sec:periodic-beta}
%_______________________________________________________________________________
%---------------------------------------------------------------------------------------------------------------
We now want to extend our treatment to general periodic magnetic lattices.
As discussed before, periodicity implies periodic boundary conditions as stated in Eqs.~(\ref{eq:periodic-beta})--(\ref{eq:periodic-gamma}). In this case, the transfer matrix for one period can as well be represented by its Twiss form (the Twiss matrix, cf.\ Eq.~(\ref{eq:Twiss-matrix})).
If we start at a symmetry point $s_0$ where $\alpha_0=0$, we obtain by comparing the transfer matrix $\mathbf{M}(s_0, s_0+L)$, derived from multiplying the transfer matrices of the individual elements, with its Twiss form
%------------------
\begin{equation}
\mathbf{M}(s_0, s_0+L) = \begin{pmatrix}r_{11} & r_{12} \\ r_{21} & r_{22} \end{pmatrix} =
\begin{pmatrix} \cos\mu & \ \beta_0 \sin\mu \\ -\gamma_0\sin\mu & \cos\mu \end{pmatrix}~,
\end{equation}
%------------------
and get the Twiss parameters at position $s_0$
%------------------
\begin{equation}
\alpha_0 = 0~,~~~~~~ \beta_0 = \frac{r_{12}}{\sqrt{1-r_{11}^2}}~, ~~~~~~\gamma_0 = \frac{1}{\beta_0} = \frac{-r_{21}}{\sqrt{1-r_{11}^2}}~, ~~~~~~ \cos\mu = r_{11}~.
\end{equation}
%------------------
Applying the formalism based on the transformation of the beta matrix, these parameters can be transformed to any position $s$.
This only requires to first compute the transformation matrix $\mathbf{M}(s_0, s)$ which then has to be applied to the beta matrix according to Eq.~(\ref{eq:beta-matrix-transformation})
%------------------
\begin{equation}
\begin{pmatrix}\beta(s) & -\alpha(s) \\ -\alpha(s) & \gamma(s) \end{pmatrix} =
\mathbf{M}(s_0,s) \cdot \begin{pmatrix}\beta_0 & 0 \\ 0 & \frac{1}{\beta_0} \end{pmatrix}\cdot {^T}{\mathbf{M}(s_0,s)}~,
\label{eq:eigenfunctions-machine}
\end{equation}
%------------------
thus revealing the evolution of the optical functions $\beta(s), \alpha(s), \gamma(s)$.

Again, we obtained a very important result: for a circular accelerator (or in general a periodic structure) the optical functions are only determined by the magnetic lattice!
This was caused by periodicity and Floquet's theorem requiring the beta functions to be periodic as well.

The optical functions derived by Eq.~(\ref{eq:eigenfunctions-machine}) define what is named a {\em machine ellipse}.
More precisely: the orientation and shape of the machine ellipse is uniquely defined by the optical functions, its area is usually set to $\pi\epsilon$ where $\epsilon$ is the emittance of the circulating beam.
But take care: the properties of a~circulating beam are represented by its beam matrix (cf.\ Eq.~(\ref{eq:beam-matrix})) which reads for the plane $u$:
%------------------
\begin{equation}
\mathbf{\Sigma}_\text{beam} =\begin{pmatrix}\sigma_u^2 & \sigma_{u u^\prime} \\ \sigma_{u u^\prime} & \sigma_{u^\prime}^2 \end{pmatrix}~,
\end{equation}
%------------------
and define the beam's r.m.s.\ emittance by
%------------------
\begin{equation}
\epsilon_u = \sqrt{\det\left( \mathbf{\Sigma}_\text{beam}\right)}~,
\end{equation}
%------------------
%---------------------------------------------------------------------------------
\begin{figure}[!ht]
\begin{center}
\includegraphics[width=15cm]{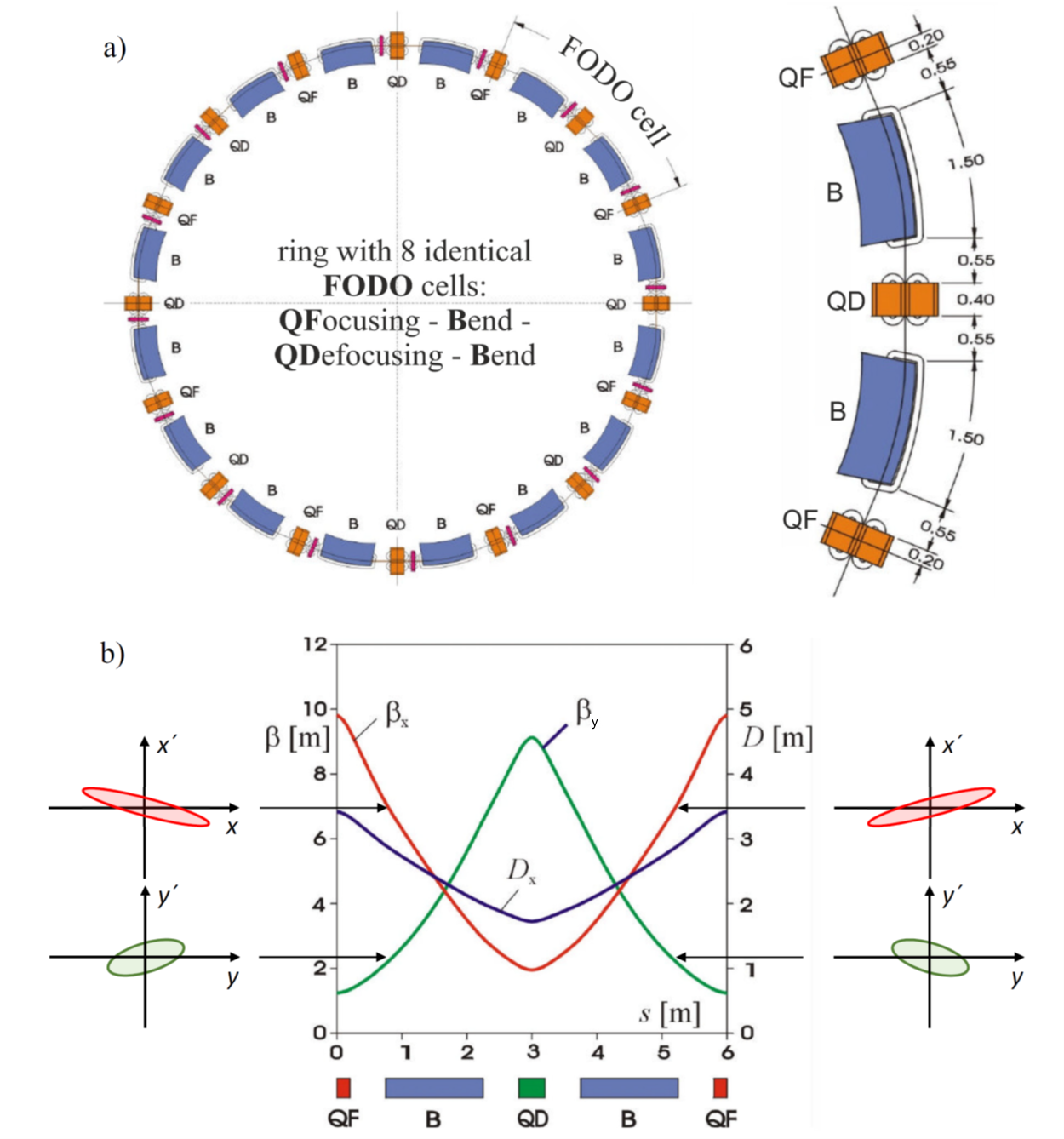}
\caption{a) Toy model of a simple circular accelerator consisting of 8 identical FODO cells in which rectangular dipole magnets between the quadrupoles deflect the beam on a circular orbit. 
b) Evolution of the horizontal and vertical beta functions in a FODO cell of the toy model ring. In addition, two machine ellipses at different positions are shown. A third displayed function $D_x$, the horizontal dispersion, will be discussed in Section \ref{sec:off-momentum-particles}.}
\label{fig:toy-ring-complete}
\end{center}
\end{figure}
%--------------------------------------------------------------------------------
Using Eqs.~(\ref{eq:def-beta})--(\ref{eq:def-alpha}), the beam matrix can as well be written as
%------------------
\begin{equation}
\mathbf{\Sigma}_\text{beam} =\epsilon_u\cdot\mathbf{B}_\text{beam}=\epsilon_u\cdot
\begin{pmatrix}\beta_u & -\alpha_u \\ -\alpha_u & \gamma_u \end{pmatrix}~,
\end{equation}
%------------------
thus defining optical parameters $\beta_u, \alpha_u, \gamma_u$ of the beam.It is important to note that these optical parameters $\beta_u, \alpha_u, \gamma_u$ must not be equal to the optical functions derived from the lattice.
If this is the case, we have a {\em matched beam}.
In general, the beam might have any distribution not in accordance with the~machine ellipse!
This will be discussed in more detail in Subsection \ref{sec:filamentation}.

As an example, we will apply this to a simple toy model ring having a FODO lattice with rectangular dipole magnets between the quadrupoles (cf.\ \cite{Wille} and Fig.~\ref{fig:toy-ring-complete}(a)).
According to the procedure described above, the evolution of the beta functions can be derived. It has been computed for a bending radius $\rho=3.8197\,\text{m}$ and $k_{QD}=-k_{QF}=1.20\,\text{m}^{-2}$ and is presented in Fig.~\ref{fig:toy-ring-complete}(b).
In addition, the~horizontal and vertical machine ellipses are shown for 2 different positions in the ring.
The shape and orientation of these ellipses vary along the ring and are determined by the optical functions.
In the~center of the quadrupole magnets, the ellipses are symmetric with respect to the trace space axes since these positions are symmetry points of the lattice.
There, the beta functions have a local extremum:
\begin{itemize}
\item $\beta_x$ will have a local maximum in a (horizontally) focusing quadrupole $QF$ and a local minimum in a defocusing quadrupole $QD$,
\item $\beta_y$ will just show the opposite: a local minimum in a focusing quadrupole $QF$ and a local maximum in a defocusing quadrupole $QD$.
\end{itemize}

%_______________________________________________________________________________
%---------------------------------------------------------------------------------------------------------------
\subsection{Betatron tune}
\label{sec:betatron-tune}
%_______________________________________________________________________________
%---------------------------------------------------------------------------------------------------------------
Each particle orbiting in a circular accelerator is moving in trace space on a machine ellipse with area $\pi A^2$ defined by the Courant-Snyder invariant $A^2$ of the particle.
The shape and orientation of this ellipse is determined by the optical functions of the machine lattice and is fixed at a given longitudinal position $s$.
Since the particle's orbit oscillates around the reference orbit (the particle performs betatron oscillations), the actual position of the particle's point in trace space on this ellipse is determined by the phase of its betatron oscillation.
The number of oscillations per turn is called the {\em betatron tune} $Q_u$ and can be computed via
%------------------
\begin{equation}
Q_u = \frac{1}{2 \pi} \oint \frac{1}{\beta_u}\mathrm{d}s~.
\label{eq:def-tune}
\end{equation}
%------------------
In case of a mono-energetic beam and a linear lattice, the tune is the same for all particles.
Thus, for each revolution of a particle its point $(u, u^\prime)$ in trace space makes $Q_u$ turns along the ellipse equivalent to a phase advance per turn of $\mu_u = 2\pi Q_u$.
%---------------------------------------------------------------------------------
\begin{figure}[ht]
\begin{center}
\includegraphics[width=7cm]{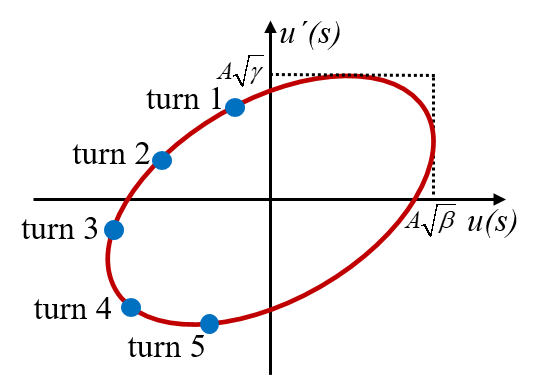}
\caption{Trace space ellipse for one particle at a fixed longitudinal position in the ring. For each revolution of the~particle, the point $(u,u^\prime)$ makes $Q_u$ turns along the machine ellipse with area $\pi A^2_u$ determined by the Courant-Snyder invariant $A^2_u$ for the coordinate $u$ of the particle.}
\label{fig:ellipse-point-hopping}
\end{center}
\end{figure}
%--------------------------------------------------------------------------------
This is illustrated in Fig.~\ref{fig:ellipse-point-hopping}, representing the trace space at a fixed longitudinal position $s$.
From turn to turn the particle hops on the machine ellipse by an angle given by $2\pi$ times the fractional tune.

%_______________________________________________________________________________
%---------------------------------------------------------------------------------------------------------------
\subsection{Filamentation}
\label{sec:filamentation}
\begin{figure}[ht]
\begin{center}
\includegraphics[width=5cm]{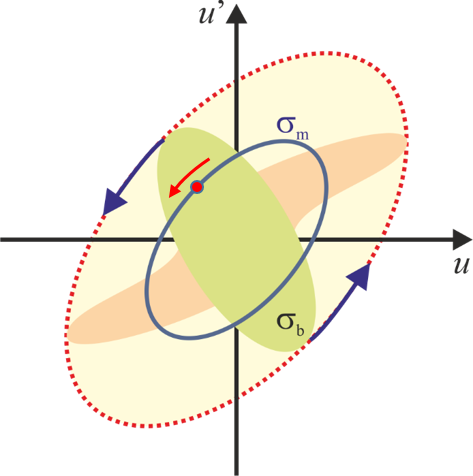}
\caption{In case of an unmatched beam, the envelope ellipse of the~beam starts to rotate since each particle of the~beam moves on an ellipse whose orientation and shape is determined by the machine ellipse. Non-linear processes lead to an amplitude-dependent phase advance per turn and finally to filling a machine ellipse with larger area. This process is called  filamentation.}
\label{fig:filamentation}
\end{center}
\end{figure}

%_______________________________________________________________________________
%---------------------------------------------------------------------------------------------------------------
Let us now consider a beam injection into a circular accelerator.
If the optical parameters of the beam at the injection point differ from the optical functions of the lattice at this location, the beam's individual ellipse is not matched to the machine ellipse.
If we track an individual particle of the beam at the~point of injection turn by turn, its trace space coordinates will follow the machine ellipse.
Since this happens for all particles, the beam ellipse will start to rotate with a phase advance per revolution of $2 \pi Q_u$.
Due to effects of higher order the quadrupole strengths and therewith the phase advance depends on the~amplitude and thus the Courant-Snyder invariant of the particle.
In case of a mismatch, the~beam trace space distribution starts to filament as indicated in Fig.~\ref{fig:filamentation}.
%---------------------------------------------------------------------------------
%--------------------------------------------------------------------------------
After a large number of turns, the distribution smears out and may be enclosed by a larger ellipse with the shape and orientation of the~machine ellipse.
Hence the emittance of the injected beam has therewith been enlarged (or ``spoiled''), which will not happen if the injected beam is matched to the machine parameters at the injection point.

Please note that Liouville's theorem is not violated by this process since at any point in time, the~phase space area covered by the particle distribution is preserved.
In fact non-linear and non-conservative effects which are always present lead to mixing and coupling the particle's motion which finally damps the correlated oscillations and fills the entire new equilibrium phase space. Now Liouville's theorem is no longer applicable, the emittance has suffered a real growth.

%_______________________________________________________________________________
%---------------------------------------------------------------------------------------------------------------
\subsection{Normalized coordinates}
\label{sec:normalized-coordinates}
%_______________________________________________________________________________
%---------------------------------------------------------------------------------------------------------------
It is often useful to transform the oscillatory solution with varying amplitude and frequency to a solution which looks exactly like that of a harmonic oscillator.
So far, we had
%------------------
\begin{eqnarray}
u(s) &=& A\sqrt{\beta(s)} \cos\left( \mu(s)+\varphi_0\right)~, \\
u^\prime(s) &=& -A\left[\frac{\alpha(s)}{\sqrt{\beta(s)}}\cos\left( \mu(s)+\varphi_0\right) - \frac{1}{\sqrt{\beta}}sin\left( \mu(s)+\varphi_0\right)\right]~.
\end{eqnarray}
%------------------
We now introduce new coordinates $u_n(\psi)$ defined by
%------------------
\begin{eqnarray}
\label{eq:def-u-norm}
u_n &=& \frac{u}{\sqrt{\beta}}~, \\
\psi &=& \frac{\mu}{Q}~.
\end{eqnarray}
%------------------
The angle $\psi$ advances by $2 \pi$ every revolution. It coincides with the geometric angle $\theta$ (which is defined by $s = \frac{L\theta}{2\pi}$) at each $\beta_\text{max}$ and $\beta_\text{min}$ location and does not depart very much from $\theta$ in between. 
We can as well use $u_n(\mu)$ which only differs by the different phase advance of $2\pi Q$ per revolution.

\noindent Considering $\mu^\prime = \frac{1}{\beta}$ and $\beta^\prime = -2\alpha$ we get for $\dot u_n(\mu)$ where ``dot'' represents the derivative with respect to $\mu$ while ``prime'' is the derivative with respect to $s$
%------------------ 
\begin{equation}
\dot u_n(\mu) = \frac{\mathrm{d}u_n}{\mathrm{d}\mu} = \frac{\mathrm{d}u_n}{\mathrm{d}s} \frac{\mathrm{d}s}{\mathrm{d}\mu} = \frac{\mathrm{d}u_n}{\mathrm{d}s} \beta =
\left(\frac{u^\prime}{\sqrt{\beta}}-\frac{1}{2}u\beta^{-3/2}\beta^\prime\right) =
u^\prime\sqrt{\beta}+\frac{a}{\sqrt{\beta}}u~,
\end{equation}
%------------------
and together with Eq.~(\ref{eq:def-u-norm}) we obtain the required transformation
%------------------ 
\begin{equation}
\begin{pmatrix}u_n \\ \dot u_n \end{pmatrix} = 
\begin{pmatrix}\frac{1}{\sqrt{\beta}} & 0 \\ \frac{\alpha}{\sqrt{\beta}} & \sqrt{\beta} \end{pmatrix} \cdot
\begin{pmatrix}u \\ u^\prime \end{pmatrix}~,
\end{equation}
%------------------
or in short form
%------------------ 
\begin{equation}
\vec u_n(\mu) = \mathbf{T} \cdot \vec u(s)~, ~~~~~~~~~ \vec u(s) = \mathbf{T}^{-1} \cdot \vec u_n(\mu)~,
\end{equation}
%------------------
with the transformation matrices
%------------------ 
\begin{equation}
\mathbf{T}(s) = \begin{pmatrix}\frac{1}{\sqrt{\beta}} & 0 \\ \frac{\alpha}{\sqrt{\beta}} & \sqrt{\beta} \end{pmatrix},
~~~~~~~~~ 
\mathbf{T}^{-1}(s) = \begin{pmatrix}\sqrt{\beta} & 0 \\ -\frac{\alpha}{\sqrt{\beta}} & \frac{1}{\sqrt{\beta}} \end{pmatrix}~.
\end{equation}
%------------------
Please note that the transformation matrix $\mathbf{T}$ is explicitly depending on the longitudinal position $s$ since the optical functions are explicitly depending on $s$ as well.

\noindent The transformation to $u_n(\psi)$ is derived completely analogue to the procedure above.
We get
%------------------ 
\begin{equation}
\vec u_n(\psi) = \mathbf{\tilde T} \cdot \vec u(s)~, ~~~~~~~~~ \vec u(s) = \mathbf{\tilde T}^{-1} \cdot \vec u_n(\psi)~,
\label{eq:trans-to-psi}
\end{equation}
%------------------
with
%------------------ 
\begin{equation}
\mathbf{\tilde T}(s) = \begin{pmatrix}\frac{1}{\sqrt{\beta}} & 0 \\ Q\frac{\alpha}{\sqrt{\beta}} & Q\sqrt{\beta}
\end{pmatrix},
~~~~~~~~~ 
\mathbf{\tilde T}^{-1}(s) = \begin{pmatrix}\sqrt{\beta} & 0 \\ -\frac{\alpha}{\sqrt{\beta}} & \frac{1}{Q\sqrt{\beta}} \end{pmatrix}~.
\label{eq:matrix-trans-to-psi}
\end{equation}
%------------------
The new coordinates $\vec u_n$ are often called {\em Floquet coordinates} and the transformation $\mathbf{T}$ {\em Floquet transformation}.
Since normalized coordinates transform through the lattice as
%------------------ 
\begin{equation}
\vec u = \mathbf{T}^{-1}\cdot\vec u_n =\mathbf{M}\cdot\vec u_0 = \mathbf{M}\cdot\left(\mathbf{T}^{-1}\cdot\vec u_{n,0}\right)
~~~~~~\rightarrow~~~~~~\vec u_n = \mathbf{T}\cdot\mathbf{M}\cdot\mathbf{T}^{-1}\cdot u_{n,0}~,
\end{equation}
%------------------ 
the one-turn matrix $\mathbf{M}$ has to be replaced replaced by $\mathbf{R}$
%------------------ 
\begin{equation}
\mathbf{R} = \mathbf{T}\cdot\mathbf{M}\cdot\mathbf{T}^{-1} = 
\begin{pmatrix}\cos(2\pi Q) & \sin(2\pi Q) \\ -\sin(2\pi Q) & \cos(2\pi Q) \end{pmatrix}~,
\end{equation}
%------------------ 
which is a pure rotation matrix when transforming normalized coordinates.
Using these normalized coordinates, the equation of motion is simplified to
%------------------ 
\begin{equation}
\frac{\mathrm{d}^2 u_n}{\mathrm{d}\mu^2} + u_n =0~, ~~~~~~~\text{or}~~~~~~~
\frac{\mathrm{d}^2 u_n}{\mathrm{d}\psi^2} + Q^2 u_n =0~.
\end{equation}
%------------------ 
The ellipse equation transforms to
%------------------ 
\begin{equation}
A = \gamma u^2 + 2\alpha u u^\prime + \beta {u^\prime}^2 = u_n^2 + {\dot u_n}^2~,
\end{equation}
%------------------ 
and the ellipse in the trace space transforms to a circle in the normalized trace space,
%---------------------------------------------------------------------------------
\begin{figure}[ht]
\begin{center}
\includegraphics[width=10cm]{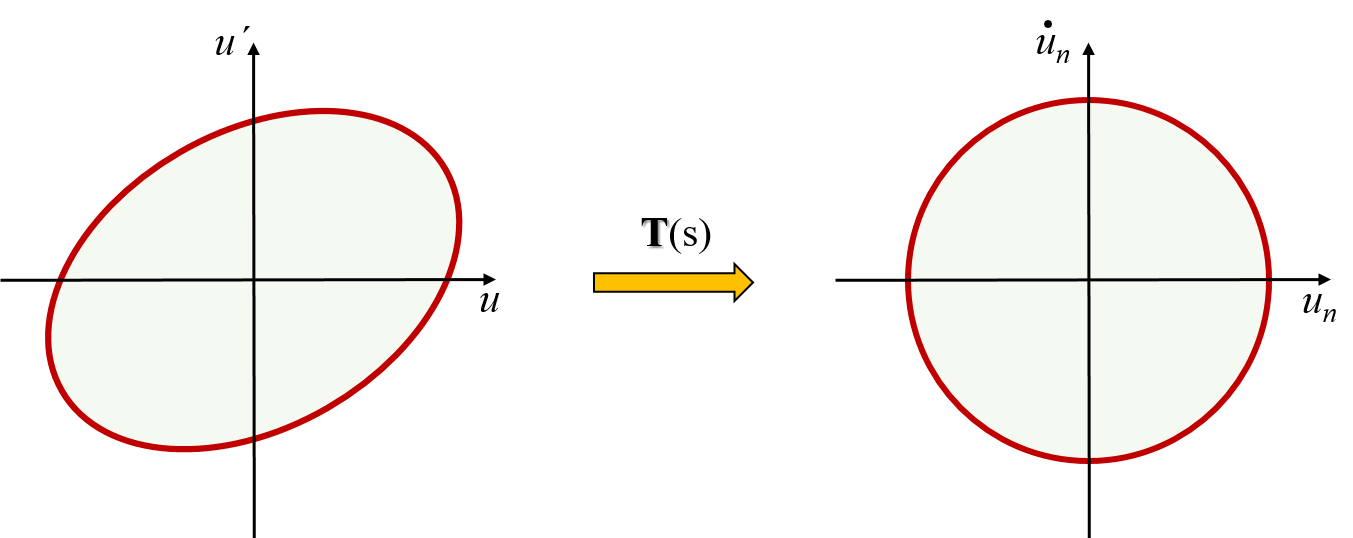}
\caption{Floquet transformation to normalized coordinates changes the machine ellipse in the trace space to a circle in the normalized trace space.}
\label{fig:ellipse-circle}
\end{center}
\end{figure}
%--------------------------------------------------------------------------------
which is visualized in Fig.~\ref{fig:ellipse-circle}.
We thus have
%------------------ 
\begin{equation}
\langle u_n^2 \rangle = \langle {\dot u_n}^2 \rangle = \epsilon~, ~~~~~~~~~~~~~ \langle u_n\cdot \dot u_n \rangle = 0~.
\end{equation}
%------------------

%_______________________________________________________________________________
%---------------------------------------------------------------------------------------------------------------
\subsection{Dipole errors and closed orbit distortions}
\label{sec:closed-orbit-distortions}
%_______________________________________________________________________________
%---------------------------------------------------------------------------------------------------------------
%_______________________________________________________________________________
\subsubsection{Closed orbit}
\label{sec:closed-orbit}
%_______________________________________________________________________________
Periodicity, which in circular accelerators will appear at least after one turn, leads to a periodic amplitude function $\beta_u(s)$ which reproduces itself after one turn according to Floquet's theorem.
This implies, that the charge center of the beam moves on a closed trajectory, which is called the {\em closed orbit}.
Please note that this does not apply for the trajectories of individual particles as well (which would have disastrous consequences as we will discover in Section~\ref{sec:optical-resonances}).
The shape of the closed orbit is determined by the~magnets and can, due to errors and misalignments, significantly deviate from the design orbit.
In the~following, we will investigate the impact of dipole errors on the closed orbit causing deviations which are called closed orbit distortions.
Dedicated steerer magnets (e.g.\ small dipole magnets or correction coils), which have to be installed around the ring, are typically used to correct these distortions.

%_______________________________________________________________________________
\subsubsection{Dipole error}
\label{sec:dipole-error}
%_______________________________________________________________________________
Let us assume a single erroneous dipole magnet.
It can be approximated by a adding a small extra dipole field $\delta B$ to an error-free dipole magnet, produced by a short dipole.
Such a short dipole of length $l$ will additionally bend the beam by a small angle $\varphi$ on a circular orbit with radius $r = \frac{p_0}{q\delta B}$.
%---------------------------------------------------------------------------------
\begin{figure}[ht]
\begin{center}
\includegraphics[width=3.5cm]{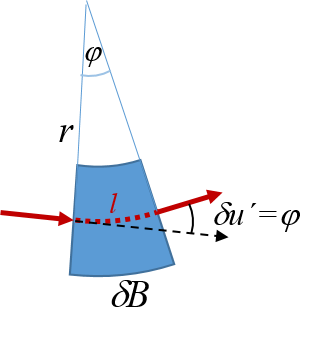}
\caption{A short dipole generates a small kick $\delta u^\prime$ in divergence.}
\label{fig:dipole-error}
\end{center}
\end{figure}
%--------------------------------------------------------------------------------
It will therefore generate a small angular kick $\delta u^\prime$ in divergence of a beam whose momentum is defined by its magnetic rigidity $p_0 = q\rho B$:
%------------------ 
\begin{equation}
l = r\cdot\varphi = \frac{p_0}{q \delta B}\delta u^\prime = \frac{\rho B}{\delta B}\delta u^\prime~.
\end{equation}
%------------------
We thus get for the angular kick changing the angular displacement $u^\prime$
%------------------ 
\begin{equation}
\delta u^\prime = \frac{\delta \left( B l \right)}{\rho B}~.
\end{equation}
%------------------
However, the transverse displacement $u$ will remain unchanged in good approximation.

%_______________________________________________________________________________
\subsubsection{Closed orbit distortions}
\label{sec:co-distortions}
%_______________________________________________________________________________
We will start considering the following simplified situation: a single erroneous dipole is located at longitudinal position $s_0$ causing an extra angular kick $\delta u^\prime$. Elsewhere we have an unperturbed lattice, for which the orbit obeys the unperturbed Hill's differential equation.

\begin{figure}[ht]
\begin{center}
\includegraphics[width=10cm]{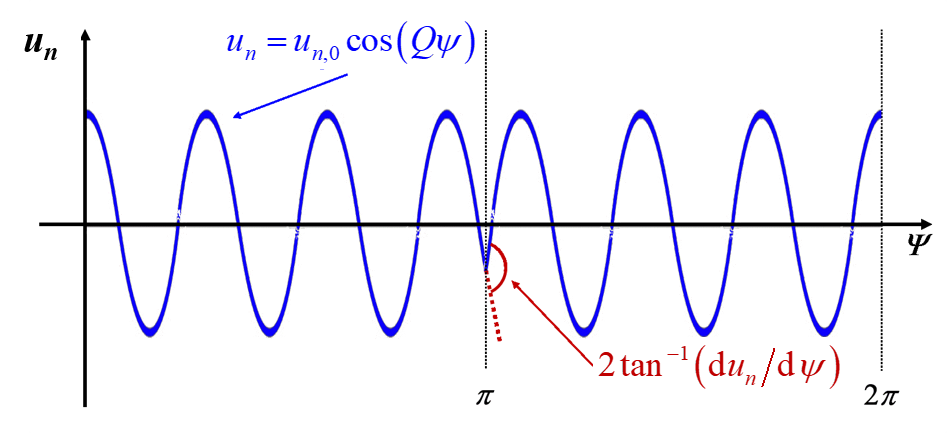}
\caption{Perturbed closed orbit in normalized coordinates. A single dipole error at $\psi=\pi$ causes a harmonic oscillation of $u_n(\psi)$ around the reference orbit. The orbit is symmetric with respect to the position of the field error where it has a ``cusp''.}
\label{fig:perturbed-co-nc}
\end{center}
\end{figure}

Let us first analyze the situation using normalized coordinates $u(\psi)$.
Hill's differential equation then simplifies to
%------------------ 
\begin{equation}
\frac{\mathrm{d}^2 u_n}{\mathrm{d}\psi^2} + Q^2 u_n =0~,
\end{equation}
%------------------ 
which is solved by
%------------------ 
\begin{equation}
u_n(\psi) = u_{n,0}\cos(Q\psi + \psi_0)~.
\end{equation}
%------------------
Please note, that for the unperturbed closed orbit we have $u_{n,0} = 0$.
We know have to include the~impact of the erroneous dipole, where Hill's differential equation is no longer valid.
This dipole will cause an extra kick $\delta u^\prime$ at position $s_0$.
We choose $\psi = 0$ to be diametrically opposite to the~position of the~erroneous dipole magnet.
Then by symmetry $\psi_0 = 0$ and the perturbed closed orbit makes harmonic oscillations around the ideal path (the unperturbed closed orbit), cf.\ Fig.~\ref{fig:perturbed-co-nc}.
%---------------------------------------------------------------------------------
%--------------------------------------------------------------------------------
What remains is to compute the amplitude $u_{n,0}$ of the oscillations, which requires to transform the kick $\delta u^\prime$ to normalized coordinates.
Since at the position of the kick $\beta = \beta_0$ and $u$ does not change in the short dipole, we get with Eqs.~(\ref{eq:trans-to-psi}) and (\ref{eq:matrix-trans-to-psi})
%------------------ 
\begin{equation}
\delta \dot u_n = \delta \left( Q\sqrt{\beta_0} u^\prime + \frac{Q\alpha}{\sqrt{\beta_0}} u \right) =
Q\sqrt{\beta_0} \delta u^\prime~.
\end{equation}
%------------------
Since the kick changes the normalized angular displacement by $\delta \dot u_n$, by symmetry $\dot u_n(s_0) = \dot u_{n,0} = -\frac{1}{2}\delta \dot u_n$ must hold right before the kick.
Thus
%------------------ 
\begin{equation}
-\frac{1}{2} \delta \dot u_n = -\frac{1}{2}Q\sqrt{\beta_0}\cdot\delta u^\prime = \dot u_n(\pi) =-Q\cdot u_{n,0}\sin(\pi Q)~,
\end{equation}
%------------------
giving
%------------------ 
\begin{equation}
u_{n,0} = \frac{\sqrt{\beta_0}}{2\sin(\pi Q)} \delta u^\prime = \frac{\sqrt{\beta_0}}{2\sin(\pi Q)} \frac{\delta(B l)}{B \rho}~.
\end{equation}
%------------------
Finally, we have to express this result in unnormalized coordinates.
Since the kick is at $s_0$, we get for the~phase advance $\mu(s)-\mu(s_0) = Q\cdot\psi-Q\pi$ yielding a closed orbit distortion
%------------------ 
\begin{equation}
u_c(s) = \sqrt{\beta(s)} u_{n,0}\cos(Q\psi) = \left( \frac{\sqrt{\beta(s)\beta(s_0)}}{2\sin(\pi Q)}\frac{\delta(B l)}{B\rho}
\right) \cos \left(\mu(s)-\mu(s_0)+Q\pi \right)~,
\label{eq:co-dipole-kick}
\end{equation}
%------------------
where the index ``c'' denotes the displacement of the closed orbit.

The same result is obtained using the matrix formalism.
For doing so, we use the Twiss matrix representation of the transfer matrix $\mathbf{M}_\text{turn}$ for one turn.
The displacement of the closed orbit $u_c$ at the~position $s_0$ of the dipole error can be calculated from the displacements just before and after the~erroneous dipole since the closed orbit including the extra kick $\delta u^\prime$  must be periodic:
%------------------ 
\begin{equation}
\begin{pmatrix}u_{c,0} \\ u_{c,0}^\prime - \delta u^\prime \end{pmatrix} = \mathbf{M}_\text{turn}\cdot
\begin{pmatrix} u_{c,0} \\ u_{c,0}^\prime \end{pmatrix} =
\begin{pmatrix}\cos\mu+\alpha_0\sin\mu & \beta_0\sin\mu \\ -\gamma_0\sin\mu & \cos\mu-\alpha_0\sin\mu \end{pmatrix}
\cdot \begin{pmatrix}u_{c,0} \\ u_{c,0}^\prime \end{pmatrix}~,
\end{equation}
%------------------
which yields with $\mu=2\pi Q$
%------------------ 
\begin{eqnarray}
u_{c,0} &=& \frac{\beta_0\delta u^\prime}{2\sin(\pi Q)}\cos(\pi Q)~, \\
u_{c,0}^\prime &=& \frac{\delta u^\prime}{2\sin(\pi Q)}\left[\sin(\pi Q)-\alpha_0\cos(\pi Q) \right]~.
\end{eqnarray}
%------------------
The closed orbit displacement $u_c(s)$ is calculated from $\vec u_c(s) = \mathbf{M}(s_0,s)\cdot\vec u_{c,0}$.
Again using the Twiss parameter dependent representation of $\mathbf{M}(s_0,s)$ (cf.\ Eq.~(\ref{eq:transfer-matrix-twiss})), we finally get Eq.~(\ref{eq:co-dipole-kick}) for the displacement $u_c$ of the closed orbit at $s$, caused by a dipole kick at $s_0$.
The effect of a random distribution of dipole errors can be estimated from the r.m.s.\ average, weighted with the square root of the beta functions at the positions of the kicks.
When defining the ``kick density'' $\frac{\mathrm{d}\delta (B l)}{\mathrm{d}s}(s_0)$, we can write
%------------------ 
\begin{equation}
\boxed{
u_c(s) = \frac{\sqrt{\beta(s)}}{2\sin(\pi Q)} \oint\sqrt{\beta(s_0)}\cdot\frac{1}{B\rho}
\frac{\mathrm{d}\delta (B l)}{\mathrm{d}s}(s_0)\cdot
\cos \left(\mu(s)-\mu(s_0)+Q\pi \right)\cdot\mathrm{d}s_0
}~.
\label{eq:co-perturbation}
\end{equation}
%------------------
From Eq.~(\ref{eq:co-perturbation}) we extract a very important finding:
\begin{center}
\textbf{
For integer tunes the closed orbit displacements will grow indefinitely!
Thus, integer tunes have to be avoided in circular accelerators, $Q \ne n$!
}
\end{center}

This instability can be easily understood.
If the betatron tune is integer, a small perturbation always kicks a particle in the same direction and at the same phase of its betatron oscillation. The individual kicks add up coherently leading to an unlimited increase of the amplitude of the particle's transverse oscillation.

%_______________________________________________________________________________
%---------------------------------------------------------------------------------------------------------------
\subsection{Quadrupole errors, tune shift, and beta beating}
\label{sec:quad-errors}
%_______________________________________________________________________________
%---------------------------------------------------------------------------------------------------------------
Similar to the approach chosen to treat dipole errors, we will first concentrate on the situation dealing with a single erroneous small quadrupole magnet of length $\mathrm{d}s$ added to an error-free quadrupole magnet.
Its additional focusing or defocusing, often called a gradient error, will cause a small change in the~angular displacement but will leave the orbit displacement unchanged.
Thus, the quadrupole error can be described by an additional matrix $\mathbf{\delta Q}$
%------------------ 
\begin{equation}
\mathbf{\delta Q} = \begin{pmatrix}1 & 0 \\ -\frac{1}{f} & 1 \end{pmatrix} =
\begin{pmatrix} 1 & 0 \\ -\delta K\mathrm{d}s & 1 \end{pmatrix}~.
\end{equation}
%------------------
Please note that we have used the convention of Eq.~(\ref{eq:K-generalized}), thus $K_x = -k, K_y = k$ and $K > 0$ means focusing.

%_______________________________________________________________________________
\subsubsection{Tune shifts}
\label{sec:tune-shifts}
%_______________________________________________________________________________
We start right before the faulty element at $s_0$, apply $\mathbf{\delta Q}$ and continue for one turn with the one-turn matrix $\mathbf{M}_0$ in Twiss representation.
This reveals the modified one-turn matrix $\mathbf{\tilde M}_0$ including the gradient error
%------------------ 
\begin{equation}
\begin{split}
\mathbf{\tilde M}_0 &= \mathbf{\delta Q}\cdot\mathbf{M}_0 \\
&= \begin{pmatrix}\cos\mu_0+\alpha_0\sin\mu_0 & \beta_0\sin\mu_0 \\
-\delta K\mathrm{d}s\left(\cos\mu_0+\alpha_0\sin\mu_0\right) -\gamma_0\sin\mu_0 & 
-\delta K \mathrm{d}s\beta_0\sin\mu_0+\cos\mu_0-\alpha_0\sin\mu_0
\end{pmatrix}~.
\end{split}
\end{equation}
%------------------
It has to be mentioned that the optical functions will in principle be slightly changed by the changed focusing.
Since here we are only interested in the the change of the phase advance, this effect (which is of higher order) will be neglected.
Its investigation is postponed to the next section.

\noindent The phase advance over one turn can be be calculated from the trace of the matrix:
%------------------ 
\begin{equation}
\begin{split}
\frac{1}{2}\mathrm{Tr}\left(\mathbf{\tilde M}_0\right) &= \cos\mu = \cos(\mu_0+\Delta\mu) \approx \cos\mu_0-\Delta\mu\cdot\sin\mu_0 \\
&= \cos\mu_0 -\frac{1}{2}\delta K \mathrm{d}s\beta_0\sin\mu_0~,
\end{split}
\end{equation}
%------------------
yielding with $\Delta\mu = 2\pi\Delta Q$ the change of the betatron tune or the {\em tune shift}
%------------------ 
\begin{equation}
2\pi\Delta Q = \Delta\mu = \frac{1}{2}\beta_0\delta K\mathrm{d}s~.
\end{equation}
%------------------
Now putting the gradient error to an abitrary position $s$ and integrating over all existing gradient errors $\delta K(s)$, 
one obtains
%------------------ 
\begin{equation}
\boxed{
\Delta Q = \frac{1}{4\pi} \oint \beta(s)\delta K(s) \mathrm{d}s
}~,
\label{eq:delta-q-int}
\end{equation}
%------------------
which is the quadrupole error induced (linear) tune shift.
This formula comprises the effect of a distribution of quadrupole errors $\delta K(s)$ along the ring on the betatron tune.
It states that quadrupole errors $\delta K$ at locations with large beta functions will create larger tune shifts than quadrupole errors at locations with small beta functions.

\noindent \underline{Remark:} {\em Forcing the tune shift by deliberatly changing the strength of a single quadrupole by a known $\delta K$ and measuring the resulting tune shift is a common method to determine the beta function at the~location of the quadrupole.}

%_______________________________________________________________________________
\subsubsection{Beta beating}
\label{sec:beta-beating}
\begin{figure}[ht]
\begin{center}
\includegraphics[width=10cm]{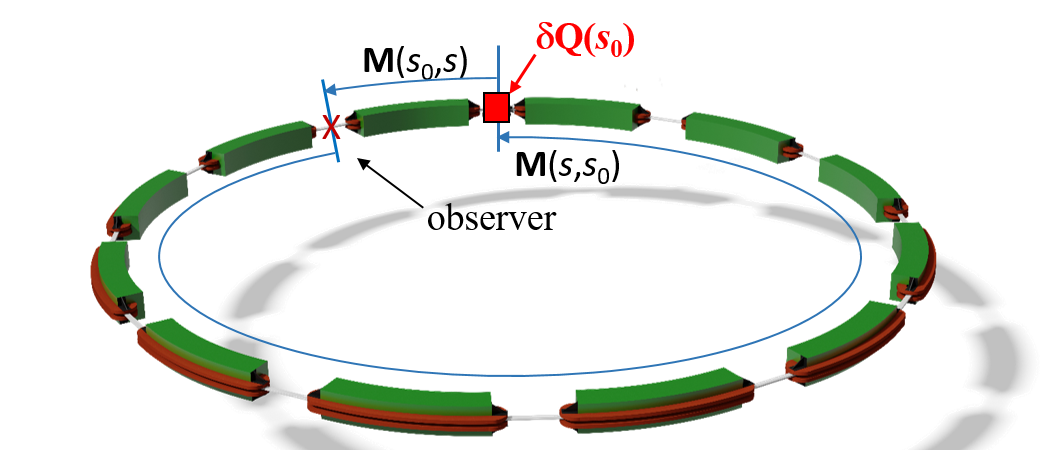}
\caption{A gradient error changes the one-turn matrix, which always depends on the starting position. When starting at $s$, the one-turn matrix $\mathbf{\tilde M}$ can be obtained by matrix multiplication of $\mathbf{M}(s,s_0)$, $\mathbf{\delta Q}(s_0)$ and $\mathbf{M}(s_0, s)$ as indicated.}
\label{fig:beta-beating-mult}
\end{center}
\end{figure}

%_______________________________________________________________________________
We now want to look at the impact of gradient errors on the beta function.
For a beam passing all quadrupoles in their centers a gradient error will not influence the closed orbit.
However a gradient error will change the betatron function of the lattice.
In order to calculate the modulation of the betatron oscillation amplitude, we have to determine the modified single turn matrix $\mathbf{\tilde M}_s$ starting at a given observer position $s$.
This is can be achieved by matrix multiplication, cf.\ Fig.~\ref{fig:beta-beating-mult}.
%---------------------------------------------------------------------------------
%--------------------------------------------------------------------------------
Starting at $s$, the matrix $\mathbf{M}(s, s_0)$ will transform the coordinates to the position $s_0$ where a small gradient error is located.
There, the error matrix $\mathbf{\delta Q}$ has to be applied.
The remaining part of the turn is added by finally applying the~matrix $\mathbf{M}(s_0, s)$.
Thus, we have
%------------------ 
\begin{equation}
\begin{split}
\mathbf{\tilde M}_s &= \begin{pmatrix}\square & \tilde r_{12} \\ \square & \square \end{pmatrix} = 
\mathbf{M}(s_0,s)\cdot\mathbf{\delta Q}(s_0)\cdot\mathbf{M}(s,s_0)\\
&= \begin{pmatrix}b_{11} & b_{12} \\ b_{21} & b_{22} \end{pmatrix}\cdot
\begin{pmatrix}1 & 0 \\ -\delta K \mathrm{d}s_0 & 1 \end{pmatrix}\cdot
\begin{pmatrix}a_{11} & a_{12} \\ a_{21} & a_{22} \end{pmatrix}~.
\end{split}
\end{equation}
%------------------
Since we are only interested in the change of the beta function which can be determined from the matrix element $\tilde r_{12} = \Delta(\beta\cos\mu)$, it is only necessary to calculate this element.
We get
%------------------ 
\begin{equation}
\tilde r_{12} = b_{11}a_{12}+b_{12}\left( -\delta K\mathrm{d}s_0\cdot a_{12}+a_{22}\right) =
r_{12}-\delta K\mathrm{d}s_0\cdot a_{12} b_{12}~,
\end{equation}
%------------------
where $r_{12}$ from the unperturbed matrix can be found by putting $\delta K\mathrm{d}s_0 =0$.
Using the Twiss representation of $\mathbf{M}$ (Eq.~(\ref{eq:transfer-matrix-twiss})) we obtain $a_{12}=\sqrt{\beta\beta_0}\sin(\mu-\mu_0)$, $b_{12}=\sqrt{\beta\beta_0}\sin(\mu_0-\mu)$, and the variation in the $r_{12}$ term due to the gradient error is
%------------------ 
\begin{equation}
\begin{split}
\Delta \left[ \beta(s)\sin(2\pi Q) \right] 
&= -\delta K \mathrm{d}s_0 \beta(s)\beta(s_0)\cdot\sin\left(\mu(s)-\mu(s_0)\right)\cdot\sin\left(\mu(s_0)-\mu(s)\right) \\
&= -\delta K \mathrm{d}s_0 \beta(s)\beta(s_0)\cdot\sin\left(\mu(s)-\mu(s_0)\right)\cdot\sin\left[2\pi Q - \left(\mu(s)-\mu(s_0)\right)\right]~.
\end{split}
\end{equation}
%------------------
Using $\sin x\cdot\sin y=\frac{1}{2}\left[\cos(x-y) - \cos(x+y)\right]$, the left-hand and right-hand sides can be expanded to give
%------------------ 
\begin{equation}
\begin{split}
\Delta\beta(s)\sin(2\pi Q_0) +
\underbrace{
 \beta(s) 2\pi\Delta Q
}_{\substack{~=}}
\cdot \cos&(2\pi Q) = \\
\overbrace{
\frac{1}{2}\delta K \mathrm{d}s_0\beta(s_0)\beta(s)
}
&\left\{\cos(2\pi Q) - \cos\left[ 2\left( \mu(s)-\mu(s_0)-\pi Q\right)\right]\right\}~.
\end{split}
\end{equation}
%------------------
We therewith get for the change $\Delta\beta(s)$ of the beta function at position s
%------------------ 
\begin{equation}
\Delta\beta(s) = \frac{\beta(s)}{2\sin(2\pi Q_0)}\delta K\mathrm{d}s_0\beta(s_0)
\cos\left[ 2\left( \mu(s)-\mu(s_0)-\pi Q\right)\right]~.
\end{equation}
%------------------
Now again putting the gradient error to an abitrary position $s_0$ and integrating over all quadrupole perturbations, one obtains
%------------------ 
\begin{equation}
\boxed{
\Delta\beta(s) = \frac{\beta(s)}{2\sin(2\pi Q_0)} \oint
\delta K(s_0)\beta(s_0)\cos\left[ 2\left( \mu(s)-\mu(s_0)-\pi Q\right)\right]\mathrm{d}s_0
}~.
\label{eq:beta-beating}
\end{equation}
%------------------
Quadrupole errors will lead to a position dependent change of the beta function which is called {\em beta beating}.
This beta beating has no effect on the transverse position of the beam, it rather describes the change of the beam size caused by focusing errors.
It is worth to note that the actual value $\Delta\beta(s)$ linearly depends on both the beta function $\beta(s_0)$ at the position of the gradient error and the beta function $\beta(s)$ at the observer's position.
Thus, gradient errors occuring at positions with large beta functions are more harmful than those at positions with small beta functions.
In circular accelerators, field errors of focusing quadrupoles typically will have a larger impact on the horizontal beam size whereas field errors of defocusing quadrupoles will more affect the vertical beam size.
In addition, we extract another very important finding from Eq.~(\ref{eq:beta-beating}):
\begin{center}
\textbf{
For half integer tunes the beta function and therewith the beam size will grow indefinitely!
Thus, half integer tunes have to be avoided in circular accelerators, $Q \ne \frac{n}{2}$!
}
\end{center}

Again, this instability can be easily understood.
If the betatron tune is half integer, a focusing error $\delta K$ will always kick a transversly displaced particle inwards or outwards depending on the sign of $\delta K$.
The individual kicks add up coherently leading to an unlimited increase of the amplitude of the~betatron oscillation.
However, a particle moving on the design orbit will not experience any kicks since it passes the quadrupoles at their field-free centers.
Thus, only the beam size is affected, the close orbit will remain unchanged.

%_______________________________________________________________________________
%---------------------------------------------------------------------------------------------------------------
\subsection{Optical resonances}
\label{sec:optical-resonances}
\begin{figure}[ht]
\begin{center}
\includegraphics[width=16cm]{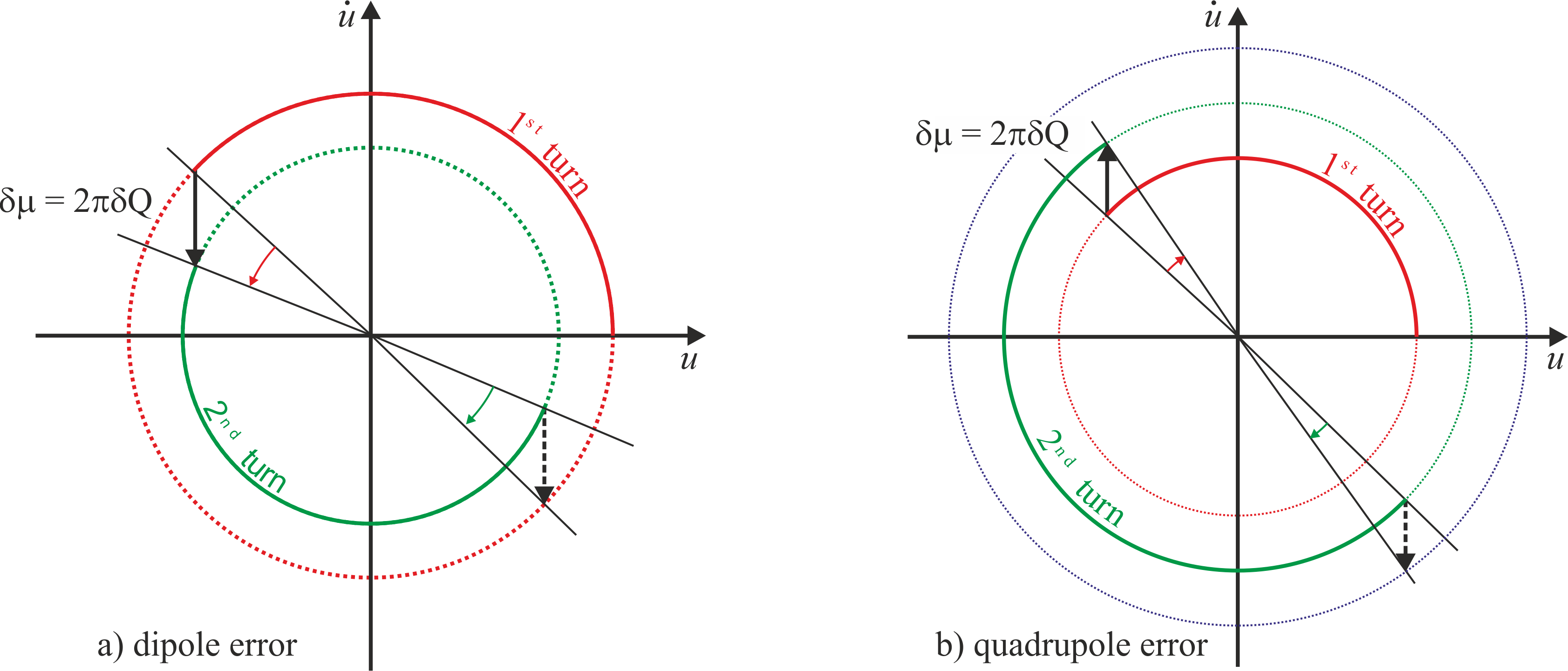}
\caption{Normalized trace space of a single particle at the position of an erroneous magnetic element. The point representing the particle rotates along along a circle and performs $2\pi Q$ rotations per revolution. Left: Impact of a~small dipole error on a particle with half integer unperturbed tune. Right: Impact of a~small quadrupole error on a~particle with an unperturbed tune close to a half integer which has already been locked into the resonance.}
\label{fig:optical-resonances}
\end{center}
\end{figure}

%_______________________________________________________________________________
%---------------------------------------------------------------------------------------------------------------
We obtained the following important impact of field errors in circular accelerators:
\begin{itemize}
\item Dipole errors will cause closed orbit distortions. The displacement of the closed orbit will grow indefinitely when the tune approaches an integer value.
\item Gradient errors will produce an average tune shift $\Delta Q$. They will not affect the closed orbit but change the beta function. The beam size will grow indefinitely when the tune approaches half integer values.
\end{itemize}
These phenomena are called {\em optical resonances} and can be studied best when regarding the normalized trace space where the particles' ellipses transform to circles, cf.\ Fig.~\ref{fig:optical-resonances}.
%---------------------------------------------------------------------------------
%--------------------------------------------------------------------------------
Let us for reasons of simplicity assume a circular accalerator with a single faulty element only: a small dipole or quadrupole error.
In addition, we will concentrate on a single particle represented by a point in the normalized trace space.
The evolution of the particle's coordinates can be followed turn by turn by looking at the movement of its point in the trace space.
Since we are interested in the impact of the error, we select the trace space at the position of the faulty element.
Without an erroneous element, the particle's point stays on a circle and its actual position is only dependent on its betatron phase described by an initial phase $\varphi_0$ and the~phase advance $\mu(s)$.
It will perform $2\pi Q$ rotations per revolution.

In case of a dipole error, every turn the particle will get a constant kick $\delta \dot u$ changing the radius of the circle.
This kick may cause an additional phase advance $\delta\mu$ depending on the particle's actual betatron phase when passing the faulty element.
This additional phase advance changes the tune by $\delta Q = \frac{\delta\mu}{2\pi}$.
It is important to note that $\delta Q$ will change its sign with the sign of $u$.
Thus, on average there will be no tune shift since both signs appear equally likely.
However, there will be a modulation of the~tune with an~amplitude $\mathrm{d} Q = \delta Q_\text{max}$.
So, if the particle's unperturbed tune $Q$ deviates less than $\mathrm{d} Q$ from an integer value, the tune modulation will cause that $Q$ will be occasionally changed to an integer value.
Since this will be a stable situation being reproduced every turn, the particle's tune will remain at this integer value.
From then on all kicks will add coherently and the particle's coordinates will increase indefinitely, a process which is called ``auto locking'' into the resonance.
This type of resonance is called an integer resonance since it occurs at integer tunes.
 
 In case of a quadrupole error, the sign of the kick will depend on the sign of $u$ and its strength will scale with $u$.
 Thus, the particle will always be kicked in the same ``direction'' (outwards or inwards, depending on the sign of $\delta K$) leading to an average tune shift of $\Delta Q = \frac{1}{4\pi}\beta\delta(K l)$.
 Again, the kicks will cause a change in the phase advance depending on the actual betatron phase when passing the faulty element.
 This will change the tune turn by turn accordingly and lead to a tune modulation with amplitude $\mathrm{d} Q = \delta Q_\text{max}$.
If the particle's unperturbed tune deviates less than $\mathrm{d} Q$ from a half-integer value, auto locking into the resonance will take place and all kicks will add coherently.
This type of resonance is called a half-integer resonance since it occurs at half integer tunes.
 
We therewith obtain regions of instability, called {\em stop bands}, around integer and half-integer tunes.
The width of these stop bands are given by the tune modulation amplitude which depends on the strength of the field error.
Each particle whose unperturbed $Q$ lies in the stop band will lock into resonance and will finally be lost.
The observation that dipole errors lead to integer resonances and quadrupole errors to half-integer resonances indicates that sextupole fields excite resonances at third-integer $Q$ values and in general a magnetic multipole of order $n$ will excite a $\frac{1}{n}$ resonance.
This is in fact the case but out of the scope of the linear treatment applied in this course.
Anyhow, we conclude that field errors cause stop bands around fractional tunes which therefore have to be avoided.
The corresponding resonance condition reads
%------------------ 
\begin{equation}
n Q \neq N
\label{eq:1D-resonance}~,
\end{equation}
%------------------
where $n$ and $N$ are integers.
Moreover, one has in general a coupling between horizontal and vertical betatron oscillations due to higher order multipoles or magnet misalignments.
This leads to the more general resonance condition
%------------------ 
\begin{equation}
\boxed{
m Q_x + n Q_y \neq N
}~,
\label{eq:2D-resonance}
\end{equation}
%------------------
where $m$, $n$, and $N$ are integers.
$m+n$ is called the {\em order} of the resonance.
Since the~widths of the stop bands decrease with increasing $m, n$, only the ``lower'' orders are important.
Which orders have to be considered depend on the actual requirements (e.g.\ only the first $\approx$5 orders are important for a~fast ramping synchrotron whereas higher orders up to $>$10 may be important in case of a long store in a~circular collider or a synchrotron radiation source).
The situation is often visualized in the {\em tune diagram} where $Q_y$ is plotted versus $Q_x$ and the leading order resonance lines are indicated.
%---------------------------------------------------------------------------------
\begin{figure}[ht]
\begin{center}
\includegraphics[width=7cm]{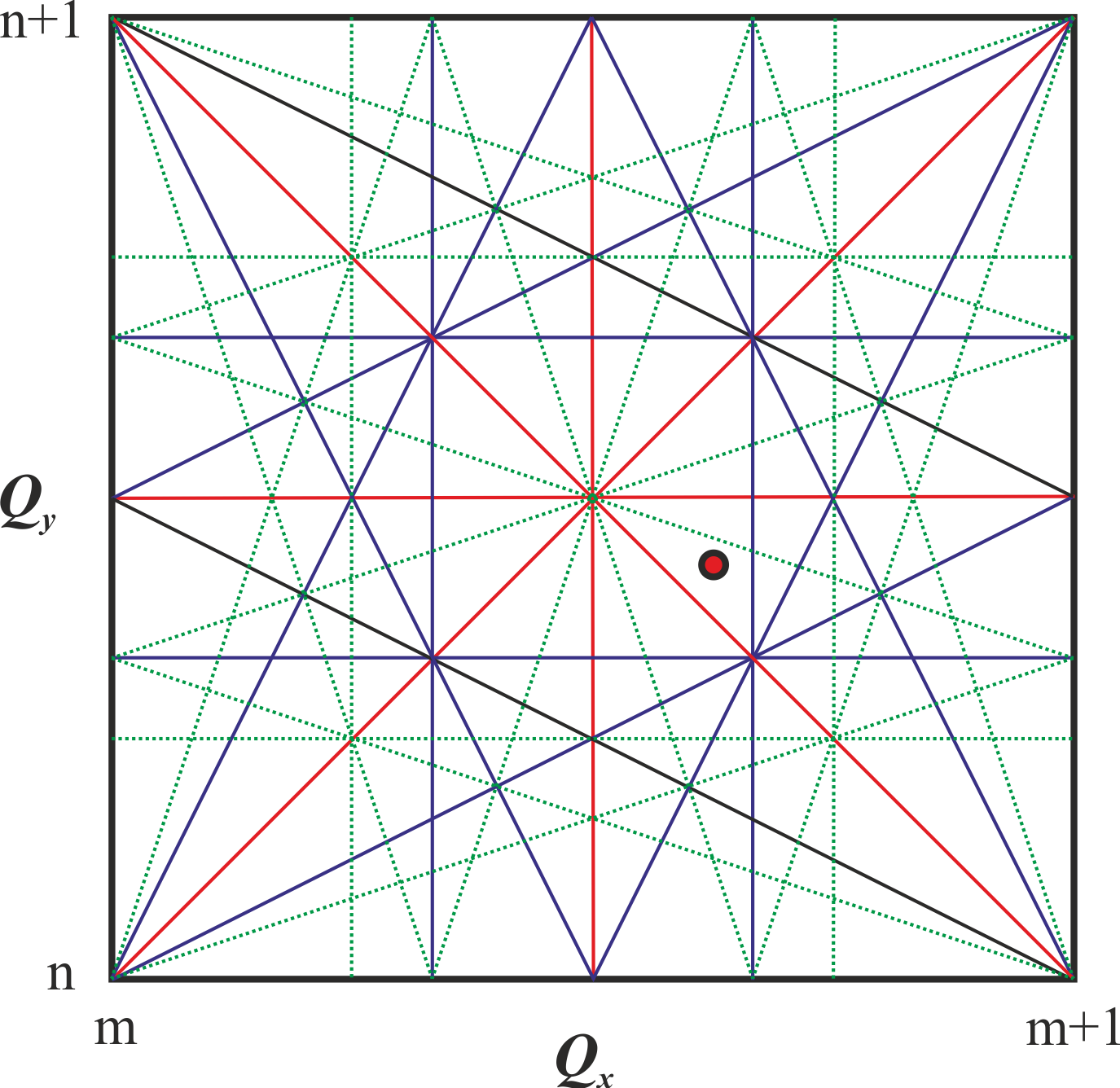}
\caption{Tune diagram displaying the intervals $m \le Q_x \le m+1$, $n \le Q_y \le n+1$. All resonance lines up to 5th order are shown in different colors. The working point chosen is indicated by a red filled circle.}
\label{fig:tune-diagram}
\end{center}
\end{figure}
%--------------------------------------------------------------------------------
Such a tune diagram in which all reonance lines up to 5th order in the intervals $m \le Q_x \le m+1$, $n \le Q_y \le n+1$ are drawn is shown in Fig.~\ref{fig:tune-diagram}.
The horizontal and vertical tunes and therewith the working point $(Q_x, Q_y)$ have to be chosen in a reasonable distance from the resonance lines.

%_______________________________________________________________________________
%---------------------------------------------------------------------------------------------------------------
\subsection{Beam dynamics with acceleration}
\label{sec:dynamics-acceleration}
\begin{figure}[ht]
\begin{center}
\includegraphics[width=12cm]{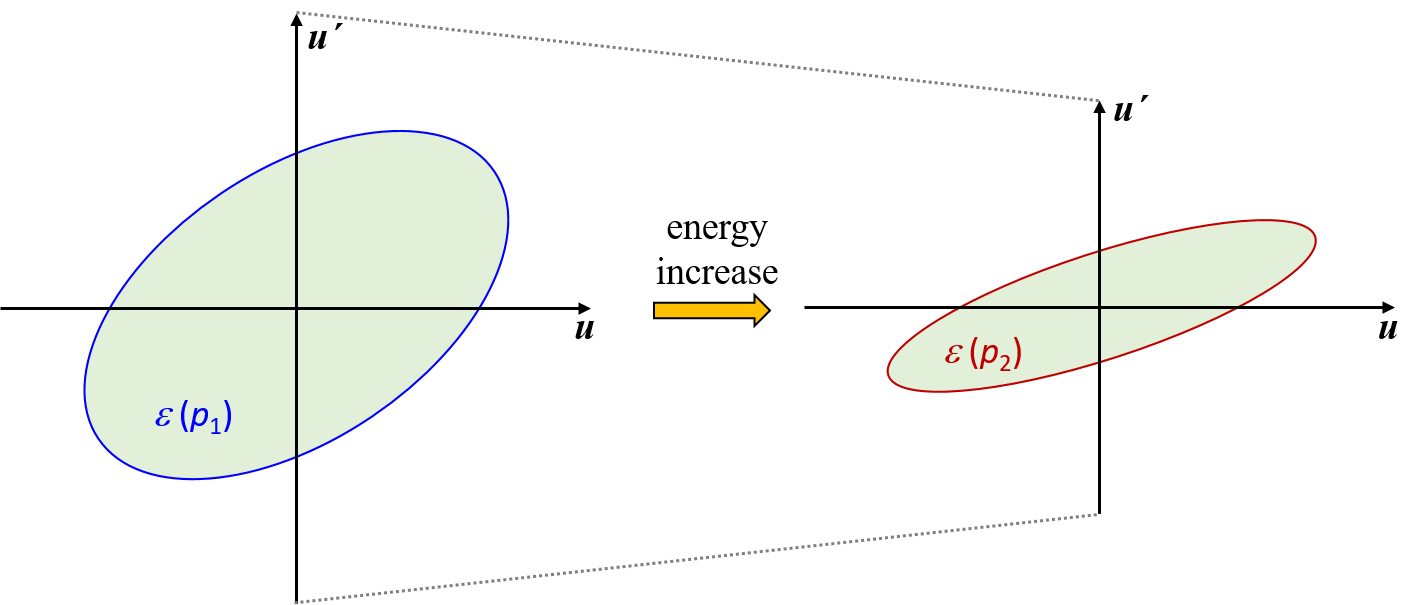}
\caption{Trace spaces for different beam energies. Since the angular displacement $u^\prime$ scales with the beam energy, an energy increase causes a shrinkage of the $u^\prime$ dimension of the trace space and thus a reduction of the beam's emittance.}
\label{fig:adiabatic-damping}
\end{center}
\end{figure}

%_______________________________________________________________________________
%---------------------------------------------------------------------------------------------------------------
Here, beam acceleration means the increase of the beam energy by accelerating structures like rf cavities.
Typically, this is a rather slow process compared to the time scale of the transverse beam dynamics determind by the oscillation frequency of the betatron oscillations.
We thus can assume that the transverse dynamics will adiabatically follow since the multipole strengths will be kept constant by continuously adjusting the magnetic fields to the actual beam energy.
However, the changing Lorentz factors $\gamma_r$, $\beta_r$ will lead to a changing scaling of the transverse angular displacements.
Since the transverse momentum which is related to the angular displacement by
%------------------ 
\begin{equation}
p_u = m\cdot \frac{\mathrm{d}u}{\mathrm{d}t} = m\cdot \frac{\mathrm{d}s}{\mathrm{d}t} \cdot u^\prime \approx p_0\cdot u^\prime = \beta_r \gamma_r m_0 c\cdot u^\prime~,
\end{equation}
%------------------
is not changed by the beam acceleration only increasing the longitudinal momentum, the angular displacements change with changing beam energy according to
%------------------ 
\begin{equation}
\beta_r \gamma_r \cdot u^\prime = \text{const.}
\end{equation}
%------------------
Thus beam acceleration causes  a decrease of the angular displacement $u^\prime$ and therewith a compression of the $u^\prime$ axis of the trace space.
This consequently leads to a decrease of the beam emittance which is called {\em adiabatic damping} and illustrated in Fig.~\ref{fig:adiabatic-damping}.
%---------------------------------------------------------------------------------
%--------------------------------------------------------------------------------

For this more general case, a more appropriate invariant is defined by
%------------------ 
\begin{equation}
\boxed{
\epsilon_n = \beta_r \gamma_r\cdot\epsilon
}~,
\end{equation}
%------------------
and called the {\em normalized emittance} $\epsilon_n$ which is conserved independent from the beam energy.
$\epsilon$ is called the {\em geometric emittance}.

%===========================================================================
%===========================================================================
\section{Dynamics with off momentum particles}
\label{sec:off-momentum-particles}
%===========================================================================
%===========================================================================
Up to now, we assumend that all particles in the beam have the same nominal momentum $p_0$.
For real beams, this is of course not the case.
In the following, we want to discuss the influence of momentum deviations $\Delta p = p - p_0$ on the linear beam dynamics.
Coming back to the equations of motion (Eq.~(\ref{eq:linearized-equations-of-motion})) we recognize that only the equation for the horizontal plane containes a momentum-dependent term.
Thus, in flat accelerators the dynamics of off momentum particles is only affected in the horizontal plane.
We will therefore restrict our treatment to the horizontal trace space including the momentum dependence.

%_______________________________________________________________________________
%---------------------------------------------------------------------------------------------------------------
\subsection{Momentum dispersion}
\label{sec:momentum-dispersion}
%_______________________________________________________________________________
%---------------------------------------------------------------------------------------------------------------
The linearized equation of motion for the horizontal coordinate reads
%------------------ 
\begin{equation}
x^{\prime\prime}(s) + \left(\frac{1}{\rho^2(s)} - k(s) \right) \cdot x(s) = \frac{1}{\rho(s)}\frac{\Delta p}{p_0} =
\frac{\delta}{\rho(s)}~,
\end{equation}
%------------------
where we have set $\delta = \frac{\Delta p}{p_0}$.
It is called a homogeneous differental equation when $\delta = 0$ and an inhomogeneous differential equation when $\delta \neq 0$.
The solution consists of the general solution of the~homogeneous equation, the betatron oscillations $x_\beta$, plus a special solution $x_D$ of the inhomogeneous equation.
Since for quadrupole magnets $\rho \to \infty$ and the inhomogeneous term vanishes, we have only to consider dipole magnets where $\rho(s) = \rho = \text{const.}$ and $k(s) =0$.
Thus, a special solution of the~inhomogeneous equation is $x_D = \rho\cdot\delta$, leading to
%------------------ 
\begin{equation}
x(s) = x_\beta(s) + x_D(s) = a\cdot\cos\left(\frac{s}{\rho}\right) + b\cdot\sin\left(\frac{s}{\rho}\right) + \rho\delta~.
\end{equation}
%------------------
The integration constants $a$, $b$ are again derived from the initial values at $s=0$.
Since now the inhomogeneous solution $x_D$ has to be included, we get
%------------------ 
\begin{equation}
x(s=0) = x_0 = a+\rho\delta~,~~~~~~~~~~x^\prime(s=0) = x_0^\prime = \frac{b}{\rho}~,
\end{equation}
%------------------
and by defining the bending angle $\varphi = \frac{L}{\rho}$ of the dipole magnet with an arc length $L$, we obtain
%------------------ 
\begin{eqnarray}
\label{eq:disp-offset}
x(L) &=& x_0\cdot\cos\varphi + \rho x_0^\prime\cdot\sin\varphi + \rho(1-\cos\varphi)\cdot\delta~, \\
\label{eq:disp-ang-offset}
x^\prime(L) &=& -\frac{x_0}{\rho}\sin\varphi + x_0^\prime\cdot\cos\varphi + \sin\varphi \cdot\delta~.
\end{eqnarray}
%------------------
By extenting the matrix formalism in the bending plane to three variables, we can easily include the~impact of the momentum deviation.
The horizontal position vector is written as
%------------------ 
\begin{equation}
\vec x(s) = \begin{pmatrix}x(s) \\ x^\prime(s) \\ \delta \end{pmatrix}~,
\end{equation}
%------------------
and is transformed by a 3x3 transfer matrix $\mathbf{M}$:
%------------------
\begin{equation}
\vec x(s) = \mathbf{M}\cdot\vec x_0~.
\end{equation}
%------------------
For a sector dipole magnet, $\mathbf{M}_\text{dip}$ can be derived from Eqs.~(\ref{eq:disp-offset}) and (\ref{eq:disp-ang-offset}) giving
%------------------ 
\begin{equation}
\mathbf{M}_\text{dip} = \begin{pmatrix}\cos\varphi & \rho\sin\varphi & \rho(1-\cos\varphi) \\
-\frac{1}{\rho}\sin\varphi & \cos\varphi & \sin\varphi \\
0 & 0 & 1 \end{pmatrix}~.
\end{equation}
%------------------
First neglecting that the quadrupole strength is dependent on the acual momentum of the particle, the~quadrupole matrices remain ``unchanged'' but have to be extended to 3x3 matrices yielding
%------------------ 
\begin{eqnarray}
\mathbf{M}_\text{QF} = \begin{pmatrix}\cos\Omega & \frac{1}{\sqrt{\vert k \vert}}\sin\Omega & 0 \\
-\sqrt{\vert k \vert}\sin\Omega & \cos\Omega & 0 \\ 0 & 0 & 1 \end{pmatrix}~, \\
\mathbf{M}_\text{QD} = \begin{pmatrix}\cosh\Omega & \frac{1}{\sqrt{\vert k \vert}}\sinh\Omega & 0 \\
\sqrt{\vert k \vert}\sinh\Omega & \cosh\Omega & 0 \\ 0 & 0 & 1 \end{pmatrix}~,
\end{eqnarray}
%------------------
where we have used the abbreviations of Section \ref{sec:quadrupole-magnet}.
The 3x3 transfer matrix of a free drift in space of length $L$ reads 
%------------------ 
\begin{equation}
\mathbf{M}_\text{drift} = \begin{pmatrix}1 & L & 0 \\ 0 & 1 & 0 \\ 0 & 0 & 1 \end{pmatrix}~.
\end{equation}
%------------------
For more than one element, we get the total transfer matrix by matrix multiplication as explained in Section \ref{sec:matrix-formalism}.

This formalism leads to the following important result:
Whereas a drift space and a quadrupole magnet will not directly cause an impact on the trajectory of a particle with momentum deviation, a~dipole magnet will create a {\em dispersion} $D$ and its derivative $D^\prime$ by the non-vanishing matrix elements in the first two lines of the third column:
%------------------ 
\begin{eqnarray}
D &=& r_{13} = \rho(1-cos\varphi)~, \\
D^\prime &=& r_{23} = \sin\varphi~.
\end{eqnarray}
%------------------
%---------------------------------------------------------------------------------
\begin{figure}[!ht]
\begin{center}
\includegraphics[width=10cm]{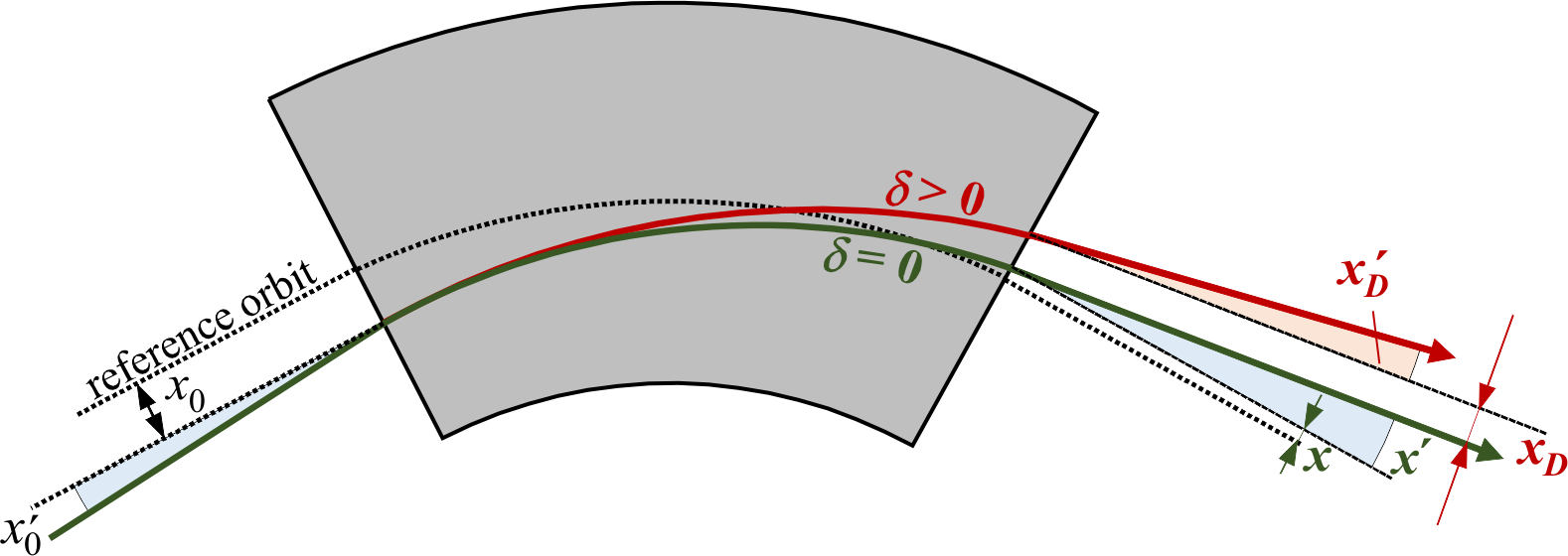}
\caption{A dipole magnet creates a dispersion $D$ and its derivative $D^\prime$ which causes an additional orbit displacement $x_D = D\cdot\delta$ and angular displacement $x_D^\prime = D^\prime\cdot\delta$ at the end of the magnet.}
\label{fig:dipole-dispersion}
\end{center}
\end{figure}
%--------------------------------------------------------------------------------
The dispersion indicates, how much the reference trajectory (the trajectory showing no betatron oscillations which is represented by a point in the center of the trace space ellipse) of a particle with momentum deviation is displaced from the design orbit defined for $\delta = 0$.
It represents this displacement for a relative momentum deviation $\delta = 1$ and contributes via the {\em dispersion orbit} $x_D$ to the particle's trajectory
%------------------ 
\begin{equation}
x(s) = x_\beta(s) + x_D(s)~, ~~~~~~~~~~~~~x_D(s) = D(s)\cdot\delta~.
\label{eq:dispersion-orbit}
\end{equation}
%------------------
Often, $D(s)$ is called the {\em dispersion function}, a solution of the equation of motion for $\delta = 1$.

It is of utmost importance to notice that, once a dispersion has been created by a dipole magnet, it will be altered by drifts and quadrupole magnets as well!
We therefore better distinguish between a~dispersion $D_\text{dip}$, which is the matrix element $r_{13}$ of the transfer matrix of a dipole magnet and thus created by the dipole magnet, and the dispersion function $D(s)$ which depends on all elements in a~magnetic lattice before the position $s$.
Building a vector
%------------------ 
\begin{equation}
\begin{pmatrix}D(s) \\ D^\prime(s) \\ 1\end{pmatrix}~,
\end{equation}
%------------------
we can compute the evolution of an already existing dispersion $D_0$ and $D_0^\prime$ by applying the 3x3 transfer matrix $\mathbf{M}$ to this vector
%------------------ 
\begin{equation}
\begin{pmatrix}D(s) \\ D^\prime(s) \\ 1\end{pmatrix} = 
\mathbf{M} \cdot \begin{pmatrix}D_0 \\ D^\prime_0 \\ 1\end{pmatrix}~,
\end{equation}
%------------------
which would explicitly read for a sector dipole magnet
%------------------ 
\begin{equation}
\begin{pmatrix}D(s) \\ D^\prime(s) \\ 1\end{pmatrix} = 
\begin{pmatrix}\cos\varphi & \rho\sin\varphi & D_\text{dip} \\ -\frac{1}{\rho}\sin\varphi & \cos\varphi &
D^\prime_\text{dip} \\ 0 & 0 & 1\end{pmatrix} \cdot \begin{pmatrix}D_0 \\ D^\prime_0 \\ 1\end{pmatrix}~.
\end{equation}
%------------------
Thus, a dipole magnet modifies an existing dispersion $D_0$ and in addition adds its own ``contribution'' $D_\text{dip}$ whereas quadrupole magnets and drift sections only modify an existing dispersion.
%---------------------------------------------------------------------------------
\begin{figure}[ht]
\begin{center}
\includegraphics[width=16cm]{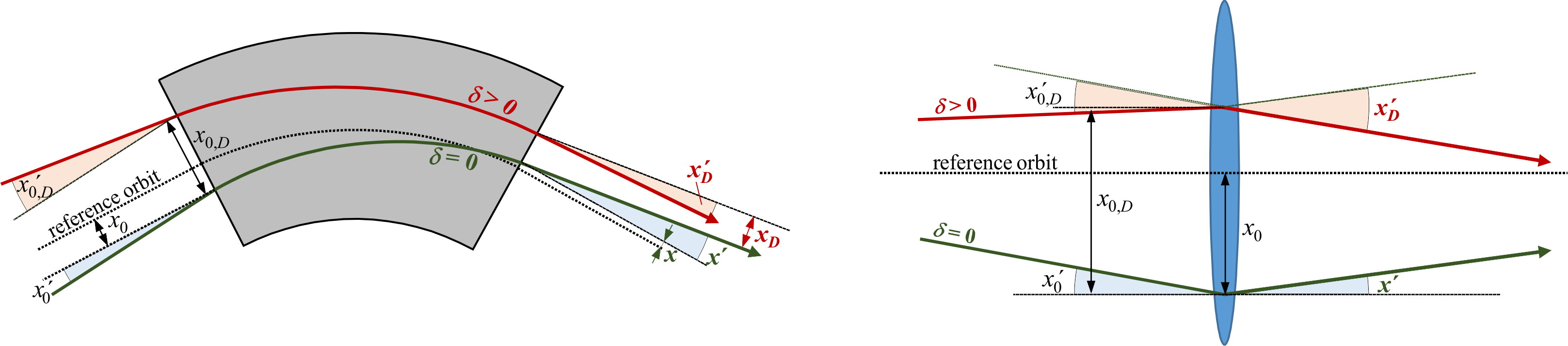}
\caption{Dipole and quadrupole magnets modify an existing dispersion causing a change in $x_D$ amd $x_D^\prime$. In a thin lens ony $D^\prime$ and thus $x_D^\prime$ is changed.}
\label{fig:dispersion-dipole-quad}
\end{center}
\end{figure}
%--------------------------------------------------------------------------------
If a beam passes through a beam transfer line, not only its optical parameters $\beta$, $\alpha$ and $\gamma$ but also its dispersive ``properties'' $D$ and $D^\prime$ are propagated starting from some initial values $\beta_0$, $\alpha_0$, $\gamma_0$, $D_0$, $D_0^\prime$ which have to be specified at the beginning of the transfer line.
We will discover in the next section that in a circulator accelerator the dispersion function $D(s)$ is again a unique, periodic function of the position along the ring which is only determined by the magnetic lattice.

The influence of a non-vanishing dispersion on the effective beam size $\tilde\sigma_x$, divergence $\tilde\sigma_{x^\prime}$ and trace space area can be calculated using the statistical definitions and replacing $\delta$ by the r.m.s.\ momentum spread $\sigma_\delta = \sqrt{\langle \delta_i^2\rangle}$:
%------------------ 
\begin{eqnarray}
\tilde\sigma_x^2 = \frac{1}{N}\sum_{i=1}^N\left( x_i + D\delta_i\right)^2 =
\frac{1}{N}\sum_{i=1}^N x_i^2 + \frac{D^2}{N}\sum_{i=1}^N \delta_i^2 + \frac{2D}{N}\sum_{i=1}^N 
\left( x_i\delta_i\right) = \sigma_x^2 + \left( D\sigma_\delta\right)^2 ~,~~~~ \\
\tilde\sigma_{x^\prime}^2 = \frac{1}{N}\sum_{i=1}^N\left( x_i^\prime + D^\prime\delta_i\right)^2 =
\frac{1}{N}\sum_{i=1}^N {x_i^\prime}^2 + \frac{{D^\prime}^2}{N}\sum_{i=1}^N \delta_i^2 + \frac{2D^\prime}{N}\sum_{i=1}^N 
\left( x_i^\prime\delta_i\right) = \sigma_{x^\prime}^2 + \left( D^\prime\sigma_\delta\right)^2~,
\end{eqnarray}
%------------------
where we have assumend that the momentum deviations will not be correlated with the transverse displacements.
Thus, we get an enlarged beam size and divergence
%------------------ 
\begin{equation}
\boxed{
\begin{split}
\tilde\sigma_x = \sqrt{\epsilon_x\beta_x + \left( D\sigma_\delta\right)^2} \\
\tilde\sigma_{x^\prime} = \sqrt{\epsilon_x\gamma_x + \left( D^\prime\sigma_\delta\right)^2}
\end{split}
}~.
\end{equation}
%------------------
The area in the horizontal trace space increases as well, since the dispersion orbit $x_D$ is added to $x_\beta$.
This may be described by an ``effective'' emittance $\tilde\epsilon_x$ which has to be derived from a lengthy calculation first defining a generalized 6-dimensional intensity distribution and then integrating over unwanted coordinates (cf.\ e.g.\ \cite{Hinterberger}).
Finally, one obtains an intensity distribution $I(x,x^\prime,\delta)$ in which the following effective emittance appears:
%------------------ 
\begin{equation}
\tilde\epsilon_x = \sqrt{\epsilon_x^2 + \epsilon_x\sigma_\delta^2\left(\gamma_x D^2 + 2\alpha_x D D^\prime + \beta_x {D^\prime}^2\right)}~.
\end{equation}
%------------------

%_______________________________________________________________________________
%---------------------------------------------------------------------------------------------------------------
\subsection{Dispersion evolution by integrals}
\label{sec:dispersion-by-integrals}
%_______________________________________________________________________________
%---------------------------------------------------------------------------------------------------------------
The dispersion function can as well be calculated using the principal solutions defined in Section \ref{sec:matrix-properties}.
Based on the definition of a so-called Green's function $G$ built from the principal solutions $C(s)$ and $S(s)$ and using the properties of the Wronskian we can find a special solution of the equation of motion where we have set $\delta=1$
%------------------ 
\begin{equation}
D^{\prime\prime}(s) + K(s)\cdot D(s) = \frac{1}{\rho(s)}~.
\label{eq:D-DGL}
\end{equation}
%------------------
Defining
%------------------ 
\begin{equation}
D(s) = \int_0^s \frac{1}{\rho(\tilde s)}\cdot
\underbrace{
\left[ S(s)\cdot C(\tilde s) - C(s)\cdot S(\tilde s)\right]
}_{\substack{= G(s,\tilde s)}}
\cdot\mathrm{d}\tilde s~,
\label{eq:D-greens}
\end{equation}
%------------------
we obtain for the first and second derivative
%------------------ 
\begin{eqnarray}
D^\prime(s) &=& S^\prime(s)\int_0^s\frac{C(\tilde s)}{\rho(\tilde s)}\mathrm{d}\tilde s - C^\prime(s)\int_0^s\frac{S(\tilde s)}{\rho(\tilde s)}\mathrm{d}\tilde s~, \\
D^{\prime\prime}(s) &=& S^{\prime\prime}(s)\int_0^s\frac{C(\tilde s)}{\rho(\tilde s)}\mathrm{d}\tilde s - C^{\prime\prime}(s)\int_0^s\frac{S(\tilde s)}{\rho(\tilde s)}\mathrm{d}\tilde s + \frac{1}{\rho(s)}
\underbrace{
\left(C(s)S^\prime(s) - S(s) C^\prime(s)\right)
}_{\substack{= \det (\mathbf{M}) = 1}}~.
\end{eqnarray}
%------------------
Since both $C(s)$ and $S(s)$ are solutions of the homogeneous equation, we have $C^{\prime\prime}(s) = -K(s) C(s)$ and $S^{\prime\prime}(s) = -K(s) S(s)$ and proofed, that $D(s)$ defined in Eq.~(\ref{eq:D-greens}) solves the inhomogeneous differential equation (\ref{eq:D-DGL}).
Thus, the dispersion function can also be derived from integration of the principal solutions
%------------------ 
\begin{equation}
\boxed{
D(s) = S(s)\int_0^s \frac{C(\tilde s)}{\rho(\tilde s)}\mathrm{d}\tilde s - C(s)\int_0^s \frac{S(\tilde s)}{\rho(\tilde s)}\mathrm{d}\tilde s
}~.
\label{eq:D-int}
\end{equation}
%------------------

%_______________________________________________________________________________
%---------------------------------------------------------------------------------------------------------------
\subsection{Periodic dispersion functions}
\label{sec:periodic-dispersion-functions}
%_______________________________________________________________________________
%---------------------------------------------------------------------------------------------------------------
In a periodic lattice with period length $L$, the dispersion function has---as well the optical functions---to fulfill periodic boundary conditions:
%------------------ 
\begin{eqnarray}
D(s_0+L) = D(s_0)~,~\\
D^\prime(s_0+L) = D^\prime(s_0)~.
\end{eqnarray}
%------------------
Thus the dispersion function can be obtained from applying the 3x3 transfer matrix $\mathbf{M}(s_0,s_0+L)$ for one period
%------------------ 
\begin{equation}
\begin{pmatrix} D_0 \\ D_0^\prime \\ 1 \end{pmatrix} =
\mathbf{M}(s_0,s_0+L)\cdot \begin{pmatrix} D_0 \\ D_0^\prime \\ 1 \end{pmatrix} =
\begin{pmatrix} r_{11} & r_{12} & r_{13} \\r_{21} & r_{22} & r_{23} \\ 0 & 0 & 1 \end{pmatrix} 
\cdot \begin{pmatrix} D_0 \\ D_0^\prime \\ 1 \end{pmatrix}~,
\end{equation}
%------------------
yielding
%------------------ 
\begin{eqnarray}
\label{eq:periodic-dispersion}
D_0 = \frac{r_{13}\left(1-r_{22}\right) + r_{12}r_{23}}{2-r_{11}-r_{22}} =
\frac{r_{13}\left(1-r_{22}\right) + r_{12}r_{23}}{2\left(1-\cos\mu\right)}~, \\
D^\prime_0 = \frac{r_{13}\left(1-r_{11}\right) + r_{21}r_{13}}{2-r_{11}-r_{22}} =
\frac{r_{13}\left(1-r_{11}\right) + r_{21}r_{13}}{2\left(1-\cos\mu\right)}~,
\end{eqnarray}
%------------------
where for the transformation of the denominator we have used the Twiss parameter representation of the~transfer matrix (cf.\ Eq.~(\ref{eq:Twiss-matrix})).
We found that a stable periodic dispersion exists, as long as the phase advance $\mu$ is not an integer multiple of $2\pi$ and therewith the tune $Q$ is not integer.
Applying Eq.~(\ref{eq:periodic-dispersion}) to our toy model ring in Section \ref{sec:periodic-beta}, we obtain the dispersion function $D(s)$ which is plotted in blue in Fig.~\ref{fig:toy-ring-complete}.
In case of a simple FODO lattice, it shows the same general behaviour as the corresponding beta function $\beta_x$.

%_______________________________________________________________________________
%---------------------------------------------------------------------------------------------------------------
\subsection{Chromaticity}
\label{sec:chromaticity}
\begin{figure}[ht]
\begin{center}
\includegraphics[width=7cm]{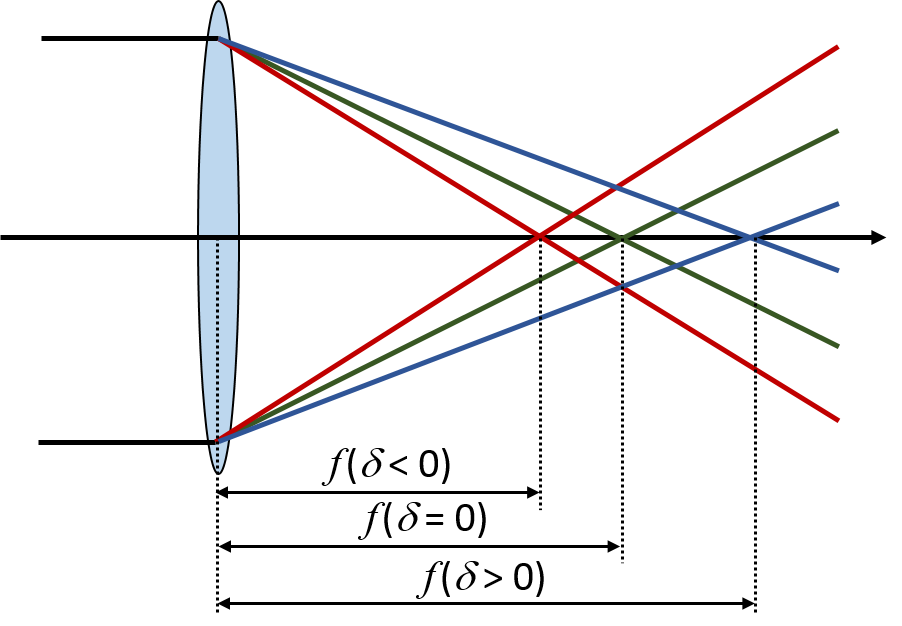}
\caption{Chromatic abberation of a focusing quadrupole treated as a thin lens. Particles with lower momentum are deflected more than particles with higher momentum. The actual quadrupole's focal length depends on the relative momentum deviation $\delta$.}
\label{fig:chromatic-abberation}
\end{center}
\end{figure}

%_______________________________________________________________________________
%---------------------------------------------------------------------------------------------------------------
Besides changing the bending in the dipole fields (creating dispersion), a momentum deviation also changes the focusing strength $k$ of the quadrupole magnets defined by Eq.~(\ref{eq:quadrupole-strength}).
So far, we neglected that off momentum particles will experience a different focusing from the quadrupole magnets.
Now, we will investigate the impact of the changed focusing on the betatron tune resulting in a tune shift $\Delta Q$.
This tune shift will explicitly depend on $\delta$, giving rise to the definition of a new quantity which is called the~{\em chromaticity} $\xi$.
It is defined by
%------------------ 
\begin{equation}
\Delta Q_u = \xi_u\cdot\delta~,
\end{equation}
%------------------
indicating the tune shift normalized to the relative momentum deviation.
It can be evaluated from the~very useful Eq.~(\ref{eq:delta-q-int}) derived in Section \ref{sec:quad-errors}.
We distinguish between {\em natural chromaticity} created by the~chromatic abberation of the quadrupole magnets (cf.\ Fig.~\ref{fig:chromatic-abberation}) and non-linear contributions produced by higher order magnetic multipoles such as e.g.\ sextupole magnets.
%---------------------------------------------------------------------------------
%--------------------------------------------------------------------------------

%---------------------------------------------------------------------------------------------------------------
\subsubsection{Natural chromaticity}
\label{sec:natural-chromaticity}
%_______________________________________________________________________________
The quadrupole strength scales with the particles' momenta as
%------------------ 
\begin{equation}
k = k_0+\Delta k = \frac{q}{p}\cdot g = \frac{q}{p_0+\Delta p}\cdot g \approx \frac{q}{p_0}\cdot g - \frac{q}{p_0^2} \Delta p \cdot g~,
\end{equation}
%------------------
where $k_0$ represents the quadrupole strength experienced by a particle with reference momentum $p_0$.
This results in a quadrupole ``error'' for off momentum particles given by
%------------------ 
\begin{equation}
\Delta k = - k_0\cdot\frac{\Delta p}{p_0} = - k_0\cdot\delta =
\left \{
\begin{array}{l l}
- \delta K, ~~~~~\text{horizontal plane} \\
+\delta K, ~~~~~\text{vertical plane}
\end{array}
\right .~,
\end{equation}
%------------------
where we have used the definition of the generalized $K$ defined in Eq.~(\ref{eq:K-generalized}).
Inserting $\delta K$ in Eq.~(\ref{eq:delta-q-int}) we find
%------------------ 
\begin{equation}
\begin{split}
\Delta Q_x = +\frac{1}{4\pi} \oint \beta_x(s) k_0(s) \mathrm{d}s \cdot\delta~, \\
\Delta Q_y = -\frac{1}{4\pi} \oint \beta_y(s) k_0(s) \mathrm{d}s \cdot\delta~,
\end{split}
\end{equation}
%------------------
yielding an expression for the natural chromaticity produced by a contribution of all quadrupole magnets in a circular accelerator
%------------------ 
\begin{equation}
\boxed{
\begin{split}
\xi_x = +\frac{1}{4\pi} \oint \beta_x(s) k_0(s) \mathrm{d}s \\
\xi_y = -\frac{1}{4\pi} \oint \beta_y(s) k_0(s) \mathrm{d}s
\end{split}
}~.
\end{equation}
%------------------
The impact of the individual quadrupole magnets on the chromaticity is weighted by the local beta function at the quadrupoles' positions.
Since in a linear lattice the horizontal beta function is always larger in a focusing quadrupole (where $k_0<0$) than in a defocusing quadrople (where $k_0 > 0$), the~horizontal natural chromaticity is always negative.
The same argument holds for the vertical chromaticity as well being always negative.
We thus have:
\begin{center}
\textbf{
The~horizontal and vertical natural chromaticities are always negative.\\
Stronger focusing leads to larger natural chromaticities.\\
Big circular accelerators will have a larger natural chromaticity than smaller rings.
}
\end{center}

It should be mentioned that both transverse planes have chromaticity while dispersion is limited to the bending plane only.

%---------------------------------------------------------------------------------------------------------------
\subsubsection{Chromaticity produced by sextupoles}
\label{sec:sextupoles-chromaticity}
%_______________________________________________________________________________
A beam moving on a dispersion orbit through a sextupole magnet is deflected by its transverse non-linear magnetic field $B_\text{sext}$:
%------------------ 
\begin{equation}
\frac{q}{p_0}\vec B_\text{sext} = m_0 x y \hat{e}_x + \frac{1}{2} m_0 \left( x^2 - y^2 \right) \hat{e}_y~,
\end{equation}
%------------------
where $m_0$ denotes the sextupole strength experienced by a particle with reference momentum $p_0$.
Building the first derivative and using Eq.~(\ref{eq:dispersion-orbit}) for the dispersion orbit, we get a position dependent focusing which reads using the ``generalized'' $K$
%------------------ 
\begin{eqnarray}
\delta K_x &=& -\delta k = \frac{q}{p_0}\cdot\frac{\partial B_{\text{sext},y}}{\partial x} = m_0\cdot x = m_0\cdot D\cdot\delta~, \\
\delta K_y &=& +\delta k = -\frac{q}{p_0}\cdot\frac{\partial B_{\text{sext},x}}{\partial y} = -m_0\cdot x = -m_0\cdot D\cdot\delta~.
\end{eqnarray}
%------------------
Again inserting these relations in Eq.~(\ref{eq:delta-q-int}) we obtain the additional contribution of sextupole magnets to the chromaticity:
%------------------ 
\begin{equation}
\boxed{
\begin{split}
\xi_x = +\frac{1}{4\pi} \oint \beta_x(s) \left[ k_0(s) + m_0(s) D(s)\right] \mathrm{d}s \\
\xi_y = -\frac{1}{4\pi} \oint \beta_y(s) \left[ k_0(s) + m_0(s) D(s)\right] \mathrm{d}s 
\end{split}
}~.
\label{eq:chromaticity-sextupoles}
\end{equation}
%------------------

%---------------------------------------------------------------------------------------------------------------
\subsubsection{Chromaticity correction}
\label{sec:chromaticity-correction}
\begin{figure}[ht]
\begin{center}
\includegraphics[width=9cm]{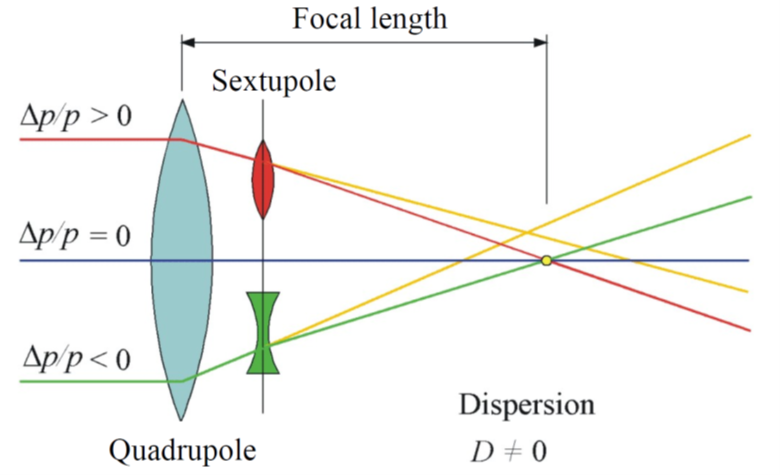}
\caption{Chromaticity correction by a sextupole magnet. Since the momentum deviation is linked with the displacement of the dispersion orbit, an appropriately chosen sextupole field can correct the momentum dependent focusing of the quadrupole magnet.}
\label{fig:chromaticity-correction}
\end{center}
\end{figure}

%_______________________________________________________________________________
The chromaticity describes the dependence of the betatron tune on the momentum deviation.
Since a real beam consists of many particles each of which having its on momentum deviation and thus an individual tune shift, the beam's momentum spread will lead to a certain tune spread.
In order to avoid a large tune spread, chromaticity has to be corrected by implementing additional sextupole magnets in the linear lattice.
From Eq.~(\ref{eq:chromaticity-sextupoles}) we obtain an important information:
\begin{itemize}
\item horizontal chormaticity is mainly affected by sextupoles positioned close to focusing quadrupoles,
\item vertical chormaticity is mainly affected by sextupoles positioned close to defocusing quadrupoles.
\end{itemize}
In addition, a non-vanishing dispersion at the positions of the sextupoles ist required.
Chosing appropriate sextupole strengths allows the compensation of both the horizontal and vertical natural chromaticity if two sets or ``families'' of sextupoles with different sextupole strengths are used: one positioned close to the focusing quadrupoles and one positioned close to the defocusing quadrupoles.
The underlying correction principle is illustrated in Fig.~\ref{fig:chromaticity-correction}.
%---------------------------------------------------------------------------------
%--------------------------------------------------------------------------------

Since sextupoles are non-linear elements, this correction will have an influence on the stability of the beam.
It will reduce the usable are in trace space to a maximum size determined by non-linear effects.
The maximum region of stability is called the {\em dynamic aperture} and has to be determined from tracking a distribution of particles through the accelerator over many thousands of turns.
Particles outside the stable dynamic aperture will get lost after many turns.
Since the required sextupole strength for chromaticity correction scales inversely proportional with the dispersion, these effects will become more severe in circular accelerators with low dispersion as 4th generation synchrotron radiation sources.
A~more detailed treatment requires a different mathematical approach using the Hamilton formalism and non-linear maps, which is out of the scope of this introductory course.

%===========================================================================
%===========================================================================
\section*{Additional literature}
%===========================================================================
%===========================================================================
J.~Le Duff, Proc. CAS-CERN Accelerator School: CAS Fifth General Accelerator Physics Course,
Jyväskylä, Finland, 7–18 September 1992 (CERN 94-01). Ed. S. Turner (CERN, Geneva, 1994), pp. 253–
311,
\url{https://doi.org/10.5170/CERN-1994-001}.\\[0.3cm]
S.~Y.~ Lee, \emph{Accelerator Physics}
(World Scientific, Singapore, 2019),\\
\url{https://doi.org/10.1080/00107514.2019.1641154}.\\[0.3cm]
H.~Wiedemann, \emph{Particle Accelerator Physics}
(Springer, Berlin, 2015),\\
\url{https://doi.org/10.1007/978-3-319-18317-6}.\\[0.3cm]
K.~Wille, \emph{The Physics of Particle Accelerators: An Introduction}
(Oxford University Press, Oxford, 2000),\\
\url{https://doi.org/10.1007/978-3-663-11850-3}.\\[0.3cm]
A.~Wolski, \emph{Beam Dynamics in High Energy Particle Accelerators}
(Imperial College Press, London, 2014),
\url{https://doi.org/10.1142/p899}.


\begin{thebibliography}{99}
\bibitem{Zolkin-PRAB}
T.~Zolkin, \emph{Phys.\ Rev.\ Accel.\ Beams} \textbf{20} (2017) 043501,\\
\url{https://doi.org/10.1103/PhysRevAccelBeams.20.043501}.

\bibitem{Rijk-CAS-warm}
G.~de~Rijk, \emph{Warm Magnets}, these proceedings.

\bibitem{Rijk-CAS-cold}
G.~de~Rijk, \emph{Superconducting Magnets}, these proceedings.

\bibitem{Courant-AG1}
E.D.~Courant, M.S.~Livingston, and H.S.~Snyder, \emph{Phys.\ Rev.} \textbf{88} (1952) 1190,\\
\url{https://doi.org/10.1103/PhysRev.88.1190}.

\bibitem{Courant-AG2}
E.D.~Courant, H.S.~Snyder, \emph{Ann. Phys.} \textbf{3} (1958) 1,\\
\url{https://doi.org/10.1016/0003-4916(58)90012-5}.

\bibitem{Hill}
G.W.~Hill, \emph{Acta Math.} \textbf{8} (1886) 1,\\
\url{https://doi.org/10.1007/BF02417081}.

\bibitem{Floquet}
G.~Floquet, \emph{Annales scientifiques de l'École Normale Superieure}, \textbf{2} Tome 12 (1883) 47,\\
\url{https://doi.org/10.24033/asens.220}.

\bibitem{Wille}
K.~Wille, \emph{The Physics of Particle Accelerators: An Introduction}
(Oxford University Press, Oxford, 2000),
\url{https://doi.org/10.1007/978-3-663-11850-3}.

\bibitem{Born}
M.~Born and E.~Wolf, \emph{Principles of Optics}
(Cambridge University Press, 1999),\\
\url{https://doi.org/10.1017/CBO9781139644181}.

\bibitem{Hinterberger}
F.~Hinterberger, \emph{Physik der Teilchenbeschleuniger und der Ionenoptik}--
(Springer, Berlin, 2008),
\url{https://doi.org/10.1007/978-3-540-75282-0}.

\end{thebibliography}
\end{document}